\documentclass[twocolumn]{aastex701}

\newcommand{\dz}{\mathrm{d}z}
\newcommand{\dr}{\mathrm{d}r}

\newcommand{\dE}{\mathrm{d}E}
\newcommand{\dd}{\mathrm{d}}
\newcommand{\Acu}{\mathscr{A}}
\newcommand{\Bcu}{\mathscr{B}}
\newcommand{\Ccu}{\mathscr{C}}
\newcommand{\Dcu}{\mathscr{D}}

\newcommand{\Fcu}{\mathscr{F}}
\newcommand{\Gcu}{\mathscr{G}}

\newcommand{\Qcu}{\mathscr{Q}}

\newcommand{\nb}{n_\mathrm{b}}

\newcommand{\el}{\mathrm{e}}
\newcommand{\Ye}{Y_\el}
\newcommand{\np}{n_\mathrm{p}}
\newcommand{\nn}{n_\mathrm{n}}
\newcommand{\mel}{m_\mathrm{e}}
\newcommand{\mpr}{m_\mathrm{p}}
\newcommand{\mn}{m_\mathrm{n}}
\newcommand{\mb}{m_\mathrm{b}}
\newcommand{\nue}{{\nu_\mathrm{e}}}
\newcommand{\nuea}{{\bar{\nu}_\mathrm{e}}}
\newcommand{\nua}{{\bar{\nu}}}
\newcommand{\nuae}{{\bar{\nu}_e}}

\def\lesssim{\mathrel{\hbox{\rlap{\hbox{\lower4pt\hbox{$\sim$}}}\hbox{$<$}}}}
\def\gtrsim{\mathrel{\hbox{\rlap{\hbox{\lower4pt\hbox{$\sim$}}}\hbox{$>$}}}}
\newcommand{\bea}{\begin{eqnarray}}
\newcommand{\eea}{\end{eqnarray}}

\newcommand{\drm}{\mathrm{d}}
\newcommand{\dt}{\mathrm{d}t}

\newcommand{\ds}{\mathrm{d}s}

\newcommand{\ddr}[1]{{\frac{\drm #1}{\dr}}}

\newcommand{\Ylep}{Y_{\mathrm{lep}}}
\newcommand{\nnue}{{n_\nue}}

\newcommand{\nnuae}{{n_\nuae}}

\usepackage{multirow}
\usepackage{mathtools}
\usepackage{mathrsfs}

\shorttitle{Accretion Regimes of Neutrino-Cooled Flows onto Black Holes}

\shortauthors{Hernández-Morales \& Siegel}
\graphicspath{{./}{./figures/}}
\begin{document}

\title{Accretion Regimes of Neutrino-Cooled Flows onto Black Holes}

\author[0000-0003-0491-6596]{Javiera Hernández-Morales}
\affiliation{Institute of Physics, University of Greifswald, D-17489 Greifswald, Germany}
\email{hernandezj@uni-greifswald.de}

\author[0000-0001-6374-6465]{Daniel M.~Siegel}
\affiliation{Institute of Physics, University of Greifswald, D-17489 Greifswald, Germany}
\affiliation{Department of Physics, University of Guelph, Guelph, ON, N1G 2W1, Canada}
\email{daniel.siegel@uni-greifswald.de}

\begin{abstract}
Neutrino-cooled accretion disks can form in the aftermath of neutron-star mergers as well as during the collapse of rapidly rotating massive stars (collapsars) and the accretion-induced collapse of rapidly rotating white dwarfs. Due to Pauli blocking as electrons become degenerate at sufficiently high accretion rates $\dot{M}$, the resulting `self-neutronization' of the dissociated accreting plasma makes these astrophysical systems promising sources of rapid neutron capture nucleosynthesis (the r-process). We present a one-dimensional general-relativistic, viscous-hydrodynamic model of neutrino-cooled accretion disks around black holes. With collapsars, super-collapsars and very massive star collapse in mind, we chart the composition of the accretion flow and systematically explore different radiatively efficient and inefficient accretion regimes with increasing $\dot M$, across a vast parameter space of $\dot{M}\sim 10^{-6}-10^6 M_\odot \,\text{s}^{-1}$, black hole masses of $M_\bullet\sim 1 - 10^4 M_\odot$ and dimensionless spins of $\chi_\bullet \in [0,1)$, as well as $\alpha$-viscosity values of $\alpha\sim 10^{-3}-1$. We show that these accretion regimes are separated by characteristic thresholds $\dot{M}_{\rm char}$ that follow power laws $\dot M_{\rm char}\propto M_{\bullet}^\alpha \alpha^\beta$ and that can be understood based on analytic approximations we derive. We find that outflows from such disks are promising sites of r-process nucleosynthesis up to $M_\bullet \lesssim 3000 M_\odot$. These give rise to lanthanide-bearing `red' super-kilonovae transients mostly for $M_\bullet \lesssim 200-500 M_\odot$ and lanthanide suppressed `blue' super-kilonovae for larger $M_\bullet$. Proton-rich outflows can develop specifically for large black hole masses ($M_\bullet \gtrsim 100 M_\odot$) in certain accretion regimes, which may give rise to proton-rich isotopes via the $\nu$p-process.

\end{abstract}






\section{Introduction}
\label{sec:introduction}

The astrophysical origin of roughly half of the heavy elements heavier than iron, those synthesized by the rapid neutron-capture process (the $r$-process; \citealt{cameron_nuclear_1957,burbidge_synthesis_1957}), represents an open, fundamental question in astrophysics and a topic of intense debate (see, e.g., \citealt{horowitz_r-process_2019,cowan_origin_2021,siegel_r-process_2022,arcones_origin_2022} for recent reviews). The detection of kilonovae \citep{li_transient_1998,metzger_electromagnetic_2010,barnes_effect_2013,metzger_kilonovae_2020} from the gravitational-wave event GW170817 \citep{abbott_multi-messenger_2017} as well as from several gamma-ray bursts (GRBs; \citealt{berger_r-process_2013,tanvir_kilonova_2013,rastinejad_uniform_2025}) establish neutron star mergers as a major if not dominant source of $r$-process elements in the Universe. 

Observations of $r$-process enhanced metal-poor stars in ultra-faint dwarf galaxies and isotopic analyses of meteorites as well as of sediments in the deep sea crust show that the main $r$-process must originate in rare events relative to core-collapse supernovae with large $r$-process yields per event both in early (i.e.~at low metallicity) and recent Galactic history \citep{ji_r-process_2016,hansen_r-process_2017,tsujimoto_enrichment_2017,beniamini_r-process_2016-1,wallner_abundance_2015,hotokezaka_short-lived_2015}. Systemic kicks of neutron star binaries, their (on average) large delay times relative to star formation with which they enrich the surrounding interstellar medium with nucleosynthesis products, as well as strong iron co-production of at least two supernovae per binary cast doubt on whether these are the dominant sources of $r$-process elements in low-metallicity environments, including (ultra-faint) dwarf galaxies and globular clusters \citep{bonetti_neutron_2019,zevin_can_2019,skuladottir_neutron-capture_2019,naidu_evidence_2022,simon_timing_2023,kirby_r-process_2023}. In particular, chemical evolution studies indicate that a high-rate, low-yield $r$-process source other than neutron-star mergers must exist\footnote{See \citet{thielemann_r-process_2020} for a compilation of caveats to this conclusion.}, which acts at low metallicities at a major if not dominant level \citep{matteucci_europium_2014,wehmeyer_galactic_2015,siegel_gw170817_2019,cote_galactic_2019,vandevoort_neutron_2020,kobayashi_can_2023,chen_inference_2025}, such as magnetorotational supernovae \citep{thompson_magnetar_2004,winteler_magnetorotationally_2012} or collapsars \citep{siegel_collapsars_2019,miller_full_2020,just_r-process_2022,issa_magnetically_2025}.

Accretion flows onto compact objects can reach sufficiently high temperatures $k_{\rm B}T\gtrsim\text{MeV}$ at which nuclei become dissociated and electron or positron capture onto individual protons and neutrons become fast compared to the accretion timescale; thus, neutrino emission is `ignited'. Furthermore, at sufficiently high accretion rates (rest-mass densities of the flow), electrons and positrons become degenerate and positron capture onto neutrons is suppressed by Pauli blocking. Consequently, the flow self-neutronizes, leading to an electron/proton-fraction $Y_e<0.5$ (\citealt{beloborodov_nuclear_2003}), a minimal requirement for $r$-process nucleosynthesis. Once in the degenerate regime, the flow maintains a high neutron richness ($Y_e\sim 0.1-0.2$) in a self-regulated process that maintains mild degeneracy $\eta_e=\mu_e/k_{\rm B} T\sim 1$, where $\mu_e$ is the electron chemical potential \citep{chen_neutrino-cooled_2007,siegel_three-dimensional_2017}. The minimum accretion rate $\dot{M}$ for neutrino emission to balance viscous heating and the onset of degeneracy (the `ignition threshold' $\dot{M}_{\rm ign}$) is estimated to be $\sim\!10^{-3}-0.1M_\odot\text{s}^{-1}$ for stellar-mass black holes, depending on the effective viscosity of the accretion flow and the black hole mass and spin \citep{chen_neutrino-cooled_2007,de_igniting_2021}. A similar value of $\gtrsim 10^{-4}M_\odot\text{s}^{-1}$ has been recently obtained for magnetized accretion onto a neutron star \citep{combi_magnetized_2025}. Once the accretion flow establishes full degeneracy, the accretion rate is typically large enough for the flow to become opaque to neutrinos (the opaque threshold $\dot{M}_\nu$). At even larger accretion rates and disk densities the accretion timescale becomes short compared to the vertical neutrino diffusion timescale, and neutrinos become trapped in the flow and get advected into the black hole (the `trapping threshold' $\dot{M}_{\nu-\text{trap}}$). 

Conditions for neutrino-cooled accretion can be realized in a number of astrophysical systems. The core-collapse of rapidly rotating, massive stars leading to a rapidly spinning black hole can give rise to an accretion disk that self-neutronizes \citep{metzger_conditions_2008-4,siegel_collapsars_2019,agarwal_ignition_2025}. Circularizing debris in neutron-star mergers forms a post-merger accretion disk around a remnant neutron star or black hole that can self-neutronize over viscous timescales \citep{metzger_time-dependent_2008,siegel_three-dimensional_2017,fernandez_long-term_2019,combi_jets_2023-2}. $r$-process elements synthesized in the outflows from post-merger accretion disks around a final black hole remnant represent the most natural explanation for the dominant ejecta source and red kilonova emission in GW170817 \citep{siegel_three-dimensional_2018,fernandez_long-term_2019,christie_role_2019,fujibayashi_mass_2020}, and earlier outflows from an accretion disk around a remnant neutron star likely generated the blue kilonova emission following GW170817 \citep{nedora_spiral-wave_2019,combi_jets_2023-2}. It has been conjectured that outflows from post-merger accretion disks also dominate the total ejecta of neutron-star mergers as a population \citep{siegel_r-process_2022}. Finally, conditions conducive for self-neutronization may be realized in the accretion-induced collapse (AIC) of rapidly rotating white dwarfs that yield a proto-neutron star surrounded by an accretion disk \citep{dessart_multi-dimensional_2006,metzger_neutron-rich_2009,batziou_nucleosynthesis_2024,cheong_gamma-ray_2025}. 

Though similarities exist, super-critical accretion onto neutron stars is fundamentally different from the black hole case due to non-absorbing inner boundary conditions at the stellar surface that alter energy and angular-momentum transport \citep{combi_magnetized_2025}. Additionally, if the neutron star is hot, as in neutron-star merger and AIC events, strong neutrino emission from the neutron star irradiates the accretion disk and its outflows and strongly impacts $r$-process nucleosynthesis \citep{lippuner_signatures_2017,fujibayashi_mass_2018,combi_jets_2023-2}. Here, we focus on neutrino-cooled accretion onto black holes, with a primary focus on collapsars, as we are interested in exploring the entire parameter space of black hole masses up to $\sim100-1000M_\odot$.

Collapsars represent the leading model for the central engine of long GRBs \citep{woosley_gamma-ray_1993,macfadyen_collapsars_1999}. Both theoretical arguments \citep{heger_presupernova_2005,gottlieb_shes_2024} and observed ejecta properties of GRB-accompanying Type Ic-BL supernovae \citep{woosley_supernova_2006} suggest that typical long GRBs result from the core collapse of progenitor stars with zero-age main sequence masses $M_{\rm ZAMS}\lesssim 40 M_\odot$ and helium cores at core-collapse of $\lesssim\!10 M_\odot$. The collapsing core may undergo a potentially brief phase of proto-neutron star formation that via luminous neutrino radiation and strong magnetic fields might power a (partial) prompt explosion, leading to the synthesis of $^{56}$Ni and the generation of a Type Ic-BL supernova. The collapsing stellar layers with $Y_e\simeq 0.5$ circularize in the equatorial plane at increasing radii with time \citep{zenati_nuclear_2020}, determined by angular momentum conservation. They feed an accretion disk that can maintain self-neutronization at $\dot{M}>\dot{M}_{\rm ign}$ during much of the early epoch in which the GRB jet is powered by a stellar-mass black hole of typically a few solar masses \citep{siegel_collapsars_2019,siegel_super-kilonovae_2022}. Magnetized accretion is likely required to enable an $r$-process in the disk outflows during this phase \citep{siegel_three-dimensional_2018,fernandez_long-term_2019,dean_collapsar_2024,issa_magnetically_2025,shibata_self-consistent_2025}. The detailed abundance distribution depends on the neutrinos' irradiation of the outflows, which increases $Y_e$ \citep{miller_full_2020,li_neutrino_2021}, and on the occurrence of neutrino fast flavor conversions, which tend to decrease $Y_e$ \citep{li_neutrino_2021}. Depending on the degree of mixing of the outflows with the infalling stellar material \citep{barnes_hydrodynamic_2023}, the high line expansion opacities of the lanthanide and actinide elements \citep{kasen_opacities_2013,tanaka_radiative_2013} trap photons in the expanding ejecta and thus redden the supernova emission; the synthesized heavy elements also contribute additional heating and thus excess emission, which eventually dominates on long timescales $\sim100$\,days after the explosion, likely when the supernova has reached the nebular phase \citep{siegel_collapsars_2019,barnes_signatures_2022}. First observational searches for such signatures based on photospheric emission are inconclusive \citep{anand_collapsars_2024,rastinejad_hubble_2024,blanchard_jwst_2024}.

Very massive stars $M_{\rm ZAMS}\gtrsim 40$ end their lives in pair-instability driven explosive oxygen burning (pair instability supernovae; PISNe) prior to reaching iron-core collapse, leaving no black hole remnant behind \citep{barkat_dynamics_1967,woosley_evolution_2002,umeda_nucleosynthesis_2002,woosley_models_2007,renzo_predictions_2020,farmer_mind_2019,woosley_pair-instability_2021}. At low metallicity, even more massive stars with $M_{\rm ZAMS}\gtrsim 250 M_\odot$ are predicted to evolve to helium core masses $\gtrsim\!130 M_\odot$ above this black hole pair-instability mass gap at core collapse. Strong photoinoization losses at the onset of explosive oxygen burning in such stars \citep{renzo_predictions_2020} or a radial gravitational instability \citep{chandrasekhar_dynamical_1964} for super massive stars $\sim\!10^4-10^5 M_\odot$ can drive very massive to super massive stars to collapse into black holes \citep{bond_evolution_1984,fryer_pair-instability_2001,nagele_stability_2022}. If such stars are rapidly rotating at collapse \citep{marchant_impact_2020,haemmerle_rotation_2018,kimura_3d_2023}, e.g.~due to continuous gas accretion
throughout their lifetime \citep[e.g.][]{cantiello_stellar_2021,jermyn_stellar_2021,dittmann_accretion_2021} or due to tidal locking in close binaries, they can give rise to collapsar-like accretion disks. Owing to the much larger helium core masses such collapsars can, in principle, synthesize much larger quantities of $\sim\!1-100 M_\odot$ of $r$-process material per event relative to ordinary collapsars and, following disk wind mass loss, give birth to massive black holes $M_{\bullet} \sim 10^{2}-10^{4}M_{\odot}$. The lower end of this mass distribution can populate the PISN black hole mass gap `from above', with rates compatible with recent detections by the gravitational-wave observatories LIGO-Virgo-Kagra (GW190521, GW231123; \citealt{abbott_gw190521_2020,the_ligo_scientific_collaboration_gw231123_2025,siegel_super-kilonovae_2022,agarwal_ignition_2025}). Estimating from an initial mass function, such events should be rare compared to ordinary collapsars, but they may represent a prolific source of $r$-process elements at low metallicity, as early as the first generation of stars (Population III stars).

Such ``super-collapsars'' may give rise to exceptionally energetic ($\sim\!10^{54}$\,erg) GRBs \citep{komissarov_supercollapsars_2010,meszaros_population_2010,suwa_can_2010,yoon_can_2015}. Since the binding energies of the progenitor envelopes by far exceed both the rotational energy of a proto-neutron star and its binding energy available to neutrino emission, they are unlikely associated with typical Type Ic-BL supernovae, yet $^{56}$Ni can be produced via jet-envelope interaction \citep{barnes_grb_2018}, an $\alpha$-process in disk outflows once self-neutronization ceases \citep{metzger_conditions_2008-4,siegel_collapsars_2019}, and nuclear burning of light nuclei in the accretion disk at later accretion epochs \citep{zenati_nuclear_2020}. Likely unaccompanied by promptly generated Ic-BL supernovae, the much larger quantities of neutron-rich disk winds compared to ordinary collapsars may give rise to ``super-kilonovae'' \citep{siegel_super-kilonovae_2022}, kilonova-like emission on timescales of weeks to months that can be detected by targeted follow-up of particularly energetic GRBs and by infrared surveys such as those planned for the Roman Space Telescope \citep{siegel_super-kilonovae_2022}.

Due to computational cost, recent multidimensional simulations in viscous hydrodynamics and magnetohydrodynamics of both isolated collapsar accretion disks \citep{siegel_collapsars_2019,miller_full_2020,agarwal_ignition_2025} and the global collapse process \citep{just_r-process_2022,dean_collapsar_2024,issa_magnetically_2025,shibata_self-consistent_2025} can only focus on at most a few models (parameter values) at a given time and have (with the exception of \citet{agarwal_ignition_2025}) thus far focused on accretion onto stellar-mass black holes and ordinary collapsars. \citet{agarwal_ignition_2025} find using three-dimensional, general-relativistic magnetohydrodynamic simulations of collapsar accretion disks with weak interactions that ignition of weak interactions and the onset of degeneracy as well as self-neutronization occurs for accretion onto black holes with an ignition accretion rate $\dot{M}_{\rm ign}\propto M_{\bullet}^{4/3}\alpha^{5/3}$ up to $M_\bullet \lesssim 3000 M_\odot$, where the disk temperature at $\dot{M}_{\rm ign}$ decreases below the proton-neutron mass difference such that electron and positron capture reactions are suppressed and the composition freezes. Yet self-neutronization can, in principle, still occur at higher accretion rates.

In this paper, we present a one-dimensional, stationary, general-relativistic viscous hydrodynamic model of neutrino-cooled accretion disks. With this computationally inexpensive framework, we explore and map out the entire parameter space of neutrino-cooled accretion onto black holes of mass $M_\bullet\sim 1-10^4 M_\odot$, dimensionless spin $\chi_\bullet\in[0,1)$, and effective $\alpha$-viscosity of $\alpha\sim10^{-3}-1$. We focus on the general physical processes that regulate the disk composition, with emphasis on neutronization as a necessary requirement for $r$-process nucleosynthesis in disk outflows. We identify various accretion regimes and provide analytic modeling to approximately delineate these regimes. At sufficiently high accretion rates and sufficiently large radii, the accretion disks become gravitationally unstable \citep{toomre_gravitational_1964,paczynski_model_1978} and fragmentation together with sufficient cooling \citep{gammie_nonlinear_2001} may lead to the formation of small, bound, neutron star like objects that can give rise to GRB variability, gravitational-wave emission, and additional electromagnetic transients \citep{piro_fragmentation_2007,metzger_fragmentation_2024,lerner_fragmentation_2025}. Since the stationary, non-self-gravitating model formally breaks down in this limit, here we focus on the gravitationally stable part of the parameter space. Parameter space exploration with this one-dimensional model, together with the analytic insight gained, can direct current and future sophisticated multi-dimensional simulations of collapsar and super-collapsar accretion processes and their observational signatures, nucleosynthesis analyses, and provides as a basis for interpreting simulation results.

This paper is structured as follows. In Section \ref{sec:model} and Appendix~\ref{app:disk_structure_equations} we describe the black hole-accretion disk model. We define characteristic accretion regimes and derive approximate analytical scaling relations for the characteristic accretion rates that delineate these regimes in Section \ref{sec:theoretical_scalings}. Section~\ref{sec:method_solution} discusses the numerical method used to solve the coupled set of disk equations. In Section~\ref{sec:results}, we present numerical results from a parameter space survey for the physics of the accretion flow, with a focus on disk composition, as well as numerical results for the characteristic accretion rates that delineate qualitatively different accretion regimes. In Section~\ref{sec:discussion}, we discuss our results in comparison to three-dimensional GRMHD simulations of collapsar accretion disks and comment on implications for ``super-kilonovae'' from collapsars that populate the PISN mass-gap as well as for exceptionally bright GRB events such as GRB221009A. Section~\ref{sec:conclusions} summarizes our conclusions. In most of this work, we use geometric units $G=c=1$, unless these constants explicitly appear.

\section{Accretion disk model}
\label{sec:model}

Accreting matter around a black hole is modeled using a variant of the one-dimensional, general-relativistic Novikov-Thorne model \citep{novikov_astrophysics_1973}, which has been extended and applied in many subsequent works (e.g.~\citealt{beloborodov_inertia_1997,beloborodov_accretion_1999,chen_neutrino-cooled_2007,kawanaka_neutrino-cooled_2007,janiuk_nucleosynthesis_2014,de_igniting_2021}). The basic underlying assumptions include (1) spacetime is given by a Kerr black hole and it is not affected by the mass of the disk nor by the mass accreted into the black hole, (2) the accretion process is steady, with a constant accretion rate (radially, and in time), (3) the disk is axisymmetric, located in the equatorial plane of the Kerr spacetime, and (4) the vertical extent is small compared to the radial extent ($z/r\ll 1)$. Disk quantities thus become time-independent, they can be azimuthally and vertically averaged and are thus only a function of radius $r$. 

Here, we closely follow the formulation by \citet{novikov_astrophysics_1973} and \citet{page_disk-accretion_1974} with corrections to zero order in $z/r$ regarding vertical pressure balance by \citet{riffert_relativistic_1995}, resulting in a model nearly identical to \citet{chen_neutrino-cooled_2007}, albeit with some modifications and additions. For completeness and reference, this section together with Appendix~\ref{app:disk_structure_equations} summarize the disk model in a self-consistent derivation. We start by defining the spacetime and deriving the structural equations of the disk from conservation laws (Section~\ref{sec:disk_structure}). Sections~\ref{sec:chemical_composition}, \ref{sec:thermodynamic_quantities}, and \ref{sec:cooling_emission} discuss the local microphysical state of the disk and neutrino cooling.

\subsection{Disk structure equations}
\label{sec:disk_structure}

We assume a non-self-gravitating disk and fix the spacetime to a Kerr black hole of mass $M_\bullet$, spin parameter $a_\bullet \in [0,M_\bullet)$ and dimensionless spin parameter $\chi_\bullet = a_\bullet/M_\bullet \in [0,1)$. We transform the metric $g_{\mu\nu}$ in Boyer-Lindquist coordinates $x^\mu=(t,r,\theta,\phi)$ with line element
\begin{align}
\label{eq:Kerr_metric_Boyer_Lindquist}
    \ds^2 = &-\dt^2 + \varrho^2 \left(\frac{\dr^2}{\Delta} + \dd\theta^2\right) + (r^2 + a_\bullet^2) \text{ sin}^2\theta \dd\phi^2 \nonumber \\
    &+\frac{2M_\bullet r}{\varrho^2}(a_\bullet \text{ sin}^2\theta \dd\phi - \dt)^2,
\end{align}
with $\Delta(r) = r^2 - 2M_\bullet r + a_\bullet^2$ and $\varrho^2(r,\theta) = r^2 + a_\bullet^2 \text{cos}^2\theta$, into cylindrical coordinates $x^\mu=(t,r,\phi,z)$, where $z = r\cos\theta$, and expand the metric coefficients up to second order $\mathcal{O}((z/r)^2)$ in height $z$, as required by the structure equation of vertical balance (Appendix \ref{app:disk_structure_equations}).

The accreting material inside the disk is characterized by its four-velocity $u^\alpha = (u^t, u^r,u^\phi, u^z)$. We henceforth exclusively refer to time-averaged quantities, assuming that local fluctuations due to fluid turbulence in the disk are adequately averaged over. According to the above model assumptions, we can then assume $u^\alpha = (u^t, u^r,u^\phi, 0)$ with nearly geodesic fluid motion, $|u^r|\ll |u^\phi|,\,|u^t|$. The local (proper) half-thickness of the disk is denoted by
\begin{equation}
    H(r) = \int_{0}^{z_{\rm d}(r)} \sqrt{g_{zz}}\,\dz, \label{eq:def_height_integration}
\end{equation}
where $z_{\rm d}(r)$ denotes the local disk coordinate height. The thin-disk approximation translates into $H/r \ll 1$, which is motivated in our case by the fact that we are interested in neutrino-dominated accretion flows (NDAFs), for which the accretion disk becomes geometrically thin ($H/r \lesssim 0.1$) as a result of neutrino cooling. The assumption is less well justified in the advection-dominated case, in which viscous heating is not balanced by neutrino cooling and the accretion disk remains geometrically thick ($H/r \gtrsim 0.4$). 

The disk structure equations follow from the conservation laws of baryon number, 
\begin{equation}
\label{eq:mass_conservation_covariant}
    \nabla_\mu(\rho u^\mu) = 0,
\end{equation}
lepton number,
\begin{equation}
\label{eq:lepton_number_conservation_covariant}
    \nabla_\mu(n_\text{lep} u^\mu) = R,
\end{equation}
and energy-momentum,
\begin{equation}\label{eq:energy_momentum_conservation_covariant}
    \nabla_\mu T^{\mu \nu} = 0.
\end{equation}
Here, $\rho = m_{\rm p}n_{\rm b}$ is the rest-mass density, with $m_{\rm p}$ the proton mass and $n_{\rm b}$ the baryon number density.\footnote{We choose the baryon mass to be the proton mass and neglect the proton-neutron rest-mass difference for purposes of computing the rest-mass density of baryonic material.} Furthermore, $ n_{\rm lep} = n_{e^-} - n_{e^+} + n_{\nu} - n_{\bar{\nu}}$ denotes the lepton number, with $n_{e^-}$, $n_{e^+}$, $n_{\nu}$ and $n_{\bar{\nu}}$ being the number densities of electrons, positrons, electron neutrinos, and electron antineutrinos, respectively. The sink term
\begin{equation}
    R = -\frac{1}{H}(\dot n_{\nue} - \dot n_{\nuae})
    \label{eq:lepton_number_source term}
\end{equation}
represents the lepton number per unit volume per unit time in the rest frame of the fluid radiated away from the accretion disk in the form of electron neutrinos and antineutrinos; $\dot n_\nu$ and $\dot n_\nua$ represent the number \emph{emission} rates per unit area of electron neutrinos and antineutrinos, respectively, hence the minus sign in Equation~\eqref{eq:lepton_number_source term} (see Section~\ref{sec:cooling_emission} for details). The energy-momentum tensor $T^{\mu \nu}$ is given by
\begin{equation}
\label{eq:stress_energy_tensor_definition}
    T^{\mu \nu} = \rho h u^\mu u^\nu + p g^{\mu \nu}  + t^{\mu \nu} + q^\mu  u^\nu + q^\nu u^\mu,
\end{equation}
with $h = 1 + (e + p)/\rho$ the relativistic specific enthalpy, $p$ the pressure, $e$ the internal energy density, $t^{\mu \nu}$ the viscous stress tensor, and $q^\mu$ the neutrino energy flux vector.

Integrating Equation~\eqref{eq:mass_conservation_covariant} over a spacetime volume between $r$ and $r+\Delta r$ and $t$ and $t+\Delta t$ yields
\begin{equation}
    \label{eq:mass_conservation}
    \dot{M} = -4\pi \sqrt{-g} \rho H u^r = \text{const.},
\end{equation}
i.e.~a constant accretion rate $\dot M$ as a function of radius.

With the help of Equation~\eqref{eq:mass_conservation_covariant}, the conservation of lepton number (Equation~\eqref{eq:lepton_number_conservation_covariant}), height-integrated according to Equation~\eqref{eq:def_height_integration}, implies (cf.~Appendix \ref{app:disk_structure_equations})
\begin{multline}
\label{eq:lepton_num_conservation}
    H u^r\left[\frac{\rho}{\mpr} \frac{\dd Y_e}{\dr} -\frac{n_{\nue} - n_{\nuae}}{\rho} \frac{\dd \rho}{\dr} + \frac{\dd (n_{\nue} - n_{\nuae})}{\dr}\right] \\ 
    = \dot{n}_{\nuae} - \dot{n}_{\nue},
\end{multline}
where $\Ye=\np/\nb$ is the electron or, more precisely, the proton-fraction. This expression is identical to Equation~(9) of \citet{chen_neutrino-cooled_2007}, except for a term proportional to the density gradient, which is absent in the formulation of \citet{chen_neutrino-cooled_2007}. In contrast to, e.g., radial pressure gradients, this term is, in principle, non-negligible given the set of assumptions made here. We find that it can lead to changes of individual disk quantities at the tens-of-percent level.

By projecting Equation~\eqref{eq:energy_momentum_conservation_covariant} onto $u^\mu$ and orthogonal to $u^\mu$ using the projection tensor $\mathcal{P}^{\alpha\beta}=u^\alpha u^\beta + g^{\alpha \beta}$, one obtains expressions for energy and momentum conservation, respectively. From the radial component ($i=1$) of $\mathcal{P}^{i\nu}\nabla_\mu T^{\mu}_{\phantom{\mu}\nu}=0$ (radial momentum balance), we obtain to lowest order in small quantities (cf.~Appendix \ref{app:disk_structure_equations})
\begin{equation}
\label{eq:angularvelocity}
    \Omega \equiv \frac{u^\phi}{u^t} = \frac{M_\bullet^{1/2}}{r^{3/2}}\frac{1}{\Bcu} = \left(\frac{r^{3/2}}{M_\bullet^{1/2}} + \chi_\bullet M_\bullet \right)^{-1},
\end{equation}
i.e.~the geodesic motion of fluid elements. In particular, radial pressure gradients have been neglected here. For the definition of the auxiliary functions $\Acu$, $\Bcu$, $\Ccu$, $\Dcu$, $J$, $S$, we refer to Appendix \ref{app:disk_structure_equations}.

From the $z$-component ($i=3$) of $\mathcal{P}^{i\nu}\nabla_\mu T^{\mu}_{\phantom{\mu}\nu}=0$ integrated over height (vertical momentum balance), we obtain to lowest order in small quantities (cf.~Appendix~\ref{app:disk_structure_equations})
\begin{equation}
\label{eq:verticalbalance}
    \left(\frac{H}{r}\right)^2 = \frac{p}{\rho}\frac{ r}{J M_\bullet},
\end{equation}
which differs from the expression of \citet{novikov_astrophysics_1973} and contains the full relativistic corrections to the tidal force in general relativity to zero order in $z/r$ as pointed out by \citet{riffert_relativistic_1995} (and used by \citealt{beloborodov_accretion_1999,chen_neutrino-cooled_2007}).

From the equation of energy conservation, $u^\nu\nabla_\mu T^{\mu}_{\phantom{\mu}\nu}=0$, integrated over height, we obtain to lowest order (cf.~Appendix \ref{app:disk_structure_equations})
\begin{equation}
\label{eq:energy_conservation}
    u^r \left(\frac{\dd (eH)}{\dr} - \frac{e+p}{\rho} \frac{\dd (\rho H)}{\dr}\right) = F^+ - F^-.
\end{equation}
Here,
\begin{equation}
    F^- \equiv q^z(z=H) = F_\nu + F_{\bar{\nu}} \label{eq:fminus_definition}
\end{equation}
is the $z$-component of the radiation flux $q^\mu$ in the rest frame of the fluid leaving the disk at $z=H$. The electron neutrino and antineutrino cooling (emission) fluxes $F_\nu$ and $F_{\bar{\nu}}$ depend directly on the microphysical state of the disk plasma. neutrino cooling is discussed in Section~\ref{sec:cooling_emission}. Together with the equation of azimuthal momentum balance (see below), one obtains for the viscous heating rate to lowest order (Appendix \ref{app:disk_structure_equations})
\begin{equation}\label{eq:heating_rate}
    F^+ \equiv - \int_0^H t^{\mu\nu}\sigma_{\mu\nu} \sqrt{g_{zz}}\dz = \frac{3\dot{M}M_\bullet}{8\pi r^3} \frac{\Dcu^2}{\Ccu^2}S, 
\end{equation}
where $\sigma_{\mu\nu}$ denotes the shear tensor. We note that the latter expression is identical to the expressions used by, e.g., \citet{beloborodov_inertia_1997}, \citet{beloborodov_accretion_1999}, and differs from that of \citet{chen_neutrino-cooled_2007} only by a factor of $h\simeq 1$ (see Appendix \ref{app:disk_structure_equations}). However, introducing the additional factor of the specific enthalpy renders the heating rate inconsistent with both the height-integrated energy balance and the height-integrated azimuthal momentum balance.

The $\phi$-component ($i=2$) of $\mathcal{P}^{i\nu}\nabla_\mu T^{\mu}_{\phantom{\mu}\nu}=0$ (azimuthal momentum balance) yields to lowest order in small quantities (cf.~Appendix~\ref{app:disk_structure_equations}) the identity
\begin{equation}
    t_{\hat r \hat \phi} = - r\rho u^r \frac{M_\bullet^{1/2}}{r^{3/2}} \frac{\Ccu^{1/2}\Qcu}{\Bcu\Dcu},
    \label{eq:t_rphi_azimuthalbalance}
\end{equation}
where $t_{\hat r \hat \phi}$ denotes the $r$-$\phi$ component of $t^{\mu\nu}$ in the comoving tetrad frame of the orbiting fluid.

We adopt the $\alpha$-viscosity ansatz \citep{shakura_black_1973}, i.e.~we assume that accretion is driven by viscous stress with a kinematic viscosity given by $\nu = \frac{2}{3}\alpha c_s H$, where $c_s\equiv\sqrt{p/\rho}$ is the isothermal sound speed and $\alpha \in (0,1)$ specifies the typical turbulent eddy velocity as a fraction of the local sound speed. Viscous stress in astrophysical accretion disks is generated by magnetohydrodynamic turbulence driven by the magnetorotational instability. Numerical simulations suggest typical effective values of $\alpha$ of $\alpha \approx 0.005-0.1$ (e.g., \citealt{balbus_instability_1998,hawley_dynamical_2002,penna_shakura-sunyaev_2013,de_igniting_2021,agarwal_ignition_2025}). 
This $\alpha$-viscosity ansatz leads to an expression of the $r$-$\phi$ component of the stress tensor in the co-moving tetrad, which is proportional to pressure (Appendix \ref{app:disk_structure_equations}),
\begin{equation}
    t_{\hat r \hat \phi} = \alpha p \frac{\Dcu}{\Ccu J^{1/2}}.\label{eq:alpha_viscosity_ansatz}
\end{equation}
The factor $\Dcu/(\Ccu J^{1/2})$ represents a relativistic correction relative to the Newtonian ansatz pointed out by \citet{riffert_relativistic_1995}, which makes the resulting expression consistent with relativistic vertical momentum balance.

Inserting the ansatz \eqref{eq:alpha_viscosity_ansatz} into Equation~\eqref{eq:t_rphi_azimuthalbalance} results in
\begin{equation}\label{eq:angularmomentum_conservation}
    u^r = -\frac{H}{r} \alpha c_s S^{-1}.
\end{equation}
This relation is identical to Equation~(3) of \citet{chen_neutrino-cooled_2007}.

\subsection{Chemical composition}
\label{sec:chemical_composition}

The plasma inside the disk is modeled as a mixture of electrons, positrons, protons, neutrons, $\alpha$-particles, neutrinos and antineutrinos in thermal equilibrium. Electrons and positrons are modeled as ideal Fermi gases and thus follow a Fermi-Dirac distribution. Their number densities are given by (cf.~Equation~\eqref{eq:n_e+-_derivation})
\begin{equation}\label{eq:electron_density}
    n_{e^{\pm}} = \frac{(\mel c)^3}{\pi^2\hbar^3}\int^\infty_0 f(\sqrt{\xi^2+1},\mp\eta_e,\theta)\xi^2\mathrm{d}\xi,
\end{equation}
with
\begin{equation}\label{eq:fermi_dirac_distr}
    f(E,\eta,\theta) = \frac{1}{\text{exp}(E/\theta - \eta)+1},
\end{equation}
where $E$ denotes energy in units of $m_e c^2$, $\eta_e = \mu_{e^-}/k_BT$ is the normalized chemical potential, and $\theta = k_BT/\mel c^2$ is the normalized temperature, $\hbar=h/2\pi$ the reduced Planck constant, $k_B$ the Boltzmann constant, and $m_e$ the electron mass. Furthermore, $\eta_{e^+} = - \eta_{e^-}$, which follows from the assumption that creation of pairs is in thermal equilibrium inside the disk, $e^- + e^+ \leftrightarrow 2 \gamma$. This yields for the chemical potentials $\mu_{e^-} + \mu_{e^+} = 2\mu_\gamma = 0$.

Charge neutrality of the disk plasma then dictates the proton number density to follow 

\begin{equation}\label{eq:np}
    \np = n_{e^-} - n_{e^+},
\end{equation}
and the neutron number density can be expressed as

\begin{equation}\label{eq:nn}
    \nn = \left(\frac{1-Y_e}{Y_e}\right) \np.
\end{equation}

The number density of $\alpha$-particles is given by
\begin{equation}\label{eq:nalpha}
    n_\alpha = \frac{1}{4} (\np + \nn)(1-X_{\rm f}).
\end{equation}
Here, $X_{\rm f}$ is the free-nucleon mass fraction, which, assuming nuclear statistical equilibrium, is given by (e.g., \citealt{meyer_r-_1994}),

\begin{multline}\label{eq:Xf} 
    \frac{3.780 \times 10^{16}}{(1.027 \times 10^{12})^{1/\theta}} \frac{ \theta^{9/4}}{\rho^{3/2}}= \\
    \left(Y_e - \frac{1-X_{\rm f}}{2}\right) \left(1-Y_e- \frac{1-X_{\rm f}}{2}\right)\frac{1}{\sqrt{1-X_{\rm f}}}.
\end{multline}

Consistent with the assumption in Section~\ref{sec:disk_structure}, we set the baryon mass $m_{\rm b}$ to the proton mass and express the plasma density as
\begin{equation}\label{eq:density}
    \rho = \mb \nb = \frac{\mpr \np}{\Ye}.
\end{equation}

Where the disk plasma becomes optically thick to electron neutrinos and antineutrinos, we assume thermodynamic equilibrium is established via
\begin{equation}\label{reaction:heatcool1}
    e^- + p \rightleftharpoons n + \nu,
\end{equation}
\begin{equation}\label{reaction:heatcool2}
    e^+ + n \rightleftharpoons p + \bar{\nu},
\end{equation}
such that they form an ideal Fermi gas with number densities given by (see Equation~\eqref{eq:number_density_nue-nuea})
\begin{equation}
\label{eq:n_nue-n_nuea}
    n_{\nue,\nuea}=\frac{(k_B T)^3}{2\pi^2 (\hbar c)^3} \int_0^\infty \drm\xi \,\frac{\xi^2}{\exp(\xi \mp \eta_{\nu})+1}.
\end{equation}

The degeneracy parameter of electron neutrinos $\eta_\nu \equiv \eta_\nue = -\eta_{\nuea}$ follows from chemical equilibrium of Equations~\eqref{reaction:heatcool1}--\eqref{reaction:heatcool2}, $\mu_{e^-} - \mu_\nue = \mu_{\rm n} - \mu_{\rm p}$, $-\mu_{e^+} + \mu_{\nuea} = \mu_{\rm n} - \mu_{\rm p}$. Further assuming that  $\nn$ and $\np$ follow Maxwellian distributions, one obtains (cf.~also \citealt{beloborodov_nuclear_2003})
\begin{equation}\label{eq:etaeetanu}
    \eta_e - \eta_\nu = \text{ln}\left(\frac{1-\Ye}{\Ye}\right) + \frac{Q}{\theta},
\end{equation}
where $Q\equiv(\mn-\mpr)/\mel \approx 2.531$. 

\subsection{Equation of state \& thermodynamic quantities}
\label{sec:thermodynamic_quantities}

We construct the equation of state in the independent thermodynamic variables $\rho$, $T$, and $Y_\el$ based on the Helmholtz free energy,
\begin{equation}
    \mathcal{F} = \mathcal{F}_{\rm b} + \mathcal{F}_\gamma + \mathcal{F}_{e^+} + \mathcal{F}_{e^-} + \mathcal{F}_\nu + \mathcal{F}_\nua,
\end{equation}
considering contributions from baryonic plasma (b), radiation ($\gamma$), electrons and positrons ($e^\pm$), and (electron) neutrinos and antineutrinos ($\nu$, $\nua$). All dependent thermodynamic variables such as pressure $p$, internal energy density $e$, and specific entropy $s$ with their partial components are consistently computed as partial derivatives of the free energy. We derive these expressions in Appendix~\ref{app:EOS}.

\subsection{Plasma cooling via neutrino emission}
\label{sec:cooling_emission}

The disk plasma cools through the fluxes of free-streaming neutrinos and antineutrinos, which, for one half disk, we define as
\begin{equation}\label{eq:cooling_transition}
    F_{\nu,\bar\nu} = (1-x_{\nu,\bar\nu})F_{\nu,\bar\nu}^{\text{transp}} + x_{\nu,\bar\nu} F_{\nu,\bar\nu}^{\text{opaque}}.
\end{equation}

The transition parameters
\begin{equation}\label{eq:xnu}
    x_{\nu,\bar\nu} = \text{exp}\left(-\frac{e^{\text{opaque}}_{\nu,\bar\nu}}{e^{\text{transp}}_{\nu,\bar\nu} }\right) 
\end{equation}
smoothly interpolate between the optically thin and optically thick limits using their respective energy densities, defined in Equations \eqref{eq:Unu_transparent} and \eqref{eq:Unu_opaque}.

In the optically thick limit, the fluxes can be written as
\begin{equation}
    F^{\text{opaque}}_{\nue,\nuae} = \frac{c}{\text{exp}(\tau_{{\nue}, \nuea})}e^{\text{opaque}}_{\nue,\nuae} .
\end{equation}
Here, $\tau_{\nue,\nuea}$ denote the total optical depths of electron neutrinos and electron antineutrinos, which we compute as described by \citet{chen_neutrino-cooled_2007}, considering the processes of neutrino absorption by nucleons, elastic neutrino-baryon scattering, and scattering of neutrinos off electrons and positrons.

In the free-streaming regime, the energy densities of the radiation fluxes are given by
\begin{equation}\label{eq:Unu_transparent}
    e^{\text{transp}}_{\nu,\nua} = 
    \frac{1}{c}F^{\text{transp}}_{\nue,\nuae} 
    + \frac{1}{c}F^{\text{transp}}_{\nu_{\mu},\nua_{\mu}}
    + \frac{1}{c}F^{\text{transp}}_{\nu_{\tau},\nua_{\tau}} .
\end{equation}
Here, $F^{\text{transp}}_{\nu}$ and $F^{\text{transp}}_\nua$ denote the vertically integrated, free-streaming neutrino fluxes.

The main source of optically thin cooling of the disk is via emission of electron neutrinos and antineutrinos as a result of electron and positron captures into nucleons (Equations~\eqref{reaction:heatcool1} and \eqref{reaction:heatcool2}). The free-streaming neutrino and antineutrino fluxes for these processes read \citep{shapiro_black_1983}
\begin{multline}\label{eq:Fnu_capture}
    F_{\nue (e^- \text{ capture})} = H (\np - 2n_\alpha) K_{e^\pm N} \mel c^2 \\
   \times \int^\infty_0 f(E+Q,\eta_e,\theta) (E+Q)^2  \sqrt{1-\frac{1}{(E+Q)^2}} E^3 \dE
\end{multline}
and
\begin{multline}\label{eq:Fantinu_capture}
     F_{\nuae (e^+ \text{ capture})} =  H(\nn - 2n_\alpha) K_{e^\pm N} \mel c^2 \\
     \times \int^\infty_{Q+1} f(E-Q,-\eta_e,\theta) (E-Q)^2  \sqrt{1-\frac{1}{(E-Q)^2}} E^3 \dE,
\end{multline}
where $K_{e^\pm N} = 6.5 \times 10^{-4} \text{s}^{-1}$, and $\np - 2 n_\alpha$ and $\nn - 2 n_\alpha$ denote the free proton and neutron number densities, respectively.

A secondary source of neutrino cooling is $e^\pm$-pair annihilation into all neutrino flavors,
\begin{equation}
    \label{reaction:eeannihilation}
    e^- + e^+ \longrightarrow \nu + \bar{\nu}.
\end{equation}
This process is most relevant in conditions of dominant radiation and/or relativistic $e^\pm$ pressure (Equations~\eqref{eq:Fnu_capture}, \eqref{eq:Fantinu_capture} and \eqref{eq:F_annihilation_nu}):
\begin{equation}
    \frac{F_{\nu(e^{\pm} \text{ annihilation})}}{F_{\nu (e^\pm \text{ capture})}} \propto \frac{T^9}{\rho T^6} \propto \frac{p_{\gamma,e^\pm}}{p_{\rm b}}.
\end{equation}
As we discuss in Section~\ref{sec:results_pressure}, radiation pressure can account for $\sim$30-60\% of the total pressure in the inner part of accretion disks at accretion rates when weak interactions become energetically significant (Figs.~\ref{fig:Pressure_conts_M3} and \ref{fig:Pressure_conts_M1000}), suggesting that this process could be relevant for at least part of the parameter space probed here. The total vertically integrated free-streaming neutrino flux of one half-disk for this process reads \citep{burrows_neutrino_2006}
\begin{multline}
\label{eq:F_annihilation_nu}
    F_{\nu(e^{\pm} \text{ annihilation})} = H K_{e^\pm\nu} \frac{m_e^7 c^8}{\hbar^6}\\
   \times  \int_0^\infty f(E,\eta_e,\theta) E^4dE \int_0^\infty f(E,-\eta_e,\theta) E^3dE,
\end{multline}
where $K_{e^\pm\nu} = K_{e^\pm\nue} \approx 3.4256 \times 10^{-37} \text{cm}^{3}\text{s}^{-1} $ for the production of electron neutrinos, and $K_{e^\pm\nu} = K_{e^\pm{\nu_{\tau,\mu}}}\approx 0.73564\times 10^{-37}\text{cm}^{3}\text{s}^{-1}$ for the production of muon or tau neutrinos. The definition of $F_{\bar{\nu}(e^{\pm} \text{ annihilation})}$ is analogous, with the replacement $\eta_e \rightarrow -\eta_e$. 

The total flux in the transparent regime is given by
\begin{equation}
\label{eq:F_transparent}
    F^{\text{transp}}_\nu = \frac{F_{\nue (e^- \text{ capture})}}{\text{exp}(\tau_{\nue})} + \frac{F_{\nue,\nu_\tau,\nu_\mu (e^{\pm}\text{ annihilation})}}{\text{exp}(\tau_{\nue,\nu_\tau,\nu_\mu})},
\end{equation}
with an analogous expression for $F^{\text{transp}}_{\bar{\nu}}$. We consider effective total optical depths $\tau_{\nu_\tau} = \tau_{\nu_\mu} = 0$ for the muon and tau neutrino flavors.

While this concludes the calculation of the cooling terms for the disk structure equations (Section~\ref{sec:disk_structure}), we compute two additional quantities of interest---the number flux of electron neutrinos and antineutrinos and their mean energies.

In the transparent regime, the number fluxes of neutrinos and antineutrinos produced by electron and positron capture are computed in an analogous way to the fluxes discussed above,
\begin{multline}\label{eq:Ndotnu_capture}
    {\dot{n}}_{\nue (e^- \text{ capture})} = H(\np - 2n_\alpha) K_{e^\pm N}  \\
    \times \int^\infty_0 f(E+Q,\eta_e,\theta) (E+Q)^2  \sqrt{1-\frac{1}{(E+Q)^2}} E^2 \dE,
\end{multline}
\begin{multline}\label{eq:Ndotantinu_capture}
     {\dot{n}}_{\nuae (e^+ \text{ capture})} =  H(\nn - 2n_\alpha) K_{e^\pm N} \\
     \times \int^\infty_{Q+1} f(E-Q,-\eta_e,\theta) (E-Q)^2  \sqrt{1-\frac{1}{(E-Q)^2}} E^2 \dE.
\end{multline}

The mean energies of neutrinos and antineutrinos are then given by
\begin{equation}\label{eq:meanenergy_capture}
    \langle E_{\nue,\nuae (e^\pm \text{ capture})} \rangle = \frac{F_{\nue,\nuae (e^\pm \text{ capture})}}{{\dot{n}}_{\nue,\nuae (e^\pm \text{ capture})}},
\end{equation}
where $F_{\nue,\nuae (e^\pm \text{ capture})}$ corresponds to the associated flux from Equation~\eqref{eq:Fnu_capture} or \eqref{eq:Fantinu_capture}. For electron neutrinos emitted due to electron-positron annihilation, the number flux and mean energy of neutrinos are given by 
\begin{multline}\label{eq:Ndotnu_annihilation}
    \dot{n}_{\nue (e^{\pm} \text{ annihilation})} = H K_{e^\pm\nue} \left(\frac{m_e c}{\hbar}\right)^6\\
   \times \int_0^\infty f(E,\eta_e,\theta) E^3dE \int_0^\infty f(E,-\eta_e,\theta) E^3dE,
\end{multline}
and 
\begin{equation}\label{eq:meanenergy_annihilation}
    \langle E_{\nue (e^\pm\text{ annihilation})} \rangle = \frac{F_{\nue (e^\pm \text{ annihilation})}}{{\dot{n}}_{\nue (e^\pm \text{ annihilation})}},
\end{equation}
respectively. Analogous expressions hold for antineutrinos, using the replacement $\eta_e \rightarrow -\eta_e$. Employing these definitions, it is straightforward to check that 
\begin{equation}
    \dot{n}_{\nue (e^{\pm} \text{ annihilation})} = \dot{n}_{\nuae (e^{\pm} \text{ annihilation})}
\end{equation}
as expected, since neutrinos and antineutrinos are emitted in pairs.

The total electron neutrino number density in the transparent regime is then given by the sum of the contributions of both processes,
\begin{equation}\label{eq:n_dot_transparent}
    \dot{n}^{\text{transp}}_\nue = \frac{\dot{n}_{\nue (e^- \text{ capture})} + \dot{n}_ {\nue (e^{\pm}\text{ annihilation})}}{\text{exp}(\tau_{\nue})}.
\end{equation}

In the opaque regime, we consider the neutrino number densities and average energy given in Equation~\eqref{eq:number_density_nue-nuea} to obtain
\begin{equation}\label{eq:n_dot_opaque}
    \dot{n}^{\text{opaque}}_\nue = \frac{c}{\text{exp}(\tau_\nue)}n_\nue.
\end{equation}
and
\begin{equation}\label{eq:meanenergy_thermal}
    \langle E_{\nue (\rm{thermalized-} \nue)} \rangle = \frac{F_{\nue}^{\rm opaque}}{{\dot{n}}_{\nue}^{\rm opaque}} = \frac{e_\nue^{\rm opaque}}{n_\nue^{\rm opaque}}.
\end{equation}

Finally, we ensure a smooth transition between the different regimes by using a formula analogous to Equation~\eqref{eq:cooling_transition}, 
\begin{equation}
    \label{eq:n_dot_transition}
    \dot{n}_\nu = (1-x_\nu)\dot{n}^{\text{transp}}_\nue + x_\nu \dot{n}^{\text{opaque}}_\nue.
\end{equation}
Analogous expressions hold for electron antineutrinos.

\section{Approximate analytic scaling relations} 
\label{sec:theoretical_scalings}

Following \citet{chen_neutrino-cooled_2007}, \citet{metzger_conditions_2008-4,metzger_time-dependent_2008}, \citet{siegel_collapsars_2019}, \citet{de_igniting_2021}, \citet{siegel_r-process_2022}, \citet{agarwal_ignition_2025}, we define various physical regimes of neutrino-cooled accretion. Below we derive approximate scaling relations that prove useful in the analysis of numerical results (Section~\ref{sec:results}), based on the relativistic disk model laid out in Section~\ref{sec:model}. 

\subsection{Ignition threshold}
\label{sec:ignition_threshold}

Whereas photons are trapped in the accretion flow across all regimes of interest in this paper, neutrinos may efficiently cool the plasma once it dissociates into individual nucleons and becomes sufficiently dense. We refer to the physical state in which cooling through weak interactions, predominantly electron and positron capture ($e^- + p \rightarrow n + \nu_e$; $e^+ + n \rightarrow p + \bar{\nu}_e$), balance viscous heating of the accretion flow as the ``ignition threshold''.

The local disk density is given by (combine Equations~\eqref{eq:mass_conservation}, \eqref{eq:verticalbalance}, and \eqref{eq:angularmomentum_conservation})

\begin{equation}
\label{eq:ign_thresh_step1.5}
     \rho = \frac{S}{4\pi J^{1/2}} \left( \frac{H}{r} \right)^{-3}    r^{-\frac{3}{2}} \alpha^{-1} M_\bullet^{-\frac{1}{2}} \dot{M}  .
\end{equation}

Assuming that $p\propto T^4$, the disk temperature is obtained from vertical equilibrium (Equation~\eqref{eq:verticalbalance}):
\begin{equation}
    \label{eq:ign_thresh_step3}
    T \propto J^\frac{1}{4}\left( \frac{H}{r} \right)^\frac{1}{2} r^{-\frac{1}{4}} \rho^\frac{1}{4}  M_\bullet^\frac{1}{4}.
\end{equation}

At the ignition threshold, radiation pressure (Equation \eqref{eq:pressure_photons}) and relativistic $e^\pm$ pressure (cf.~Equation~\eqref{eq:pressure_neutrinos} for the ultra-relativistic case) dominate (Section~\ref{sec:results_pressure}), making $p\propto T^4$ a well justified assumption.

Furthermore, we make the assumption that electron and positron capture are the dominant cooling channels at the ignition threshold, and that the flow is still optically thin to neutrinos: $F^- \propto H \rho T^6$ (see Equations~\eqref{eq:Fnu_capture}, \eqref{eq:Fantinu_capture}). Using Equation~\eqref{eq:ign_thresh_step3} to replace the temperature and Equation~\eqref{eq:ign_thresh_step1.5} to replace the density, we obtain
\begin{equation}
    \label{eq:ign_thresh_step5}
    F^- \propto J^\frac{1}{4}S^{\frac{5}{2}}\left( \frac{H}{r} \right)^{-\frac{7}{2}} r^{-\frac{17}{4}}\alpha^{-\frac{5}{2}} M_\bullet^\frac{1}{4} \dot{M}^{\frac{5}{2}} .
\end{equation}

Combining the last expression with Equation~\eqref{eq:heating_rate} for viscous heating $F^+$, we find
\begin{equation}\label{eq:ign_thresh_step6}
    \frac{F^-}{F^+} \propto J^\frac{1}{4}S^{\frac{3}{2}}\frac{\Ccu^2}{\Dcu^2}\left( \frac{H}{r} \right)^{-\frac{7}{2}} r^{-\frac{5}{4}} \alpha^{-\frac{5}{2}}  M_\bullet^{-\frac{3}{4}}\dot{M}^{\frac{3}{2}}.
\end{equation}

We define the ignition threshold by the condition $F^-/F^+ = 0.5=\text{const.}$, at which $H/r \approx \text{const.}$ (since $H/r\sim \mathcal{O}(1)$ when advective cooling dominates), and find that the radius interior to which neutrino cooling  becomes significant (the ``ignition radius'') scales as
\begin{equation}
    \label{eq:ign_thresh_step8}
    r_{\text{ign}} \propto  J^{\frac{1}{5}} S^{\frac{6}{5}} \frac{\Ccu^{\frac{8}{5}}}{\Dcu^{\frac{8}{5}}} \alpha^{-2} M_\bullet^{-\frac{3}{5}}   \dot{M}^{\frac{6}{5}}.
\end{equation}

Neglecting the general-relativistic structure functions and normalizing to the gravitational radius, one approximately has
\begin{equation}
\label{eq:ign_thresh_step9}
    \frac{r_{\text{ign}}}{r_{\rm g}} \propto \alpha^{-2} M_\bullet^{-\frac{8}{5}}\dot{M}^{\frac{6}{5}}.
\end{equation}
The neutrino-cooled region of a disk thus decreases in radial size with increasing viscosity or black hole mass, whereas it grows in size with increasing accretion rate. 

The requirement for a neutrino-cooled region of the accretion flow to exist ($r_{\rm ign} \ge r_{\rm ISCO}$) translates into a minimum accretion rate ($\dot{M}\ge \dot{M}_{\rm ign}$; the ``ignition threshold'' accretion rate),
\begin{equation}
\label{eq:ign_scaling}
    \dot{M}_{\text{ign}} \propto \alpha^\frac{5}{3}M_\bullet^{\frac{4}{3}},
\end{equation}
where we have used $r_{\rm ISCO}\propto r_{\rm g}\propto M_\bullet$ in Equation~\eqref{eq:ign_thresh_step9}. This result has been derived in a similar way by \citet{de_igniting_2021} and agrees with the analogous Newtonian result \citep{siegel_r-process_2022,agarwal_ignition_2025}.

The ratio of radiation and/or relativistic $e^\pm$ pressure to baryon pressure at the ignition threshold $(r=r_{\rm ign})$ increases moderately as a function of black hole mass and $\alpha$-viscosity at a given accretion rate $\dot M$ (using Equations~\eqref{eq:ign_thresh_step1.5} and \eqref{eq:ign_thresh_step3}):
\begin{eqnarray}
\label{eq:pressure_ratio_rign}
    \frac{p_\gamma}{p_{\rm b}}\bigg\vert_{r_{\rm ign}} \propto \frac{T^3}{\rho}\bigg\vert_{r_{\rm ign}} \propto 
        \alpha M_\bullet^{\frac{11}{10}} \dot{M}^{-\frac{7}{10}}.
\end{eqnarray}
At $\dot{M}=\dot{M}_{\rm ign}$, one finds with Equation~\eqref{eq:ign_scaling}
\begin{eqnarray}
\label{eq:pressure_ratio_Mdotign}
    \frac{p_\gamma}{p_{\rm b}}\bigg\vert_{\dot{M}_{\rm ign}} \propto \frac{T^3}{\rho}\bigg\vert_{\dot{M}_{\rm ign}} \propto \alpha^{-\frac{1}{6}} M_\bullet^{\frac{1}{6}}.
\end{eqnarray}

\subsection{Neutrino opaque threshold}
\label{sec:opaque_threshold}

Above a critical density, the accretion flow becomes opaque to neutrinos and optically thin cooling transitions to optically thick cooling. The vertical optical depth for electron neutrinos or antineutrinos through the disk with opacity $\kappa$ scales as
\begin{equation}
\label{eq:opaque_thresh_step2}
    \tau_{\nu} \propto \rho \kappa H \propto \rho T^2 H,
\end{equation}
where we have used that the cross section depends quadratically on temperature, $\sigma \propto T^2$, for all relevant sources of opacity (absorptions by nucleons, elastic scattering off baryons, scattering off electrons and positrons). 

As discussed in Secs.~\ref{sec:results_pressure} (Figs.~\ref{fig:Pressure_conts_M3} and \ref{fig:Pressure_conts_M1000}), the pressure at the transition radius $r_\nu$ to the optically thick regime is dominated by baryon pressure for sufficiently small black hole masses $M_\bullet \lesssim 100 M_\odot$ and sufficiently small values of the viscosity $\alpha \lesssim 0.1$, in which case $p\propto p_\text{b} \propto \rho T$ in Equation~\eqref{eq:verticalbalance} leads to
\begin{equation}
\label{eq:opaque_thresh_step1}
    T \propto J \left( \frac{H}{r} \right)^2  r^{-1}  M_\bullet,
\end{equation}
in lieu of Equation~\eqref{eq:ign_thresh_step3}. However, for large black hole masses and values of $\alpha$, Equation~\eqref{eq:ign_thresh_step3} applies. Combining Equations~\eqref{eq:ign_thresh_step1.5}, \eqref{eq:ign_thresh_step3}, \eqref{eq:opaque_thresh_step2}, and \eqref{eq:opaque_thresh_step1}, we obtain
\begin{equation}\label{eq:opaque_thresh_step4}
    \tau_{\nu_e} \propto \left\{\begin{array}{ll}
       J^{\frac{3}{2}}S\left(\frac{H}{r}\right)^2 r^{-\frac{5}{2}}  \alpha^{-1} M_\bullet^\frac{3}{2} \dot{M}, & p\propto p_{\rm b}\\
        J^{-\frac{1}{4}}S^{\frac{3}{2}}\left(\frac{H}{r}\right)^{-\frac{5}{2}} r^{-\frac{7}{4}}  \alpha^{-\frac{3}{2}} M_\bullet^{-\frac{1}{4}} \dot{M}^{\frac{3}{2}}, & p\propto p_{\gamma,e^\pm}
        \end{array}\right. .
\end{equation}
The accretion flow thus becomes opaque ($\tau_{\nu_e} \ge 1$) to neutrinos interior to a radius
\begin{equation}
    \label{eq:opaque_thresh_step5}
    r_{\nu} \propto \left\{\begin{array}{ll}
        J^{\frac{3}{5}}S^{\frac{2}{5}} \left(\frac{H}{r}\right)^{\frac{4}{5}} \alpha^{-\frac{2}{5}} {M_\bullet^{\frac{3}{5}}} \dot{M}^{\frac{2}{5}} , & p\propto p_{\rm b}\\
        J^{-\frac{1}{7}}S^{\frac{6}{7}}\left(\frac{H}{r}\right)^{-\frac{10}{7}}  \alpha^{-\frac{6}{7}} M_\bullet^{-\frac{1}{7}} \dot{M}^{\frac{6}{7}}, & p\propto p_{\gamma,e^\pm}
    \end{array}\right.,
\end{equation}
which upon setting $H/r \sim \text{const.}$ as above and neglecting the general-relativistic structure functions $J$ and $S$, yields the approximate result
\begin{equation}
    \label{eq:opaque_thresh_step6}
    \frac{r_{\nu}}{r_{\rm g}} \propto \left\{\begin{array}{ll}
        \alpha^{-\frac{2}{5}} {M_\bullet^{-\frac{2}{5}}} \dot{M}^{\frac{2}{5}}, &  p\propto p_{\rm b}\\
        \alpha^{-\frac{6}{7}} M_\bullet^{-\frac{8}{7}} \dot{M}^{\frac{6}{7}}, & p\propto p_{\gamma,e^\pm}
    \end{array}\right. .
\end{equation}

For such an optically thick region of the accretion flow to exist ($r_\nu \ge r_{\rm ISCO}$), the accretion rate must be above the critical value
\begin{equation}
    \label{eq:Mdotnu_scaling}
    \dot{M}_{\nu} \propto \left\{\begin{array}{ll}
        \alpha M_\bullet, &  p\propto p_{\rm b}\\
        \alpha M_\bullet^{\frac{4}{3}}, & p\propto p_{\gamma,e^\pm}
    \end{array}\right.,
\end{equation}
which one obtains from Equation~\eqref{eq:opaque_thresh_step6} using $r_{\rm ISCO}\propto r_{\rm g}\propto M_\bullet$ as above. Analogous relations $r_{\nua}$ and $\dot{M}_{\nua}$ hold for electron antineutrinos, which only differ in their absolute normalization.

We expect the transition between the two pressure regimes considered above to be gradual as a function of the free parameters of the problem. Indeed, from the above expressions (Equation~\eqref{eq:ign_thresh_step1.5}, \eqref{eq:ign_thresh_step3}, \eqref{eq:opaque_thresh_step1}), we find that the ratio $p_{\gamma,e^\pm}/p_{\rm b}$ at the neutrino opaque threshold $(r=r_{\nu})$ increases moderately as a function of black hole mass and $\alpha$-viscosity at a given accretion rate $\dot M$:
\begin{eqnarray}
\label{eq:pressure_ratio_rnu}
    \frac{p_\gamma}{p_{\rm b}}\bigg\vert_{r_\nu} \propto \frac{T^3}{\rho}\bigg\vert_{r_\nu} \propto \left\{\begin{array}{ll}
        \alpha^{\frac{8}{5}} M_\bullet^{\frac{13}{5}} \dot{M}^{-\frac{8}{5}}, &  p\propto p_{\rm b}\\
        \alpha^{\frac{4}{7}} M_\bullet^{\frac{13}{14}} \dot{M}^{-\frac{4}{7}}, & p\propto p_{\gamma,e^\pm}
    \end{array}\right. .
\end{eqnarray}
At $\dot{M}=\dot{M}_{\nu}$, one finds with Equation~\eqref{eq:Mdotnu_scaling}
\begin{eqnarray}
\label{eq:pressure_ratio_Mdotnu}
    \frac{p_\gamma}{p_{\rm b}}\bigg\vert_{\dot{M}_\nu} \propto \frac{T^3}{\rho}\bigg\vert_{\dot{M}_\nu} \propto \left\{\begin{array}{ll}
        \alpha^{0} M_\bullet, &  p\propto p_{\rm b}\\
        \alpha^{0} M_\bullet^{\frac{1}{6}}, & p\propto p_{\gamma,e^\pm}
    \end{array}\right. .
\end{eqnarray}

\subsection{Neutrino trapping threshold}
\label{sec:trapped_threshold}

Another characteristic accretion regime emerges when neutrinos become trapped in the accretion flow and are advected into the black hole. This occurs when the local vertical diffusion time of neutrinos through the disk, $t_{\text{diff}} \sim (H/c) \tau_{\nu_e}$, becomes greater than the local accretion time
\begin{equation}
    t_{\text{acc}} = \int_0^r \frac{\drm l}{V},
\end{equation}
where $\drm l = \sqrt{g_{rr}}\dr$ is the proper radial distance element and $V=\tilde{v}/\sqrt{1+\tilde{v}^2}$, with $\tilde{v}^2\equiv u^r u_r = g_{rr}|u^r|^2$, is the radial fluid velocity in a co-rotating frame at radius $r$. Using $|u^r|\ll |u^\phi|\sim 1$, the above condition translates into
\begin{equation}
    \tau_\nu \gtrsim \frac{c}{|u^r|}\left(\frac{H}{r}\right)^{-1}.
\end{equation}
Making use of Equation~\eqref{eq:mass_conservation}, we find that trapping occurs within a radius
\begin{equation}
    \label{eq:trapping_radius}
    r_{\nu-\text{trap}} \propto \left\{\begin{array}{ll}
        J^{\frac{2}{3}} \left(\frac{H}{r}\right)^{\frac{5}{3}} \alpha^0 {M_\bullet^{\frac{2}{3}}} \dot{M}^{\frac{1}{3}} , & p\propto p_{\rm b}\\
        J^{\frac{1}{9}}S^{\frac{2}{9}}\left(\frac{H}{r}\right)^{\frac{2}{9}}  \alpha^{-\frac{2}{9}} M_\bullet^{\frac{1}{9}} \dot{M}^{\frac{2}{3}}, & p\propto p_{\gamma,e^\pm}
    \end{array}\right..
\end{equation}
Assuming $H/r\sim\text{const.}$ as before and dropping the general-relativistic structure functions, one finds approximately
\begin{equation}
    \label{eq:trapping_radius_rg}
    \frac{r_{\nu-\text{trap}}}{r_{\rm g}} \propto \left\{\begin{array}{ll}
        {\alpha^0 M_\bullet^{-\frac{1}{3}}} \dot{M}^{\frac{1}{3}} , & p\propto p_{\rm b}\\
        \alpha^{-\frac{2}{9}} M_\bullet^{-\frac{8}{9}} \dot{M}^{\frac{2}{3}}, & p\propto p_{\gamma,e^\pm}
    \end{array}\right..
\end{equation}
From this expression we deduce that for a neutrino trapped regime to exist ($r_{\nu-\text{trap}}\ge r_{\rm ISCO}$), the accretion rate must exceed the critical value
\begin{equation}
    \label{eq:Mdotnu-trap_scaling}
    \dot{M}_{\nu-\text{trap}} \propto \left\{\begin{array}{ll}
        \alpha^0 M_\bullet, &  p\propto p_{\rm b}\\
        \alpha^{\frac{1}{3}} M_\bullet^{\frac{4}{3}}, & p\propto p_{\gamma,e^\pm}
    \end{array}\right.,
\end{equation}
where we have again used the fact that $r_{\rm ISCO}\propto r_{\rm g}\propto M_\bullet$. Analogous relations $r_{\nua-\text{trap}}$ and $\dot{M}_{\nua-\text{trap}}$ hold for electron antineutrinos, which only differ in their absolute normalization.

The ratio of radiation and/or relativistic $e^\pm$ pressure to baryon pressure at the neutrino trapping threshold $(r=r_{\nu-\text{trap}})$ increases moderately as a function of black hole mass and $\alpha$-viscosity at a given accretion rate $\dot M$ (combining Equations~\eqref{eq:ign_thresh_step1.5}, \eqref{eq:ign_thresh_step3}, \eqref{eq:opaque_thresh_step1}, and \eqref{eq:trapping_radius}):
\begin{eqnarray}
\label{eq:pressure_ratio_rtrap}
    \frac{p_\gamma}{p_{\rm b}}\bigg\vert_{r_{\nu-\text{trap}}}\mskip-10mu \propto \frac{T^3}{\rho}\bigg\vert_{r_{\nu-\text{trap}}} \mskip-10mu\propto \left\{\begin{array}{ll}
        \alpha M_\bullet^{\frac{5}{2}} \dot{M}^{-\frac{3}{2}}, &  p\propto p_{\rm b}\\
        \alpha^{\frac{1}{3}} M_\bullet^{\frac{5}{6}} \dot{M}^{-\frac{1}{2}}, & p\propto p_{\gamma,e^\pm}
    \end{array}\right. .
\end{eqnarray}
At $\dot{M}=\dot{M}_{\nu-\text{trap}}$, we find with Equation~\eqref{eq:Mdotnu-trap_scaling}
\begin{eqnarray}
\label{eq:pressure_ratio_Mdotnu-trap}
    \frac{p_\gamma}{p_{\rm b}}\bigg\vert_{\dot{M}_{\nu-\text{trap}}} \mskip-10mu\propto  \frac{T^3}{\rho}\bigg\vert_{\dot{M}_{\nu-\text{trap}}} \mskip-10mu \propto \left\{\begin{array}{ll}
        \alpha M_\bullet, &  p \propto  p_{\rm b}\\
        \alpha^{\frac{1}{6}} M_\bullet^{\frac{1}{6}}, & p\propto p_{\gamma,e^\pm}
    \end{array}\right. .
\end{eqnarray}

\section{Numerical disk model solver}
\label{sec:method_solution}

The structure and thermodynamic state of the accretion disk at a given radius can be entirely determined from the local values of the three independent quantities ${Y_e}$, $\theta$ and ${\eta_e}$. Employing these as independent variables to integrate the disk structure equations has the benefit that $\theta$ and $\eta_e$ are direct arguments of the Fermi function and that all three quantities are of order unity in the regimes of interest to be explored here numerically. 

Due to the complications arising from various Fermi integrals, it is not possible to integrate the disk structure equations explicitly with standard techniques. Instead, similar to but somewhat different from \citet{chen_neutrino-cooled_2007}, we integrate Equations~\eqref{eq:lepton_num_conservation} and \eqref{eq:energy_conservation}, together with the constraint equation
\begin{equation}
\label{eq:pressure_analytic_expression}
    p(r) = \left(\frac{G\dot{M}M_\bullet}{4\pi \alpha} \frac{S(r)J(r)}{r^3}\right)^{2/3} \rho(r)^{1/3},
\end{equation}
which is obtained by combining Equations~\eqref{eq:mass_conservation}, \eqref{eq:verticalbalance} and \eqref{eq:angularmomentum_conservation}. 

Choosing suitable boundary conditions at the outermost radius (see below), we discretize the radial domain $[r_\text{in},r_\text{out}]$ with $N+1$ points \{$r_N=r_{\rm ISCO}, r_{N-1}, \ldots,r_1,r_0=r_{\rm out}$\} and integrate the above equations from the outside in, starting at $r_{[0]} = r_{\text{out}}$. At each radial step $i$, given ${Y_e}_{[i]}$, $\theta_{[i]}$ and ${\eta_e}_{[i]}$, we compute all dependent quantities and find ${Y_e}_{[i+1]}$, $\theta_{[i+1]}$, and ${\eta_e}_{[i+1]}$ at the next radial step by solving a three-dimensional minimization problem to minimize the error function
\begin{equation}\label{eq:numerical_error}
    \mathcal{L}_\text{Error}({Y_e}_{[i+1]},\theta_{[i+1]},{\eta_e}_{[i+1]}) = \sqrt{e_1^2 + e_2^2 + e_3^2}.
\end{equation}
The normalized (relative) error terms correspond to the algebraic,
\begin{multline}\label{eq:numerical_error_1}
    e_1({Y_e}_{[i+1]},\theta_{[i+1]},{\eta_e}_{[i+1]}) = \\ 
    p_{[i]}^{-1} \left[p_{[i+1]} - \left(\frac{G\dot{M}M_\bullet S(r)J(r)}{4\pi \alpha r^3}\right)^{2/3}_{[i+1]} \rho^{1/3}_{[i+1]}\right],
\end{multline}
and differential equations to be solved,
\begin{multline}\label{eq:numerical_error_2}
    e_2({Y_e}_{[i+1]},\theta_{[i+1]},{\eta_e}_{[i+1]}) = (eH)^{-1}_{[i]}\\
    \times \bigg[ (eH)_{[i+1]} - (eH)_{[i]} - (r_{[i+1]} - r_{[i]}) \\
      \times\bigg\{ \bigg(\frac{F^{+} - F^{-}}{u^{r}}\bigg)_{[i+1]} \\
      + \bigg(\frac{e + p}{\rho}\bigg)_{[i+1]} \frac{(\rho H)_{[i+1]} - (\rho H)_{[i]}}{r_{[i+1]} - r_{[i]}} \bigg\} \bigg],
\end{multline}
\begin{multline}\label{eq:numerical_error_3}
    e_3({Y_e}_{[i+1]},\theta_{[i+1]},{\eta_e}_{[i+1]}) = ({Y_e}_{[i]})^{-1} \\
    \times \bigg[ {Y_e}_{[i+1]} - {Y_e}_{[i]} - \frac{m_p}{\rho_{[i]}}(r_{[i+1]} - r_{[i]}) \\
    \times \bigg\{ \left( \frac{\dot{n}_{\bar{\nu}_e} - \dot{n}_{\nu_e}}{H u^r} \right)_{[i]} + (n_{\nu_e} - n_{\bar{\nu}_e})_{[i]} \frac{\text{log}(\rho_{[i+1]}) - \text{log}(\rho_{[i]})}{r_{[i+1]} - r_{[i]}} - \\
    \frac{(n_{\nu_e} - n_{\bar{\nu}_e})_{[i+1]} - (n_{\nu_e} - n_{\bar{\nu}_e})_{[i]}}{r_{[i+1]} - r_{[i]}}\bigg\} 
    \bigg]
\end{multline}
Here, we have discretized radial derivatives using first-order forward finite differences. With the resulting parameters ${Y_e}_{[i+1]}$, $\theta_{[i+1]}$, and ${\eta_e}_{[i+1]}$ that minimize Equation~\eqref{eq:numerical_error}, all dependent quantities at $r_{[i+1]}$ are computed and the scheme is repeated at the next radial step $r_{[i+1]}$. Our solver makes use of the limited Broyden–Fletcher–Goldfarb–Shanno bounded algorithm routine of \texttt{Scipy} \citep{virtanen_scipy_2020} to minimize the error function \eqref{eq:numerical_error}. To ensure stability of the scheme, we implement adaptive brackets that restrict ${Y_e}$, $\theta$, and ${\eta_e}$ to vary by less than 5\% between consecutive radial steps. Our three-dimensional root-finding scheme differs from the approach of \citet{chen_neutrino-cooled_2007}, who only needed to perform two-dimensional root finding in $\theta$ and $\eta_e$, due to a simplified evolution equation for the electron-fraction.

The boundary conditions are chosen at $r_{\text{out}} = 2000 r_g$. This distance from the black hole is sufficient to completely cover the neutrino-cooled region of the disk. In the outermost region of the numerical solution, the accretion flow is radiatively inefficient, energy transport is advection dominated, and the plasma is predominantly composed of $\alpha$-particles and only of a small number of free nucleons. Therefore, an initial value for the proton-to-baryon ratio of ${Y_e}(r_\text{out}) = 0.5$ is well justified. In the advective flow, thermal energy is virialized, such that the specific internal energy density approximately equals the specific virial energy,
\begin{equation}
    \label{eq:virial_theorem}
    \frac{e}{\rho} = \frac{1}{2}\frac{GM_\bullet}{r}.
\end{equation}
The initial values $\theta_{[0]}$ and $\eta_{e[0]}$ at $r=r_{\rm out}$ are computed such that they simultaneously satisfy Equations~\eqref{eq:virial_theorem} and \eqref{eq:pressure_analytic_expression}.

Other boundary conditions characterizing an advective accretion flow, such as a specific ratio of $H/r$, could be specified instead. The numerical results obtained in the regime of interest here, $r_{\rm ISCO}<r\sim r_{\rm ign}$, however, are insensitive to the details of the outer boundary conditions specified in the advective part of the accretion flow.

At sufficiently large radii, the accretion flow becomes gravitationally unstable. For a Keplerian disk in Newtonian gravity, the instability sets in where \citep{toomre_gravitational_1964,paczynski_model_1978,gammie_nonlinear_2001} 
\begin{equation}
    \label{eq:gravitational_instability}
    Q_{\rm grav} = \frac{c_s \Omega}{2 \pi G H \rho} \simeq \frac{H}{r} \frac{M_\bullet}{M_{\rm disk} (r)} < 1.
\end{equation}
Here, $M_{\rm disk}(r)$ is the cumulative disk mass. We monitor this criterion when computing the numerical disk solutions. Whereas our disk model and numerical algorithm still lead to solutions in this regime, these become physically questionable, since self-gravity of the accretion flow would need to be taken into account and the onset of gravito-turbulence would lead to non-stationary, non-axisymmetric dynamics that also modifies (increases) the effective $\alpha$-viscosity \citep{gammie_nonlinear_2001}.

\section{Numerical results}
\label{sec:results}

\subsection{Survey of parameter space}
\label{sec:results_survey_disk_composition}

\begin{figure*}
\centering
\includegraphics[width=\linewidth]{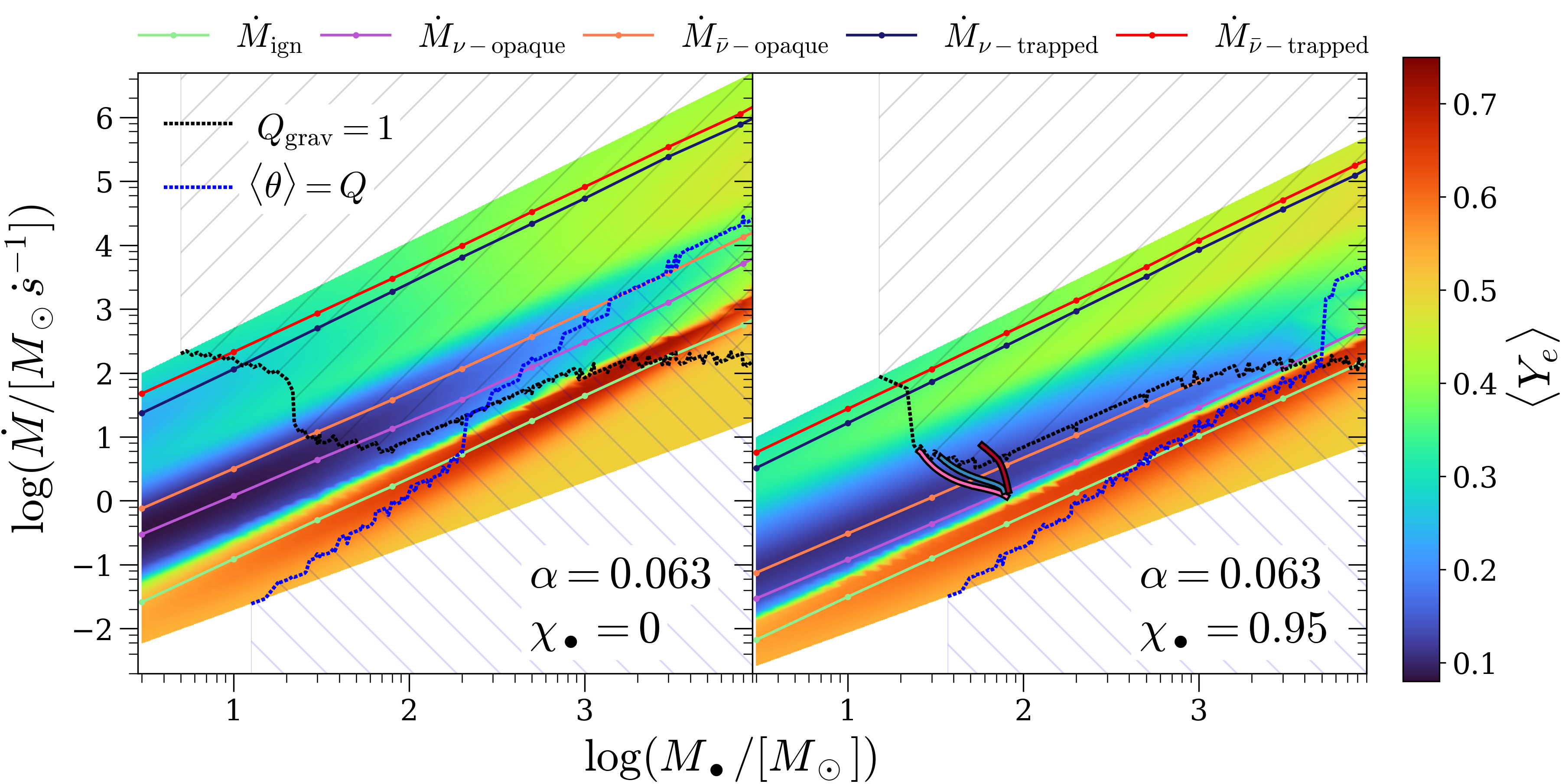} 
 \caption{Mass averaged proton-fraction $\langle Y_e\rangle$ in the inner part of the disk ($r \leq 20 r_g$) for accretion flows around non-spinning ($\chi_\bullet = 0$; left) and rapidly spinning ($\chi_\bullet = 0.95$; right) black holes with $\alpha = 0.063$ as a function of black hole mass $M_\bullet$ and accretion rate $\dot M$. Dots connected by solid lines delineate the numerically identified accretion regimes (Secs.~\ref{sec:theoretical_scalings}, \ref{sec:results_scaling_relations_Mdot}). A valley of very neutron-rich disk composition ($Y_e\lesssim 0.1-0.2$) is apparent for accretion rates between $\dot{M}_{\text{ign}}$ and $\dot{M}_{\bar{\nu}-\text{opaque}}$. The hashed region above the black dotted line corresponds to gravitationally unstable disks ($Q_{\rm grav}(r=20r_g)<1$; Equation~\eqref{eq:gravitational_instability}). The hashed region below the blue dotted line marks disks that do not reach an average temperature larger than the proton-neutron mass difference ($\langle \theta \rangle (r\leq20r_g)< Q = 2.531$). The pink, green, and red solid lines in the right panel indicate the approximate time evolution of collapsar accretion processes in super-kilonova events leading to final black holes in the pair instability supernova mass gap (see Section~\ref{sec:results_superKNe}; \citealt{siegel_super-kilonovae_2022}).
 }
 \label{fig:ye_map}
\end{figure*}

\begin{figure*}
\includegraphics[width=\textwidth]{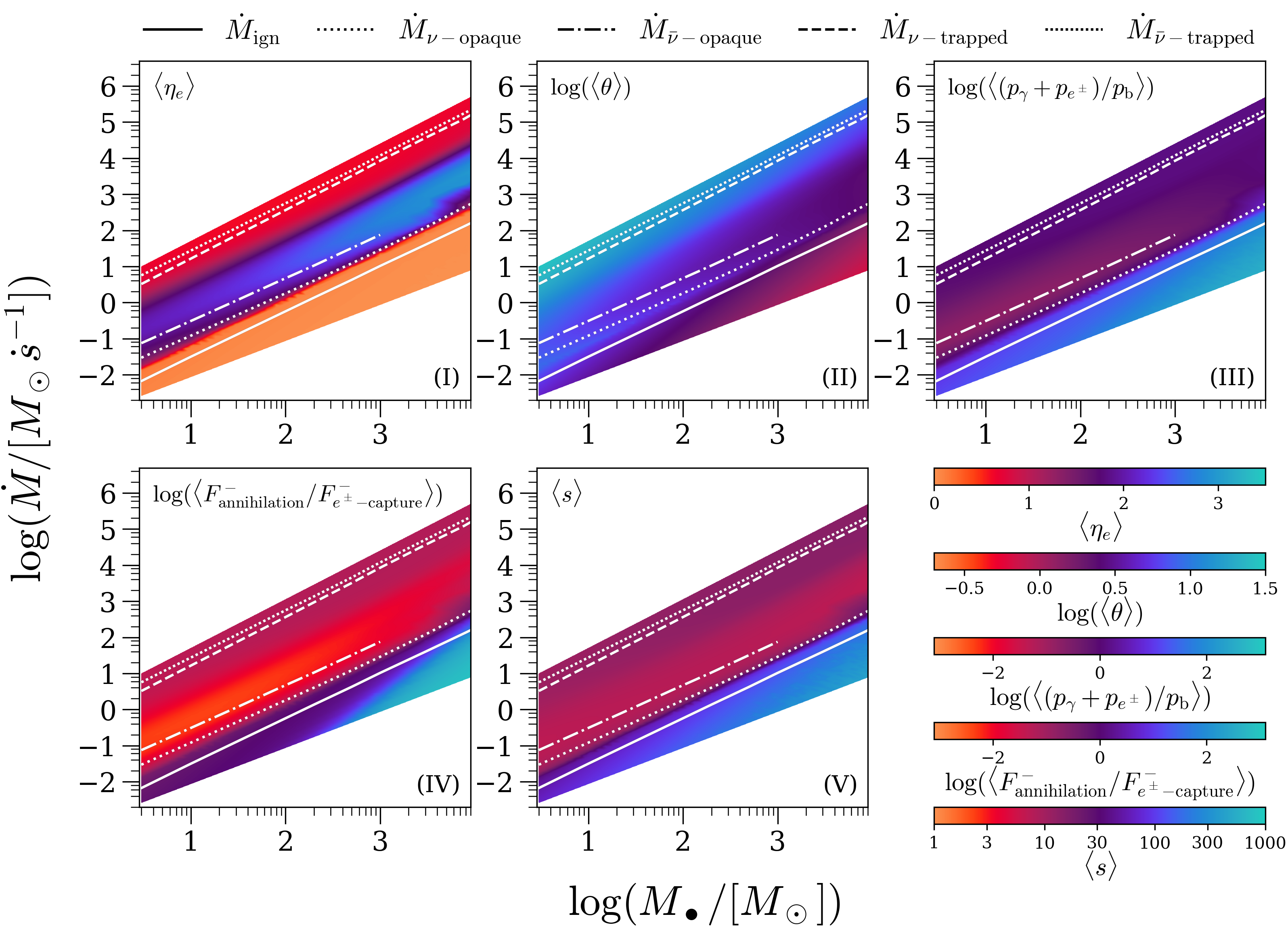}
 \caption{Parameter space results for rest-mass averaged electron degeneracy $\langle\eta_e\rangle$, temperature $\langle\theta\rangle$, radiation pressure to baryon pressure ratio $\langle(p_\gamma + p_{e^\pm})/p_{\mathrm{b}}\rangle$, annihilation cooling to $e^\pm$- capture cooling ratio $\langle F^-_{\mathrm{annihilation}} / F^-_{e^\pm\mathrm{-capture}}\rangle$ and entropy per baryon in units of $k_B$, $\langle s\rangle$, for the case of fast-rotating black holes ($\chi_\bullet = 0.95$) with masses in the range $M_\bullet = 3-3.000 M_\odot$ and viscosity parameter $\alpha = 0.063$. Averages are taken in the inner part of the disk ($r \leq 20 r_g$; Equation~\eqref{eq:def_average_disk_quantity}). The colorbars are centered around characteristic values to delineate different physical regimes, in particular $\langle \theta \rangle = Q= 2.531$ to differentiate sufficiently hot disks that ignite weak interactions from those that do not strongly cool via neutrinos, as well as dominant regimes of pressure and neutrino cooling.}
\label{fig:multi_map}
\end{figure*}

The one-dimensional disk model described in the previous sections allows us to scan the vast parameter space of steady-state, neutrino-cooled accretion onto black holes from the lightest stellar-mass black holes to intermediate-mass black holes, without the need for detailed, computationally expensive, three-dimensional magnetohydrodynamic simulations. Figures \ref{fig:ye_map} and \ref{fig:multi_map} present exemplary results of a parameter space survey in terms of several characteristic quantities, in the case of an $\alpha$-viscosity of $\alpha=0.063$, a value similar to effective $\alpha$-viscosities arising in general-relativistic magnetohydrodynamic simulations of neutrino-cooled accretion disks (e.g., \citealt{siegel_three-dimensional_2018,fernandez_long-term_2019,siegel_collapsars_2019,de_igniting_2021,agarwal_ignition_2025}). Another reason for the choice of this fiducial value of $\alpha$ is the fact that a wide range of qualitatively different behaviors of disk solutions can be illustrated in this case (see the following subsections). 

All panels of Figures~\ref{fig:ye_map} and \ref{fig:multi_map} show rest-mass averages of characteristic quantities $\psi$ within the innermost 20 gravitational radii, defined as 
\begin{equation}
    \langle \psi \rangle = \frac{\int_{r_{\rm ISCO}}^{20 r_g}  \psi(r) \rho(r)\sqrt{-g}\, \dr}{\int_{r_{\rm ISCO}}^{20 r_g} \rho(r)\sqrt{-g}\,\dr}, \label{eq:def_average_disk_quantity}
\end{equation}
with $\sqrt{-g} = r$ in our coordinates $x^\mu = (t,r,\phi,z)$. White (unpopulated) space in the figure panels corresponds to regions of the parameter space that are not of interest for the present discussion. The inner region of a few tens of gravitational radii focused on here is most characteristic of the physical state of the accretion flow. Additionally, being deepest in the gravitational potential, this region shows a sharp rise in temperature, which causes an imbalance between viscous heating and neutrino cooling off the midplane, leading to massive disk outflows, in which nucleosynthesis takes place \citep{siegel_three-dimensional_2018,siegel_collapsars_2019}. Such outflows are not considered in the present model. However, the disk composition sets a lower limit of $Y_e$ for the composition of the outflows. As long as reabsorption of neutrinos from the accretion disk by the outflowing plasma winds can be neglected (typically for $\dot{M}\lesssim \dot{M}_{\bar\nu}$), the disk wind composition is expected to be approximately that of the disk material from which the winds originate. Because of the relevance for nucleosynthesis, here we focus chiefly on the disk composition.

At sufficiently small accretion rates $\dot{M}\ll \dot{M}_{\rm ign}$, the composition remains unchanged ($Y_e\simeq 0.5$) with respect to the outer boundary condition of the accretion flow. We find a proton-rich regime $Y_e\gtrsim 0.5$ at accretion rates close to the ignition threshold $\dot{M}\lesssim \dot{M}_{\rm ign}$, at least with increasing black hole mass ($M_\bullet \gtrsim 10 M_\odot$). Between $\dot{M}_{\rm ign}\lesssim \dot{M}\lesssim M_{\bar{\nu}-\text{opaque}}$, a pronounced valley of very neutron-rich composition of $Y_e\lesssim 0.1-0.2$ emerges, with an increasing valley floor as $M_\bullet$ increases. At $\dot{M}\gtrsim \dot{M}_{\nu-\text{trap}}$, the valley floor rises to larger values of $Y_e\lesssim 0.3- 0.5$, which confine the valley floor `from above'. In the following subsections, we discuss these main features and qualitatively different regimes of disk composition in more detail.

\begin{figure*}[tb]
\centering
\includegraphics[width=\linewidth]{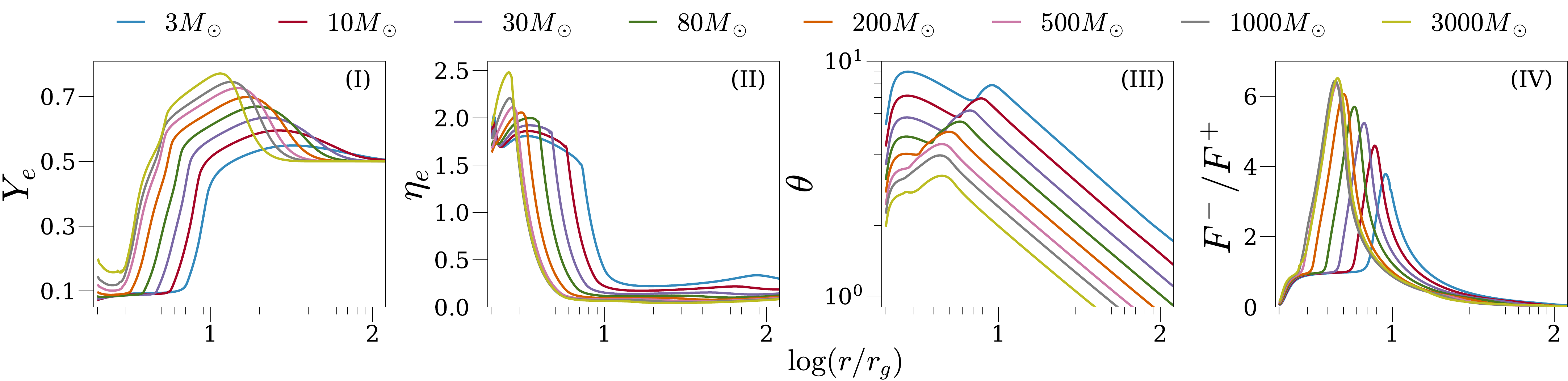}
\caption{Radial profiles (in units of the gravitational radius $r_g$) of proton-fraction $Y_e$, the normalized electron degeneracy parameter $\eta_e=\mu_e/k_{\mathrm{B}}T$, the normalized temperature $\theta=k_{\rm B}T/\mel c^2$, and the ratio $F^-/F^+$ of the rate of neutrino cooling to viscous heating for accretion flows with $\alpha = 0.063$ at the neutrino opaque threshold $\dot{M} = \dot{M}_{\nu}$ around rapidly spinning ($\chi_\bullet=0.95$) black holes across three orders of magnitude in mass.}
\label{fig:M_dot_nu}
\end{figure*}

\begin{figure*}
\centering
\includegraphics[width=\linewidth]{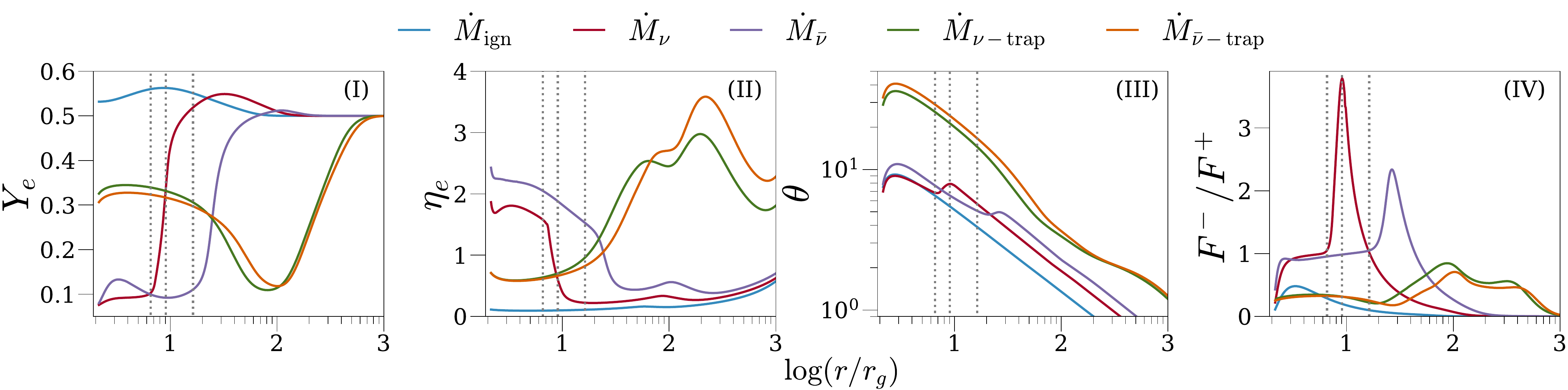}
\caption{Radial profiles (in units of the gravitational radius $r_g$) of proton-fraction $Y_e$, the normalized electron degeneracy parameter $\eta_e=\mu_e/k_{\mathrm{B}}T$, the ratio $F^-/F^+$ of the rate of neutrino cooling to viscous heating, and the normalized temperature $\theta=k_{\rm B}T/\mel c^2$ for accretion flows with $\alpha = 0.063$ around a rapidly spinning ($\chi_\bullet=0.95$) black hole of mass $M_\bullet = 3 M_\odot$ at various characteristic accretion rates (Section~\ref{sec:theoretical_scalings}). Vertical dotted lines indicate the radial range, in which the disk composition at $\dot{M} = \dot{M}_{\nu}$ transitions from the outer proton-rich to the inner neutron-rich regime.}
\label{fig:M_dot_diff}
\end{figure*}

\subsection{The neutron-rich regime}
\label{sec:results_neutron-rich_regime}

As the accretion rate exceeds the ignition threshold $\dot{M}\gtrsim \dot{M}_{\rm ign}$, the inner accretion flow starts to neutronize (Figures~\ref{fig:ye_map}, \ref{fig:M_dot_diff}). This is a result of the midplane density (Equation~\eqref{eq:ign_thresh_step1.5}) increasing to a level at which electrons become degenerate, such that $\eta_e\lesssim 1$ (compare, for example, the blue and red lines in Figure~\ref{fig:M_dot_diff} (II)). Figure~\ref{fig:multi_map} (I) also clearly shows the increase of $\eta_e$ to order unity at $\dot{M}\gtrsim\dot{M}_{\rm ign}$. As a result of Pauli-blocking, $e^\pm$-pair creation is suppressed and electron capture onto protons (Equation~\eqref{reaction:heatcool1}) is increasingly favored over positron capture onto neutrons (Equation~\eqref{reaction:heatcool2}), driving the accretion flow into a neutron-rich state ($Y_e<0.5$). 

neutrino cooling is indeed typically dominated by electron and positron capture at all characteristic accretion rates (Figure~\ref{fig:multi_map} (IV)). Generally, the ratio $F^-_{\rm annihilation}/F^-_{\rm e^\pm-capture}$ is expected to scale as (see Equations~\eqref{eq:pressure_ratio_Mdotign}, \eqref{eq:pressure_ratio_Mdotnu} and \eqref{eq:pressure_ratio_Mdotnu-trap})
\begin{equation}\label{eq:F_ann_Fcapture}
    \frac{F^-_{\rm annihilation}}{F^-_{\rm e^\pm-capture}}\propto \frac{p_{\gamma,e^\pm}}{p_{\rm b}} \propto \left\{ \begin{array}{ll}
        \alpha^{-1/6} M_\bullet^{1/6}, & \dot{M}=\dot{M}_{\rm ign} \\
        \alpha^{0} M_\bullet, & \dot{M}=\dot{M}_{\nu,\nua} \\
        \alpha^{-1/6} M_\bullet^{1/6}, & \dot{M}=\dot{M}_{\nu,\nua-\rm{trap}}
    \end{array}\right.
\end{equation}
where we have assumed that radiation and relativistic electron-positron ($e^\pm$) pressure dominate at $\dot{M}_{\rm ign}$, whereas baryon pressure dominates at all other thresholds (Figure \ref{fig:multi_map} (III) and Section \ref{sec:results_pressure}). This increase of the pressure ratio with $M_\bullet$ can indeed be observed in Figure \ref{fig:multi_map} (III), but the ratio of cooling emissivities only increases to order unity at $\dot{M}\approx \dot{M}_{\rm ign}$ for large black hole masses $M_\bullet \gtrsim 1000 M_\odot$ (lower right corner of Figure~\ref{fig:multi_map} (IV)), where this affects the composition (Sec.~\ref{sec:results_proton-rich_regime}). The onset of strong neutrino cooling is also reflected in the entropy per baryon and $k_B$, which shows a characteristic drop from typically $\langle s \rangle \gtrsim 30-100$ at $\dot{M}\lesssim \dot{M}_{\rm ign}$ to $\langle s \rangle \approx 10-20$ at $\dot{M}\gtrsim \dot{M}_{\rm ign}$ (Figure~\ref{fig:multi_map} (V)).

As the accretion rate surpasses $\dot{M}_{\rm ign}$ and approaches $\dot{M}_\nu$, we find that the inner part of the disk regulates itself to a low-$Y_e$ state of $Y_e\approx 0.05-0.1$ (Figures~\ref{fig:ye_map}, \ref{fig:M_dot_nu}, and \ref{fig:M_dot_diff}). The existence of such a well-defined `asymptotic' value of $Y_e$ across the wide parameter space of accretion flows considered here, which is particularly evident from Figures ~\ref{fig:M_dot_nu} and \ref{fig:M_dot_diff}, is the result of a self-regulation mechanism based on electron degeneracy acting in the accretion flow \citep{chen_neutrino-cooled_2007,siegel_three-dimensional_2018,siegel_collapsars_2019}: As $\eta_e$ rises beyond unity due to increasing midplane densities (Figures \ref{fig:multi_map} (I) and ~\ref{fig:M_dot_nu} (II)), suppression of $e^\pm$-pair creation sets in, which reduces the overall neutrino cooling through electron and positron capture (Equations ~\eqref{reaction:heatcool1} and \eqref{reaction:heatcool2}), leading to higher temperatures, reducing $\eta_e$. The result of this negative feedback loop is a regulated degeneracy level of $\eta_e\sim 1$ for $\dot{M}\gtrsim \dot{M}_{\rm ign}$, as evident from Figures \ref{fig:multi_map}, \ref{fig:M_dot_nu} and \ref{fig:M_dot_diff}. This self-regulated regime of the accretion flow based on electron degeneracy and temperature involves an intricate balance between neutrino cooling and viscous heating of the flow, which can be identified as a horizontal ``plateau'' of $F^-/F^+\simeq 1$ in the radial profiles shown in Figures~\ref{fig:M_dot_nu} (IV) and \ref{fig:M_dot_diff} (IV).

At sufficiently high disk densities and temperatures, an estimate for the self-regulated value of $Y_e$ can be obtained from Equation \eqref{eq:etaeetanu}, which can be recast as
\begin{equation}
\label{eq:Y_e_trapped}
    Y_e = \frac{1}{1+e^{\eta_e-\eta_\nu-\frac{Q}{\theta}}} \approx \frac{1}{1+e^{\eta_e}}\approx 0.07 - 0.18.
\end{equation}
In the second step, we have assumed sufficiently high $\dot M$, such that temperatures increase to $Q/\theta \ll \text{few} \sim \eta_e$ (see Figure~\ref{fig:multi_map}), and we neglected $\eta_\nu$ relative to $\eta_e\sim\text{few}$ (semi-transparent limit). From our numerical results, typically $1.5 \lesssim \eta_e \lesssim 2.6$, which yields $0.07 \lesssim Y_e \lesssim 0.18$. This rough estimate is in good agreement with the actual numerical results obtained in the stellar-mass black hole range. At black hole masses $M_\bullet \sim 100-1000 M_\odot$, however, disk temperatures decrease (typically $T\propto M_\bullet^{-1/6}$, see Equation \eqref{eq:disk_temperature_scaling} and Figure~\ref{fig:multi_map} (II)) and thus $Y_e$ slowly increases with black hole mass until neutronization ceases to occur at $M_\bullet \gtrsim \text{few}\times 10^3 - 10^4 M_\odot$.

The onset and `mechanics' of the self-regulation mechanism can be directly witnessed over a radial range in the accretion disk. Figure~\ref{fig:M_dot_diff} illustrates the situation by vertical dotted lines for an accretion disk around a rapidly spinning $M_\bullet = 3M_\odot$ black hole at $\dot{M}=\dot{M}_{\nu}$. From right to left, the vertical lines indicate where the efficiency of neutrino cooling relative to viscous heating, $F^-/F^+$, transitions from $<1$ to $>1$ to $\simeq 1$. As the density and temperature of the disk plasma increase with decreasing radius, i.e. as matter gradually falls deeper into the gravitational potential (Figure~\ref{fig:M_dot_diff} (IV)), neutrino cooling increases sharply, $F^-\propto \rho T^6$ (Equations ~\eqref{eq:Fnu_capture}, \eqref{eq:Fantinu_capture}); Figure~\ref{fig:M_dot_diff} (III)), as does the electron degeneracy (Figure~\ref{fig:M_dot_diff} (II)). Once degeneracy is established, $\eta_e\sim 1$ (central vertical dotted line), cooling through electron and positron capture declines heavily due to exponential suppression of $e^\pm$-pair production. As a result of this negative feedback, $F^-/F^+$ starts to decrease with decreasing radius after having reached a peak value and approaches $\simeq 1$, a regime, in which positive and negative feedback on neutrino emission are balanced (left vertical dotted line). 

The height of the peak in $F^-/F^+$ strongly depends on the original proton excess ($Y_e > 0.5$; see also Section~\ref{sec:results_proton-rich_regime}) at the radii where neutrino cooling becomes significant, and thus strongly depends both on $\dot M$ (Figure~\ref{fig:M_dot_diff} (IV)) and on $M_\bullet$ (Figure~\ref{fig:M_dot_nu} (IV)). With a larger proton excess, more electrons are available for capture, and thus higher levels of degeneracy (cf.~Figure~\ref{fig:M_dot_diff} (II), Figure~\ref{fig:M_dot_nu} (II)) can be tolerated before neutrino cooling becomes strongly suppressed.

Once $F^-/F^+$ decreases to $\simeq\!1$ with decreasing radius, the self-regulated regime (with its well defined, small value of $Y_e$; see above and Figures~\ref{fig:M_dot_nu} (I) and \ref{fig:M_dot_diff} (I)) extends down to $r\simeq r_{\rm ISCO}$. The existence and extend of this low-$Y_e$ region is of pivotal importance in estimating whether given accretion flow regimes can give rise to $r$-process nucleosynthesis. The size is indicative of the amount of material unbound in disk winds, which, given the typical accretion disk wind conditions, are promising sites for $r$-process nucleosynthesis.

The radial range $r_{\rm sr}$ of this self-regulated region is governed by $F^-/F^+ \simeq \text{const.}\simeq 1$, and scales with $\alpha$, $\dot{M}$, and $M_\bullet$. The scaling is identical to that of the ignition radius $r_{\rm ign}$ (Equation~\eqref{eq:ign_thresh_step9}) if the pressure at $r_{\rm sr}$ is dominated by radiation and/or relativistic $e^\pm$ pressure. If the pressure is dominated by baryon pressure, however, one must follow the derivation in Section \ref{sec:ignition_threshold} with $p\propto \rho T$ to obtain the combined result:
\begin{equation}
    \label{eq:self-regulation-radius}
    \frac{r_{\rm sr}}{r_{\rm g}} \propto \left\{\begin{array}{ll}
        \alpha^{-2} M_\bullet^{-\frac{8}{5}} \dot{M}^{\frac{6}{5}} , &  p\propto p_{\gamma,e^\pm} \\
        \alpha^{-\frac{2}{7}} M_\bullet^{\frac{2}{7}}\dot{M}^{0}, & p\propto p_{\rm b}
    \end{array}\right. .
\end{equation}

At $r=r_{\text{sr}}$, the accretion flow transitions from the radiation and $e^\pm$-pressure dominated regime at $r=r_{\rm ign}$ to an increasingly baryon-pressure dominated regime. One thus expects $r_\text{sr}$ to scale somewhere in between the corner cases of Equations \eqref{eq:self-regulation-radius}. For example, for the models shown in Figures \ref{fig:M_dot_nu} and \ref{fig:M_dot_diff}, with $\alpha = 0.063$ and $\chi_\bullet=0.95$, we find 
\begin{equation}\label{eq:sr-scaling_numerical}
    \frac{r_\text{sr}}{r_{\rm g}} \propto  \dot{M}^{0.95\pm0.09} M_\bullet^{-0.64\pm0.01},
\end{equation}
which is indeed within the limits found in Equation \eqref{eq:self-regulation-radius}.

As the accretion rate approaches $\dot{M} \approx \dot{M}_{\nu-\mathrm{trap}}$ and neutrinos become trapped in the inner part of the accretion flow, the reverse reaction in Equation \eqref{reaction:heatcool1} is enhanced relative to that of Equation \eqref{reaction:heatcool2}. This is because neutrons are more abundant than protons in the inner part of the accretion flow, which is neutronized at larger radii. The abundance of trapped neutrinos progressively increases the proton-fraction $Y_e$ with increasing $\dot{M}$, until the latter reaches a typical value of $Y_e \approx 0.3-0.4$ at $\dot{M}\gtrsim \dot{M}_{\nu-{\rm trap}}$, evidently present in Figures~\ref{fig:ye_map} and \ref{fig:M_dot_diff} (I). As the accretion flow neutronizes progressively less in the range $\dot{M}_{\rm ign}\lesssim \dot{M}\lesssim{M}_{\nu-{\rm trap}}$ with increasing $M_\bullet$, so does $Y_e$ also increase to $0.4-0.5$ at $\dot{M} \gtrsim \dot{M}_{\nu-\mathrm{trap}}$ (Figure~\eqref{fig:ye_map}).

\subsection{Implications for nucleosynthesis of neutron-rich nuclei}
\label{sec:results_neutron-rich-nucleosynthesis}

Outflows from neutrino-cooled accretion disks have been demonstrated to be promising sources of $r$-process nuclei by three-dimensional GRMHD simulations in the stellar-mass regime $M_\odot \approx 3 M_\odot$ (e.g., \citealt{siegel_three-dimensional_2017,siegel_three-dimensional_2018,fernandez_long-term_2019,miller_full_2019,li_neutrino_2021,siegel_collapsars_2019,miller_full_2020}). Given the results on neutronization in Section~\ref{sec:results_neutron-rich_regime}, here we briefly comment on the prospects for $r$-process nucleosynthesis more broadly across the vast parameter space of neutrino-cooled accretion onto black holes considered in this paper.

We estimate the relevant parameters determining the nucleosynthetic outcome of the $r$-process using a simple disk outflow model described in Appendix~\ref{app:outflow_model}. These are $Y_e$, $s$, and the expansion timescale $\tau_{\rm exp}$ of the outflow at the onset the neutron-capture reactions (i.e. when the flow has cooled to $\approx 5$\,GK; \citealt{lippuner_r-process_2015}). In particular, we consider outflowing plasma ejected from the inner accretion disk at $R_{\rm launch} = 3r_{\rm ISCO}$ under the assumption of adiabatic expansion (constant entropy) and neglecting absorption of neutrinos during the launch of the outflows (constant $Y_e$). We focus on disks accreting at $\dot{M}\approx \dot{M}_{\bar{\nu}}$, i.e. when the inner accretion flow is most neutron-rich (Section~\ref{sec:results_neutron-rich_regime}). We discard models with $\alpha \gtrsim 0.3$, because these disks do not reach the self-regulated regime.

The proton-fraction $Y_e$ varies strongly with the black hole mass and $\alpha$-viscosity (see Section~\ref{sec:results_proton-rich_regime}). A larger $M_\bullet$ or $\alpha$-parameter lifts the equilibrium value reached in the self-regulated regime. For $\chi_\bullet = 0$ and $\alpha = 0.001-0.3$, we find $Y_e\approx 0.06-0.13$ at $M_\bullet = 3 M_\odot$ and $Y_e\approx 0.32-0.36$ at $M_\bullet = 3000 M_\odot$. For $\chi_\bullet = 0.95$ and $\alpha = 0.001-0.3$, instead, we find $Y_e\approx 0.03-0.13$ at $M_\bullet = 3 M_\odot$ and $Y_e\approx 0.25$ at $M_\bullet = 3000 M_\odot$. 

The entropy of the outflowing trajectories generally resides in a narrow range of $s \sim 4-8k_B$ per baryon for $\alpha =0.001-0.1$, $M_\bullet = 3-3000M_\odot$, and $\chi_\bullet = 0-0.95$, with only a slight increase with increasing $\alpha$-viscosity.

The expansion timescale depends on both the ejecta velocity and the distance to the black hole when the $r$-process takes place (i.e. when the wind reaches $T\sim5$GK). The former depends only on $Y_e$ (see Appendix~\ref{app:outflow_model}), and varies between $\sim0.05-0.12c$, while the latter is proportional to the black hole mass and the spin, as these two quantities determine the value of the ISCO, and thus the absolute length scale at which the outflow is launched. For $\chi_\bullet = 0$, we find the expansion timescale approximately follows $\tau_{\rm exp} \sim 35\,\mathrm{ms}\, (M_\bullet/3M_\odot)^{0.72}(\alpha/0.063)^{0.13}$, while for $\chi_\bullet = 0.95$, it follows $\tau_{\rm exp} \sim 18\,\mathrm{ms}\, (M_\bullet/3M_\odot)^{0.72}(\alpha/0.063)^{0.19}$.

The conditions estimated above are generally conducive to the $r$-process. The main question remaining is whether only light or also heavy $r$-process elements (lanthanides and actinides) can be synthesized in the outflows. Comparing to the results of \citet{lippuner_r-process_2015}, we find that, generally, outflows from the inner accretion disks at $\dot{M}\approx \dot{M}_{\bar{\nu}}$ for $M_\bullet \lesssim 200 M_\odot$ ($\chi_\bullet = 0$) and $\lesssim 500 M_\odot$ ($\chi_\bullet = 0.95$) are sufficiently neutron-rich to produce significant mass fractions of lanthanides and actinides. Associated super-kilonovae may thus appear `red', with spectra peaking in the infrared due to the large line expansion opacities of the lanthanides and actinides. For larger black hole masses, we conclude that outflows unlikely synthesize significant mass fractions of lanthanides or actinides. This is because either $Y_e \gtrsim  0.25$, the threshold for lanthanide production, or $Y_e < 0.25$, but the expansion timescales are $\tau_{\rm exp} \gtrsim 0.5-10$\,s, when late-time reheating from radioactive decays restarts the $r$-process at higher $Y_e$ \citep{lippuner_r-process_2015}. Associated super-kilonovae would then appear as `blue' transients, peaking in the optical. 

\subsection{The proton-rich regime}
\label{sec:results_proton-rich_regime}

In the range of intermediate radii $r_{\nu} < r \lesssim r_{\rm ign}$, in which the accreting matter resides in a non-degenerate to weakly degenerate regime, $\eta_e \ll 1$ or $\eta_e \lesssim 1$, but in which neutrino cooling via electron and positron capture (Equations ~\eqref{reaction:heatcool1} and \eqref{reaction:heatcool2}) still is energetically significant, the accretion flow may become proton rich ($Y_e > 0.5$). This is evident from the pronounced red and orange regions around the $\dot{M}_{\rm ign}$ threshold in Figure~\ref{fig:ye_map}, as well as in the explicit examples of radial profiles across the $\approx 3-3000\,M_\odot$ mass range in Figure~\ref{fig:M_dot_nu} (I). 

Proton excess in this physical regime is a result of the proton-neutron mass difference. In the presence of electron and positron capture under weakly degenerate conditions, $\eta_e \lesssim 1$, and in hot matter $\theta > Q=(m_n-m_p)/ m_e \approx 2.531$ (that is, $T > 1.3$\,MeV), the proton-fraction $Y_e$ can be analytically estimated from Equations~\eqref{eq:Ndotnu_capture} and \eqref{eq:Ndotantinu_capture} as (cf.~\citealt{beloborodov_nuclear_2003,siegel_three-dimensional_2018})
\begin{equation}
    Y_e = 0.5 + 0.487\left(\frac{Q}{2\theta} - \eta_e\right) \approx 0.5 + \frac{0.616}{\theta},
    \label{eq:ye_proton_rich}
\end{equation}
where in the second step, we considered the non-degenerate limit $\eta_e \ll \theta$, in which, under conditions of approximately similar numbers of electrons and positrons, positron capture is energetically favored over electron capture, due to the positive mass difference $Q$ between the neutron and the proton. However, a small degeneracy $\eta_e > Q/(2\theta)$ is sufficient to cause the accretion flow to turn neutron-rich ($Y_e < 0.5$; Section~\ref{sec:results_neutron-rich_regime}).

Under similar physical conditions of the accretion flow, i.e. for disks around varying black hole mass, but with accretion rates in a similar regime (e.g. $\dot{M}\approx \dot{M}_{\rm ign}$ or $\dot{M}\approx \dot{M}_\nu$), the proton-rich regime becomes increasingly pronounced both in radial extend and in absolute value of $Y_e$ with increasing black hole mass. Whereas a proton excess is mild and barely noticeable for stellar-mass black holes around $\sim\! 3M_\odot$ (cf.~Figure~\ref{fig:ye_map}, Figure~\ref{fig:M_dot_nu} (I) and Figure~\ref{fig:M_dot_diff} (I)), the corresponding radial region widens and reaches values of up to $Y_e\approx 0.8$ for $3000M_\odot$ black holes.

The increasing proton excess with increasing black hole mass can be understood analytically. For accreting matter dominated by radiation and relativistic $e^\pm$ pressure, as appropriate in the radial regime under consideration (see Section~\ref{sec:results_pressure}), Equations~\eqref{eq:ign_thresh_step1.5} and \eqref{eq:ign_thresh_step3} predict the disk midplane temperature as
\begin{equation}
    T \propto J^{\frac{1}{8}} S^{\frac{1}{4}}\left(\frac{H}{r}\right)^{-\frac{1}{4}} \left(\frac{r}{r_g}\right)^{-\frac{5}{8}} \alpha^{-\frac{1}{4}} M_\bullet^{-\frac{1}{2}} \dot{M}^{\frac{1}{4}},
\end{equation}
which using Equations~\eqref{eq:ign_thresh_step9}, \eqref{eq:ign_scaling}, \eqref{eq:opaque_thresh_step6}, \eqref{eq:Mdotnu_scaling} translates into
\begin{equation}
    T \propto \left\{\begin{array}{ll} \alpha^{1/6}M_\bullet^{-1/6} & (\dot{M} \approx \dot{M}_{\rm ign},\; p\propto p_{\gamma,e^\pm}) \\
    \alpha^0 M_\bullet^{-1/6}  & (\dot{M} \approx \dot{M}_\nu,\; p\propto p_{\gamma,e^\pm})
    \end{array}\right. . \label{eq:disk_temperature_scaling}
\end{equation}

Indeed, at $\dot{M} = \dot{M}_\nu$ the temperatures of our numerical results roughly scale as $\propto M_\bullet^{-1/6}$ in the range $10\lesssim r/r_g \lesssim 100$ as illustrated in Figure~\ref{fig:M_dot_nu} (III). The corresponding proton excess according to Equation~\eqref{eq:ye_proton_rich} then scales as
\begin{eqnarray}
    \label{eq:ye_proton_rich_MBH}
    Y_e - 0.5 \propto \left\{\begin{array}{ll} 
    \alpha^{-1/6} M_\bullet^{1/6} & (\dot{M} \approx \dot{M}_{\rm ign}, \; p\propto p_{\gamma,e^\pm}) \\
     \alpha^{0} M_\bullet^{1/6}  & (\dot{M} \approx \dot{M}_\nu,\; p\propto p_{\gamma,e^\pm})
     \end{array}\right. .
\end{eqnarray}

At $\dot{M}=\dot{M}_\nu$ and a maximum $Y_e$ of $\approx\!0.55$ for a $3M_\odot$ black hole, this scaling predicts a range of $Y_e\approx 0.55-0.7$ for $M_\bullet = 3-3000\,M_\odot$. This estimate is in good agreement with the numerical results shown in Figure~\ref{fig:M_dot_nu} (I).

\subsection{Implications for nucleosynthesis of proton-rich nuclei}
\label{sec:results_proton-rich-nucleosynthesis}

The presence of proton-rich regimes in the accretion flow (Sec.~\ref{sec:results_proton-rich_regime}) gives rise to proton-rich disk outflows from these regions. Here, we estimate, using a simplified disk outflow model, whether such outflows can synthesize neutron-deficient nuclei ($p$-nuclei) via the $\nu p$-process. 

The $\nu p$-process \citep{frohlich_neutrino-induced_2006,wanajo_rpprocess_2006} synthesizes neutron-deficient nuclei up to Ag in proton-rich conditions $Y_e \gtrsim 0.5$. In nuclear statistical equilibrium at $T\gtrsim 5$\,GK, all neutrons recombine with protons into $\alpha$ particles, which, via further cooling in the range $3\,\text{GK} \lesssim T \lesssim 5$\,GK, assemble $^{12}$C via the triple alpha reaction $\alpha(2\alpha,\gamma)^{12}$C. The latter are converted in ($p$,$\gamma$)--($\gamma$,$p$) equilibrium on much faster timescales into heavier elements, mostly $^{56}$Ni, $^{60}$Zn, and $^{64}$Ge. These elements with increased stability ($\beta$-decay lifetimes $\gtrsim 60$s; equal numbers of neutrons and protons) serve as `waiting points'. In the range $1.5\,\text{GK}\lesssim T \lesssim 3\,\text{GK}$, free protons conspire with electron antineutrinos to produce free neutrons, $p + \bar{\nu}_e \rightarrow n + e^+$. Via fast ($n$,$p$) and subsequent ($\gamma$, $p$) reactions on the waiting point nuclei, these neutrons then continue the reactions to heavier elements. Cooling below 1\,GK terminates the $\nu p$-process as the ($p$,$\gamma$) reactions freeze out.

A figure of merit to quantify the potential for a successful $\nu p$-process is thus the number of free neutrons created per seed nucleus as the plasma cools through $1\,\text{GK}\lesssim T \lesssim 3\,\text{GK}$ \citep{pruet_nucleosynthesis_2006}
\begin{equation}\label{eq:pruet_nup_parameter}
    \Delta_n = \frac{Y_p}{Y_{\mathrm{heavy}}} \int_{1<T_9 < 3} \lambda_{\bar{\nu}_e p} \dt,
\end{equation}
where $Y_p$ and $Y_{\mathrm{heavy}}$ denote the abundance of free protons and seed nuclei heavier than $^4\mathrm{He}$, respectively, $\lambda_{\bar{\nu}_e p}$ is the rate of capture of antineutrinos onto protons, and $T_9$ is the temperature of the outflow in units of GK. At least $\Delta_n \sim \text{few - 10}$ is required for the production of $\nu p$-process elements heavier than the waiting point nuclei.

\begin{figure}
\centering
\includegraphics[width=\columnwidth]{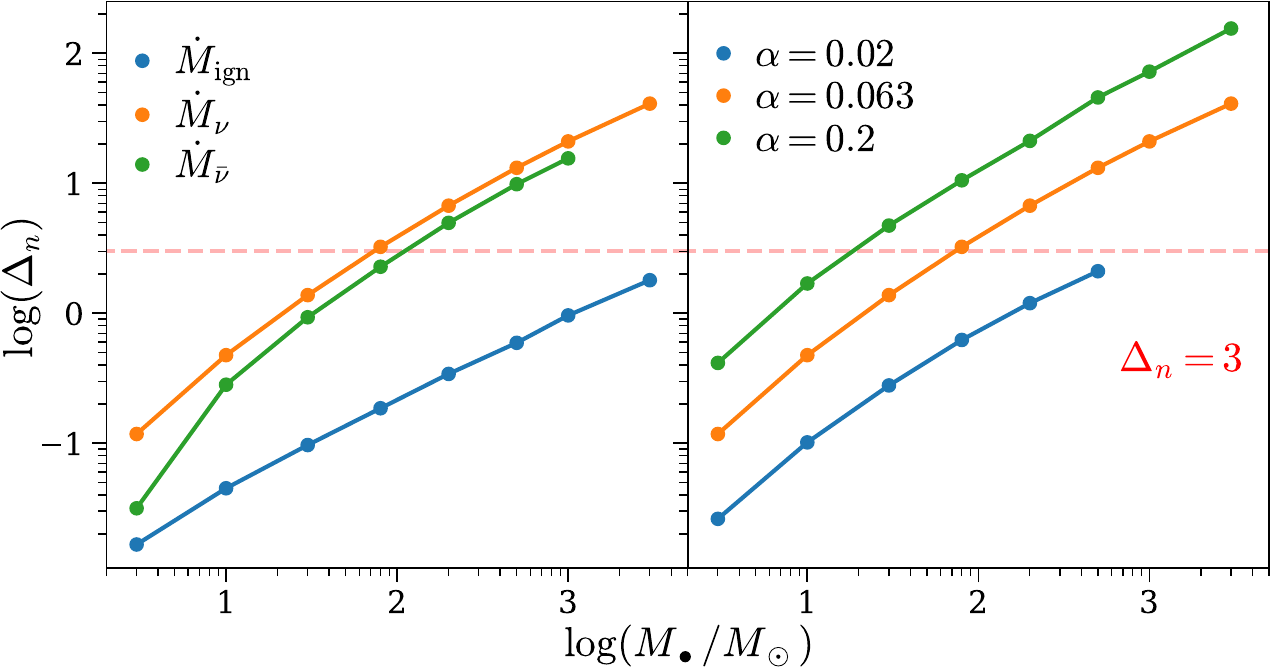}
\caption{Left: Number of free neutrons per seed nucleus $\Delta_n$ created in disk outflows by electron antineutrino absorption under proton-rich conditions, as a function of black hole mass, for models with fiducial viscosity $\alpha=0.063$, dimensionless black hole spin $\chi_\bullet = 0.95$, and various values of the accretion rate $\dot{M}$. Right: Analogous to the results in the left panel, but for models with $\dot{M}=\dot{M}_\nu$ and different values for the $\alpha$-viscosity. The red dashed line marks the minimum threshold of $\Delta_n \approx 3$ for triggering a $\nu p$-process.}
\label{fig:Delta_n}
\end{figure}

Figure \ref{fig:Delta_n} shows the results for $\Delta_n$ in disk outflows launched from the most proton-rich regions of the disk, for different values of the black hole mass, the accretion rate, and the $\alpha$-viscosity, and computed with a simplified outflow model described in Appendix \ref{app:outflow_model}. The values of $\Delta_n$ vary between several orders of magnitude, $\Delta_n \sim 10^{-2}-10^2$, and they are highly dependent on both $\dot{M}$ and $\alpha$. The latter is expected because these quantities control the efficiency of neutrino cooling and hence also modulate the proton-fraction in the outflow and the antineutrino luminosity. We find that higher values of $\alpha$ or $M_\bullet$ lead to larger $\Delta_n$, while the sensitivity to $\dot{M}$ is more complex, with highest values of $\Delta_n$ reached at $\dot{M}\approx \dot{M}_\nu$. These results are consistent with the findings that larger $M_\bullet$ leads to larger proton excess (Figure~\ref{fig:M_dot_nu}), that the proton excess is generally largest around $\dot{M}\approx \dot{M}_\nu$ (Figure~\ref{fig:M_dot_diff}), and that the proton excess at $\dot{M}\approx \dot{M}_\nu$ increases with increasing $\alpha$ (Figure~\ref{fig:Averages_opaque_trapped}).

We compare our results with those of \citet{psaltis_neutrino-driven_2024} to estimate the final nucleosynthesis products as a function of $\Delta_n$. They obtain for neutrino-driven outflows under proton-rich conditions ($Y_e\approx 0.6$) with $s/k_{\rm B} \sim 50-100$ and expansion timescales $\tau_{\rm exp} \approx 10-50$\,ms a resulting $\Delta_n \approx 3-60$, which leads to the production of elements with atomic number up to $Z\approx 50$. We find that for our models, which have values $Y_e \sim 0.5-0.75$, $s/k_{\rm B} \sim30-100$ and $\tau_{\rm exp} \sim 40\mathrm{ms} - 5\mathrm{s}$, black hole masses of $M_\bullet \gtrsim 80$ are required to trigger an $\nu p$-process ($\Delta_n \gtrsim \text{few}$), assuming $\dot{M}\approx{M}_\nu$ and the fiducial value of $\alpha \sim 0.06$. For values of the viscosity closer to those found in GRMHD simulations ($\alpha \sim 0.02$), our models just marginally enter the regime at large black hole masses $M_\bullet \gtrsim 500 M_\odot$ ($\Delta_n \sim 2.1$). These results suggest that neutrino-cooled accretion disks around massive black holes could, in principle, be promising sources of $p$-nuclei. However, given the large uncertainties involved in these estimates, more detailed hydrodynamical modeling and nuclear reaction network calculations may be necessary to assess the prospects for the $\nu p$-process more robustly.

\subsection{Partial pressure contributions}
\label{sec:results_pressure}

\begin{figure}[tb]
\centering
\includegraphics[width=\columnwidth]{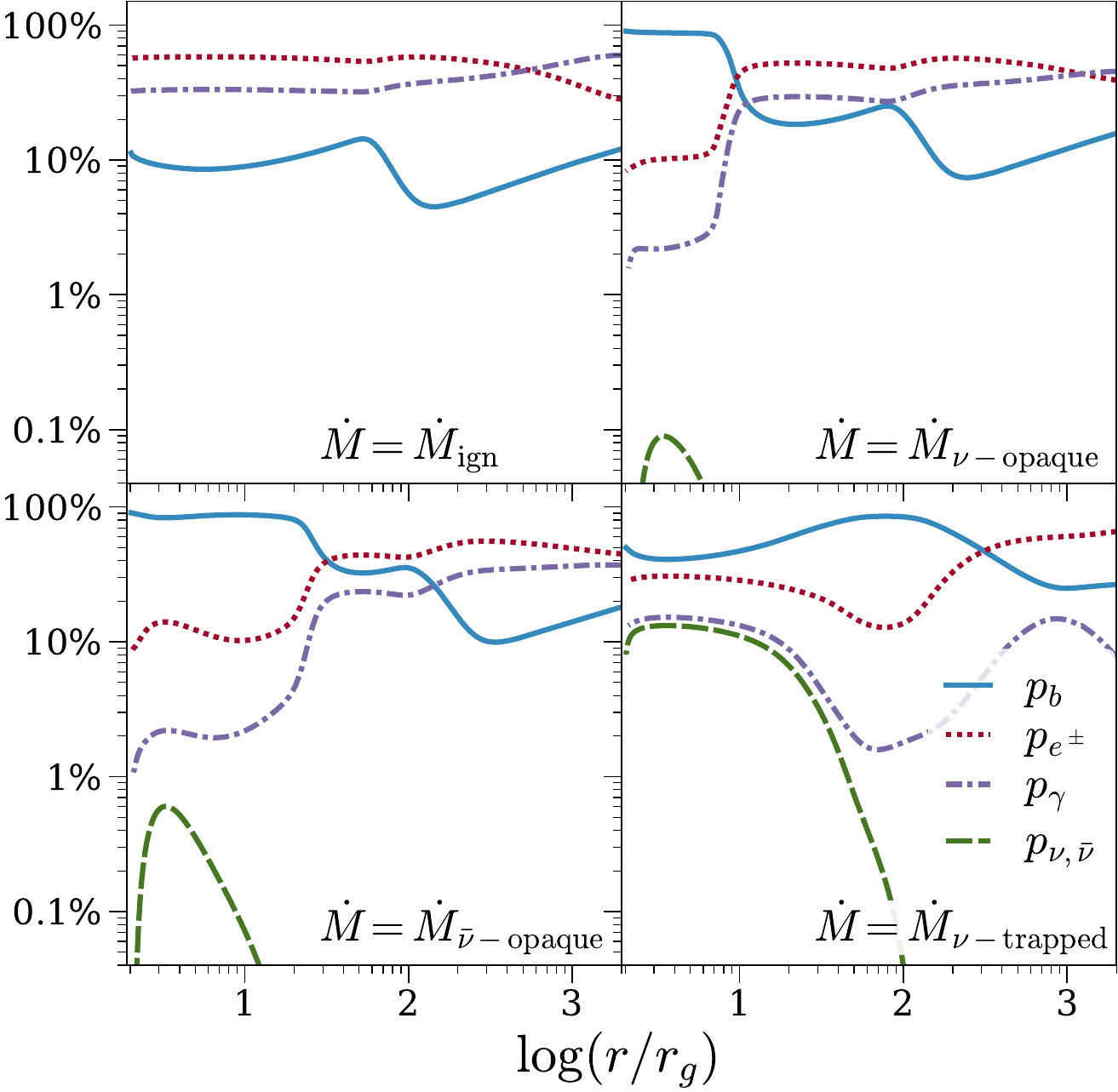}
\caption{Radial profiles (in units of the gravitational radius $r_{\rm g}$) of the percentage of the total pressure contributed by baryons $p_{\rm b}$, by electrons and positrons $p_{e^\pm}$, by radiation $p_\gamma$, and by electron neutrinos and antineutrinos $p_\nu$, for an accretion disk with $M_\bullet = 3 M_\odot$, $\chi_\bullet = 0.95$ and $\alpha = 0.063$ at various characteristic accretion rates (same models as in Figure~\ref{fig:M_dot_diff}).}
\label{fig:Pressure_conts_M3}
\end{figure}

\begin{figure}[tb]
\centering
\includegraphics[width=\columnwidth]{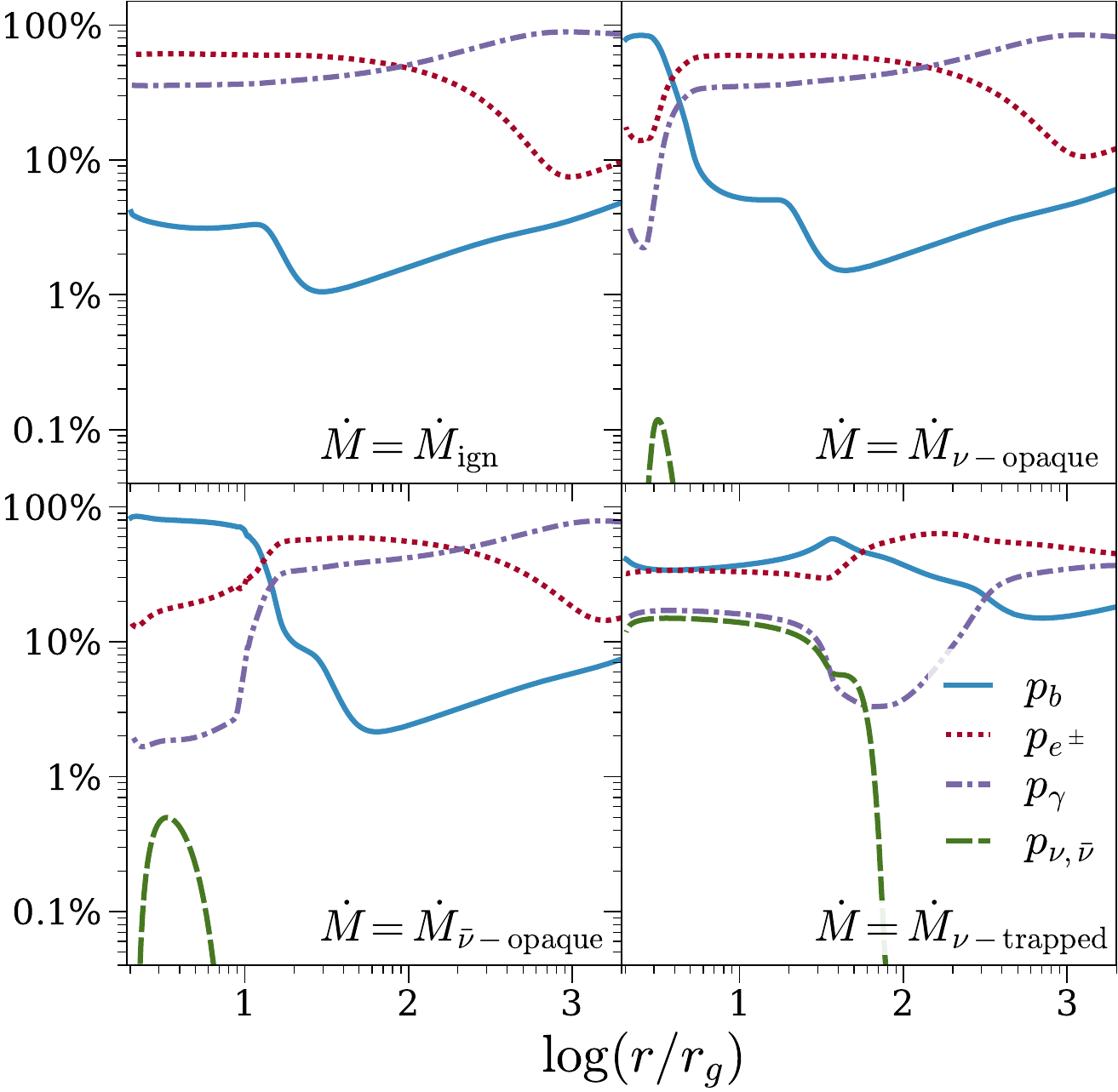}
\caption{Analogous to Figure~\ref{fig:Pressure_conts_M3} but for an accretion disk around a black hole with mass $M_\bullet = 1000 M_\odot$.}
\label{fig:Pressure_conts_M1000}
\end{figure}

The analytic scaling relations derived in Section~\ref{sec:theoretical_scalings} depend on a series of assumptions, particularly on the relation of plasma pressure with temperature. Figures~\ref{fig:Pressure_conts_M3} and \ref{fig:Pressure_conts_M1000} present examples of radial profiles of the different partial pressures of the accreting plasma considered in the model, while Figure \ref{fig:multi_map} (III) shows the average value of the ratio of radiation and relativistic electron-positron $e^\pm$ pressure versus baryon pressure for different models with $M\bullet = 3-3000 M_\odot$, $\chi_\bullet = 0.95$, $\alpha = 0.063$ and different values of the accretion rate.

The $M_\bullet=3M_\odot$ accretion disk for fiducial parameters ($\alpha = 0.063$, $\chi_\bullet = 0.95$) shows a transition from an entirely pair and radiation-pressure dominated disk at accretion rates $\dot{M} \lesssim \dot{M}_{\rm ign}$, to an increasingly baryon-pressure dominated accreting plasma with increasing accretion rate. This inversion of the hierarchy $p_{e^{\pm}} \sim p_\gamma \gg p_{\rm b} \gg p_\nu$ to $p_{\rm b} \gg p_{e^{\pm}} \gg p_\gamma \gtrsim p_\nu$ between $\dot{M} \lesssim \dot{M}_{\rm ign}$ and $\dot{M} \sim \dot{M}_{\bar{\nu}-\text{opaque}}$ occurs gradually as the accretion rate increases and proceeds radially from the inside out. Neutrino pressure remains insignificant at the $\lesssim 1\%$ level. At the ignition threshold, electrons are sufficiently relativistic ($\theta \gtrsim \text{few}-10$ for $r\lesssim 100\,r_g$, Figure~\ref{fig:M_dot_diff}), such that approximately, $p_{e^\pm}\propto T^4$ (cf.~Equation~\eqref{eq:pressure_neutrinos} for the massless (ultra-relativistic) case), as for radiation pressure, $p_\gamma \propto T^4$. The assumption $p \propto T^4$ (Section~\ref{sec:ignition_threshold}) is therefore well justified at $\dot{M} \sim \dot{M}_{\rm ign}$, whereas $p\propto \rho T$ for $r\lesssim 10-100\,r_g$ is appropriate for $\dot{M}_{\nu-\text{opaque}}\lesssim \dot{M}\lesssim \dot{M}_{\bar{\nu}-\text{opaque}}$. At $\dot{M} \gtrsim \dot{M}_{\nu-\text{trapped}}$, we typically find that $p_\text{b}\sim p_{e^\pm}$ and neither of the two contributions dominate. Furthermore, the pressure of the trapped neutrinos $p_{\nu,\nua}\sim p_\gamma$ increases to the 10\% level. 

Figure \ref{fig:Pressure_conts_M1000} shows models analogous to Figure \ref{fig:Pressure_conts_M3}, but for a black hole with mass $M_\bullet = 1000 M_\odot$. The general behavior of the pressure hierarchies are remarkably similar to that of the $M_\bullet = 3 M_\odot$ case.

The different pressure regimes in the inner accretion disk are particularly evident in Figure~\ref{fig:multi_map} (III), throughout the parameter space shown there. In particular, for a given black hole mass the ratio $(p_\gamma + p_{e^\pm})/p_\text{b}$ sharply decreases from $\gg 1$ at $\dot{M}\approx \dot{M}_{\rm ign}$ to $\ll1$, precisely where electrons become degenerate $\eta_e \sim 1$ (Figure \ref{fig:multi_map} (I)), which occurs at $\dot{M}_{\rm ign}\lesssim \dot{M}\lesssim \dot{M}_{\nu}$. With further increasing accretion rate, the inner accretion flow returns to a mildly degenerate state ($\eta_e \sim 0.3$) at $\dot{M} \gtrsim \dot{M}_{\nu-\text{trapped}}$ as the pressure ratio approaches order unity, $(p_\gamma + p_{e^\pm})/p_\text{b} \sim 1$. This is due to a decreasing cooling efficiency as neutrinos become increasingly trapped (Figure~\ref{fig:M_dot_diff} (IV)) and a corresponding rise in temperature (Figure~\ref{fig:M_dot_diff} (III)) that decreases degeneracy (Figure~\ref{fig:M_dot_diff} (II)). Figure \ref{fig:multi_map} (III) also clearly reflects the increase of the pressure ratio with black hole mass, as predicted by Equations~\eqref{eq:pressure_ratio_Mdotign}, \eqref{eq:pressure_ratio_Mdotnu}, and \eqref{eq:pressure_ratio_Mdotnu-trap}.

\subsection{Scaling relations of characteristic accretion rates}
\label{sec:results_scaling_relations_Mdot}

\begin{figure*}
\includegraphics[width=\textwidth]{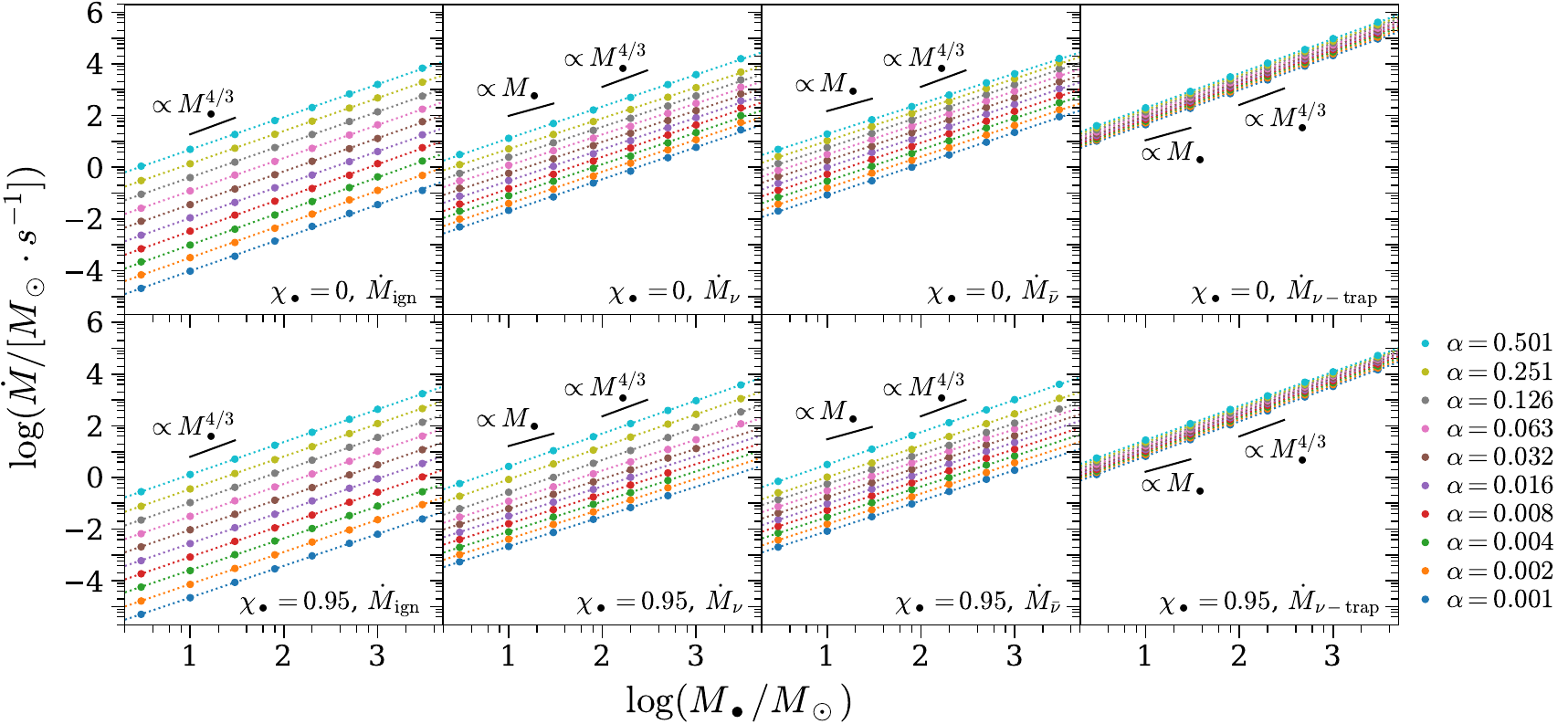}
\caption{Characteristic accretion rates as numerically extracted from disk models as a function of black hole mass $M_\bullet$ for different values of the $\alpha$-viscosity and different spin parameters $\chi_\bullet$. Dotted lines represent the best power-law fits $\dot{M} \propto M_\bullet^\beta$ to the set of disk models at a given value of $\alpha$. Black lines represent the expected slopes of the approximate analytical power laws derived in Section~\ref{sec:theoretical_scalings}.}
\label{fig:MdotvsM}
\end{figure*}

\begin{figure*}
\includegraphics[width=\textwidth]{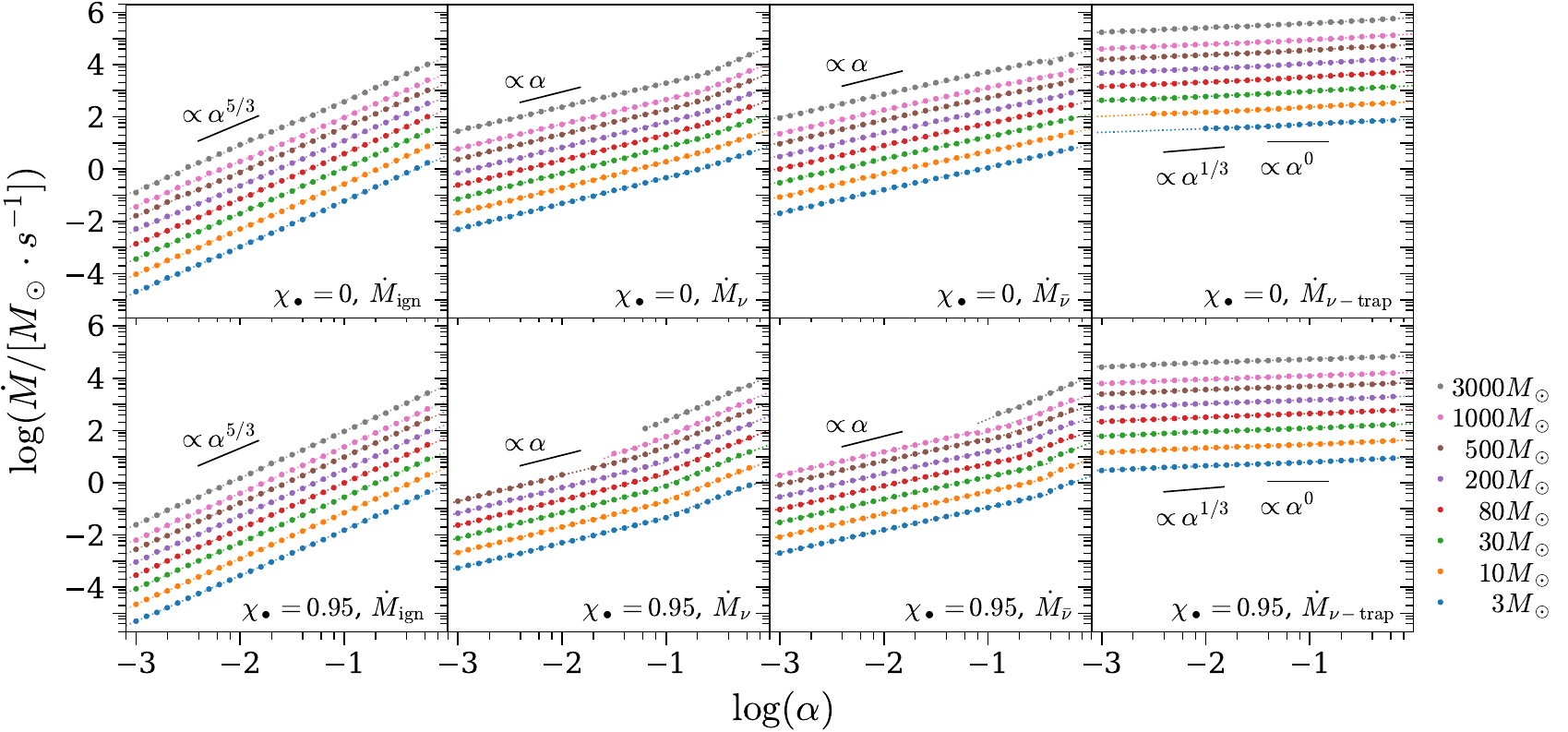}
 \caption{Characteristic accretion rates as numerically extracted from disk models as a function of the $\alpha$-viscosity for different black hole masses and spin parameters $\chi_\bullet$. Dotted lines represent the best power-law fit $\dot{M} \propto \alpha^\gamma$ to each set of models at a given black hole mass. Black lines represent the expected slope of the approximate analytical power laws derived in Section~\ref{sec:theoretical_scalings}.}
 \label{fig:MdotvsAlpha}
\end{figure*}

\begin{figure}
\includegraphics[width=\columnwidth]{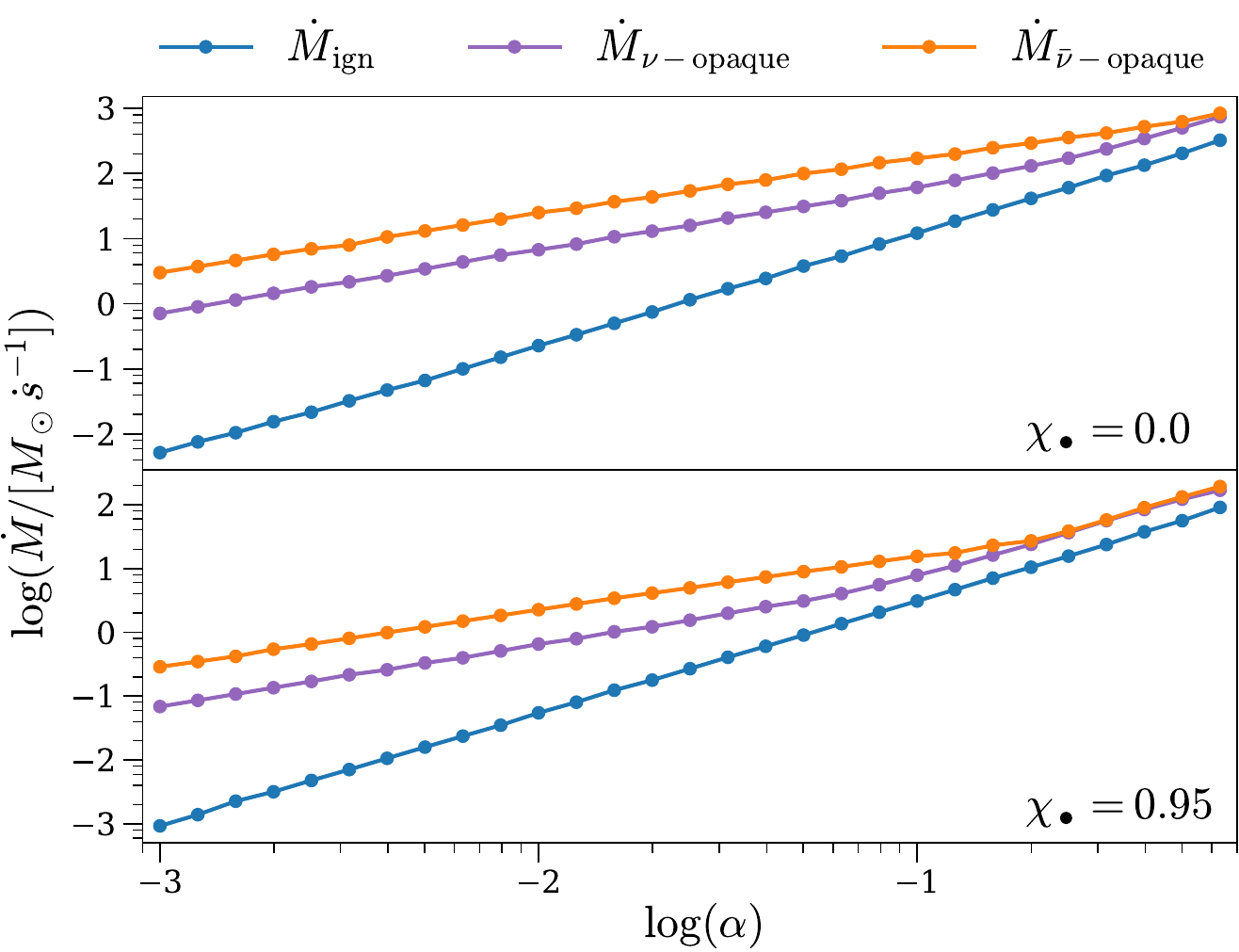}
\caption{Characteristic accretion rates as a function of $\alpha$-viscosity for non-rotating ($\chi_\bullet = 0.0$) and rapidly spinning ($\chi_\bullet = 0.95$) black holes of mass $M_\bullet = 200 M_\odot$, illustrating the occurrence of `avoided crossings' of the ignition and opaque accretion thresholds.}
\label{fig:crossing}
\end{figure}

\begin{figure*}
\includegraphics[width=\textwidth]{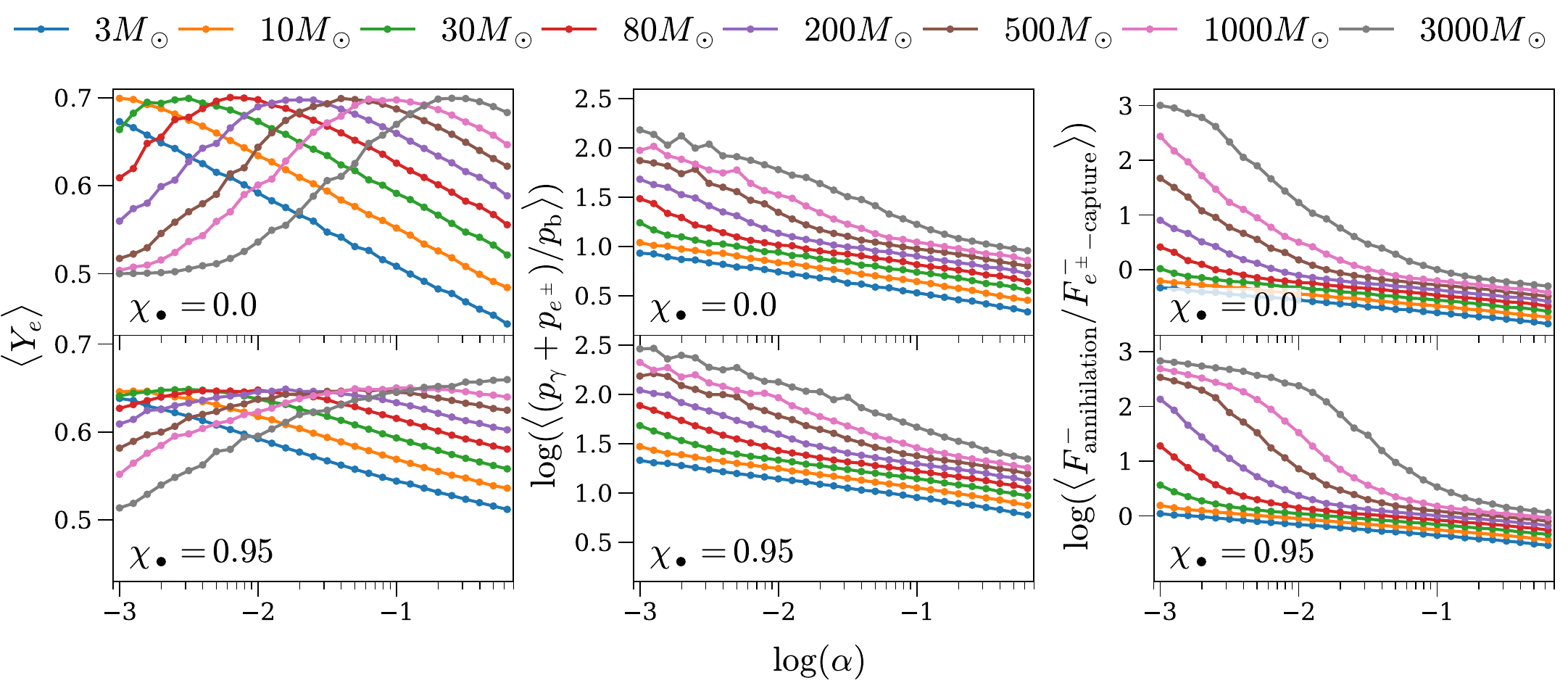}
\caption{Left: Average disk electron/proton-fraction $\langle Y_e\rangle$ within $r\le 20 r_{\rm g}$ as a function of the $\alpha$-viscosity parameter for various black hole masses $M_\bullet$ at the ignition threshold ($\dot{M}=\dot{M}_{\rm ign}$) and different values $\chi_\bullet$ of the black hole spin. Center: Corresponding pressure ratios of radiation and $e^\pm$ pressure to baryon pressure for the same set of models. Right: Corresponding ratio of the neutrino cooling fluxes due to pair annihilation and pair-capture}
\label{fig:Averages_ignition}
\end{figure*}

\begin{table*}
\centering
\label{table:scalings}
\begin{tabular}{ccccccc}

\hline
\hline
\multirow{2}{*}{Threshold} & Prediction                                                  & Spin                  & Dominant&$C_{\rm char}$& \multicolumn{2}{c}{Average fit values}  \\
                          & $\dot{M} \propto M_\bullet^\beta \alpha^\gamma$             &$\chi_\bullet$        & Pressure &$[M_\odot \rm s^{-1}]$& $\beta$   & $\gamma$        \\ 
\hline
\multirow{2}{*}{Ignition}  & \multirow{2}{*}{($p\propto p_{\gamma,e^\pm}$) \; $M_\bullet^\frac{4}{3} \alpha^\frac{5}{3}$} & 0.0                 &   $ p_{\gamma,e^\pm}$   &0.0011& 1.28      & 1.73            \\
                           &                                                                                      & 0.95                &   $ p_{\gamma,e^\pm}$ &0.0003 & 1.25      & 1.77            \\
\hline
\multirow{4}{*}{$\nu$-opaque}&\multirow{2}{*}{($p\propto p_{\mathrm{b}}$)\; $M_\bullet \alpha$}& \multirow{2}{*}{0.0} & $ p_{\mathrm{b}}$  &0.0490& 1.23      & 0.96  \\
                        &                                                                       &                   & $ p_{\gamma,e^\pm}$  &1.6877*& 1.23      & 1.49  \\
                        &\multirow{2}{*}{($ p\propto p_{\gamma,e^\pm}$) \;$M_\bullet^{\frac{4}{3}} \alpha$}& \multirow{2}{*}{0.95} & $ p_{\mathrm{b}}$  &0.0051& 1.17      & 0.97  \\
                        &                                                                       &                     & $ p_{\gamma,e^\pm}$  &0.2830*& 1.26      & 1.65  \\
\hline
\multirow{3}{*}{$\bar{\nu}$-opaque}& ($p\propto p_{\mathrm{b}}$) \; $M_\bullet \alpha$ & 0.0        &   $ p_{\mathrm{b}}$  &0.1629& 1.21      & 0.86            \\
                        &\multirow{2}{*}{($ p\propto p_{\gamma,e^\pm}$) \; $M_\bullet^{\frac{4}{3}} \alpha$}& \multirow{2}{*}{0.95} & $ p_{\mathrm{b}}$  &0.0158& 1.18      & 0.86  \\
                        &                                                                       &                      & $ p_{\gamma,e^\pm}$  &0.3318*& 1.21      & 1.57  \\
\hline
\multirow{2}{*}{$\nu$-trapped} & ($p\propto p_{\mathrm{b}}$) \; $M_\bullet \alpha^0$ & 0.0       &        $ p_{\gamma,e^\pm} \sim p_{\mathrm{b}}$                &15.468& 1.34      & 0.24            \\
                           & ($ p \propto p_{\gamma,e^\pm}$) \; $M_\bullet^\frac{4}{3} \alpha^\frac{1}{3}$  & 0.95                  &    $ p_{\gamma,e^\pm} \sim p_{\mathrm{b}}$                &2.1407& 1.35      & 0.21            \\
\hline
\multirow{2}{*}{$\bar{\nu}$-trapped}& ($p\propto p_{\mathrm{b}}$) \; $M_\bullet \alpha^0$ & 0.0                   &   $ p_{\gamma,e^\pm} \sim p_{\mathrm{b}}$  &36.291& 1.29      & 0.19            \\
                           & ($ p\propto p_{\gamma,e^\pm}$) \;  $M_\bullet^\frac{4}{3} \alpha^\frac{1}{3}$      & 0.95                  &  $ p_{\gamma,e^\pm} \sim p_{\mathrm{b}}$   &4.4729& 1.31      & 0.15            \\
\end{tabular}
\caption{Numerical results for the parameters $C_{\rm char}$, $\beta$, and $\gamma$ that characterize the scalings of the various characteristic accretion rates with black hole mass and $\alpha$-viscosity, $\dot{M}_{\rm char} = C_{\rm char} (M_\bullet/3M_\odot)^\beta (\alpha/0.01)^\gamma$ (Equation \eqref{eq:characteristic_scaling}). The values of $\beta$ and $\gamma$ represent, respectively, averages of the fits to the numerical accretion disk results for all values of $\alpha$ and $M_\bullet$ considered. The corresponding analytic approximations computed in Section~\ref{sec:theoretical_scalings} are also listed, together with the assumed respective dominant pressure component. We distinguish between different segments that arise due to a change in the dominant pressure regime at `avoided crossings' of the neutrino opaque and ignition thresholds at some value $\alpha_{\rm c}$, which depends on $\chi_\bullet$ and $M_\bullet$ (see the text for details). Asterisks * refer to cases with avoided crossings normalized to $\alpha = 0.3$.}
\end{table*}

Figures \ref{fig:MdotvsM} and \ref{fig:MdotvsAlpha} summarize results of a numerical exploration of the characteristic accretion rates across the vast parameter space of accretion flows around black holes of $M_\bullet=3-3000\,M_\odot$ with $\alpha$-viscosities between $\alpha = 0.001-0.5$. Every data point represents an accretion disk model at a given characteristic accretion rate, numerically identified by exploring the respective neighborhood in parameter space with a targeted set of numerical disk models. To each series of identified models for a fixed $\alpha$-viscosity parameter (Figure~\ref{fig:MdotvsM}) or fixed black hole mass (Figure~\ref{fig:MdotvsAlpha}) and given characteristic accretion rate $\dot{M}_{\rm char}=\{\dot{M}_{\rm ign}, \dot{M}_\nu, \dot{M}_{\bar\nu}, \dot{M}_{\nu-\text{trap}}, \dot{M}_{\bar{\nu}-\text{trap}}\}$, power-law segments of the form
\begin{equation}\label{eq:characteristic_scaling}
    \dot{M}_{\rm char} = C_{\rm char} \left(\frac{M_\bullet}{3 M_\odot }\right)^\beta \left(\frac{\alpha}{0.01}\right)^{\gamma}
\end{equation}
were fitted, the inferred parameters of which are listed in Table~\ref{table:scalings}. Figures~\ref{fig:MdotvsM}, \ref{fig:MdotvsAlpha}, and Table~\ref{table:scalings} show that all characteristic accretion rates $\dot{M}_{\rm char}=\{\dot{M}_{\rm ign}, \dot{M}_\nu, \dot{M}_{\bar\nu}, \dot{M}_{\nu-\text{trap}}, \dot{M}_{\bar{\nu}-\text{trap}}\}$ can be well described by power-law segments of this form throughout the parameter space probed. Broken power laws are a result of the critical accretion rates sweeping through different pressure regimes as `avoided crossings' of characteristic accretion rates occur (see below).

In terms of the ignition threshold, the average coefficients obtained through fits to the numerical results are remarkably close to the analytic predictions for a radiation and $e^\pm$-pressure dominated plasma, $\beta\approx 4/3$ and $\gamma \approx 5/3$, throughout the parameter space probed here. This is consistent with a weak and opposing dependence of $p_\gamma/p_{\rm b}\propto \alpha^{-1/6} M_\bullet^{1/6}$ at $\dot{M}=\dot{M}_{\rm ign}$ (see Equation \eqref{eq:pressure_ratio_Mdotign}) and is remarkable in light of the several simplifying assumptions adopted in the analytic calculations.  A similar conclusion holds for the neutrino trapping thresholds, for which $\beta\approx 4/3$ and $\gamma\lesssim 1/3$ throughout the parameter space. This is consistent with a predominantly radiation and $e^\pm$-pressure dominated regime even at small values of $\alpha$ and $M_\bullet$ (see also Figure~\ref{fig:multi_map} (III)) and an increasing but week dependence $p_\gamma/p_{\rm b}\propto \alpha^{1/6} M_\bullet^{1/6}$ at $\dot{M}=\dot{M}_{\nu-\rm{trap}}$ (see Equation \eqref{eq:pressure_ratio_Mdotnu-trap}).

The scaling behavior of the neutrino opaque thresholds is somewhat more complicated. At sufficiently small values of $\alpha$, $\dot{M}_{\nu,\nua}$ scales with $\gamma \approx 1$, as predicted by either pressure regime (Equation \eqref{eq:Mdotnu_scaling}). At some value $\alpha_{\rm c}$, which depends on $\chi_\bullet$ and $M_\bullet$, the scaling approaches $\gamma \approx 5/3$, i.e.~that of the ignition threshold (see also Figure \ref{fig:crossing}). Since the ratio $p_\gamma/p_{\rm b}$ is not expected to scale with $\alpha$ in either pressure regime (Equation \eqref{eq:pressure_ratio_Mdotnu}), this transition into a new segment is not due to a change of pressure regimes. It is rather due to an `avoided crossing' of the ignition and opaque thresholds (Figure~\ref{fig:crossing}). Due to the weaker dependence of $\dot{M}_{\nu,\nua}$ on both $\alpha$ and $M_\bullet$ relative to that of $\dot{M}_{\rm ign}$, the two thresholds are predicted to cross at some $\alpha_{\rm c}$. As a result of $\dot{M}_{\nu,\nua}$ approaching $\dot{M}_{\rm ign}$ from above, the conditions at $\dot{M}_{\nu,\nua}$ become necessarily dominated by radiation and $e^\pm$ pressure (see discussion above), and thus the scaling with $M_\bullet$ changes from $\beta\gtrsim 1$ to $\approx 4/3$ (Equation \eqref{eq:Mdotnu_scaling}), which can be noticed in Figure~\ref{fig:MdotvsM} for some $\alpha \gtrsim \alpha_c$ that depends on $\chi_\bullet$ and $M_\bullet$. The additional change in the scaling of $\dot{M}_{\nu,\nua}$ with $\alpha$ to $\gamma\approx 5/3$ becomes necessary as $\dot{M}_{\nu,\nua}$ must not cross below $\dot{M}_{\rm ign}$. Whereas in principle the accreting plasma can become dense enough to be optically thick to $\sim$MeV neutrinos at the ignition threshold, this cannot arise in practice, as the disk temperatures first need to approach the proton-neutron mass difference near the ignition threshold to generate neutrinos that can be thermalized in the flow. We call this phenomenon an `avoided crossing' of the critical accretion threshold levels.

\subsection{Disk composition at the characteristic thresholds}
\label{sec:results_disk-composition}

\begin{figure*}
\includegraphics[width=\textwidth]{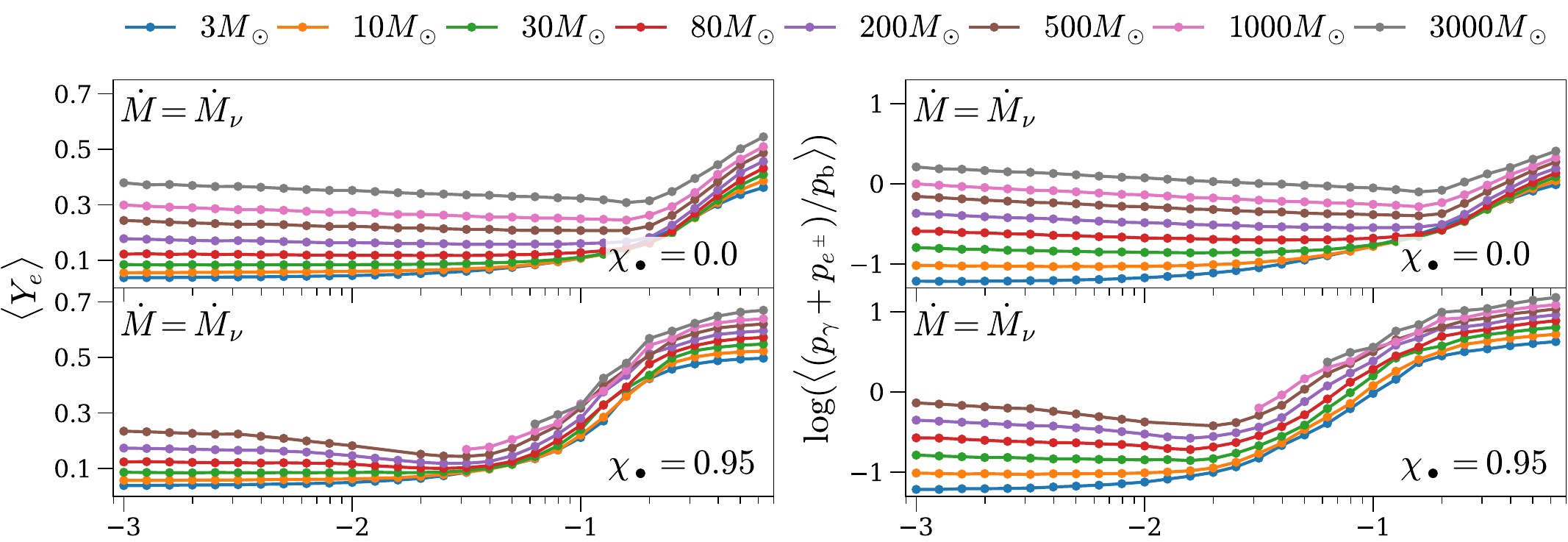}\\
\includegraphics[width=\textwidth]{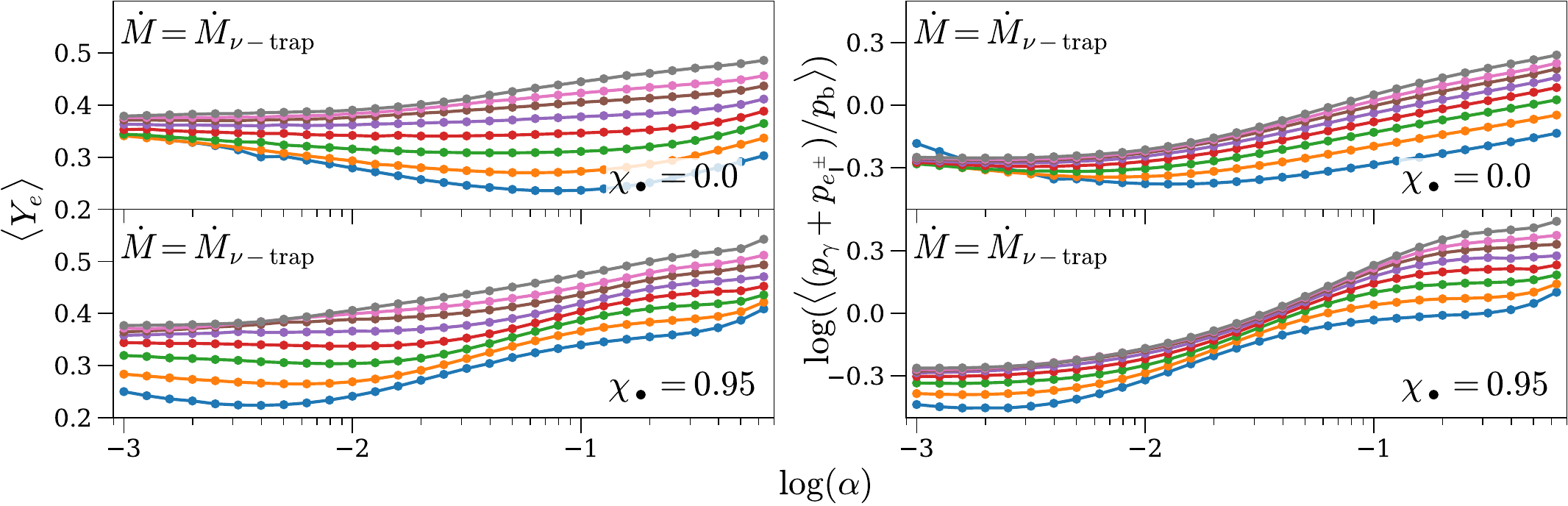}
\caption{Top: Analogous to the left and center columns of Figure \ref{fig:Averages_ignition}, for disks at the neutrino-opaque threshold ($\dot{M}=\dot{M}_{\nu}$). Bottom: Idem for disks at the neutrino-trapped threshold ($\dot{M}=\dot{M}_{\nu-\rm trap}$).}
\label{fig:Averages_opaque_trapped}
\end{figure*}

The results presented in Figures \ref{fig:M_dot_nu} and \ref{fig:M_dot_diff} illustrate that neutrino-cooled accretion disks around black holes can have a very heterogeneous composition as a function of radius. This is a consequence of different radial regions residing in different accretion regimes according to the characteristic radii $r_{\rm ign}$, $r_{\nu,\nua}$, $r_{\nu,\nua-\text{trap}}$. Therefore, disk outflows originating in different radial regimes of a neutrino-cooled accretion disk may also possess different composition. Below we discuss the rest-mass averaged composition of the innermost 20 gravitational radii of the disk at the characteristic accretion rates. We argue that this inner region is most indicative of the outflow composition insofar as realistic neutrino-cooled accretion disks in collapsars are truncated at an accretion shock at radii of $\sim\!10-100r_{\rm g}$ as the core of the progenitor star collapses onto the equatorial plane and circularizes at increasing radii with time, determined by angular momentum conservation as successively outer layers start to collpase \citep{macfadyen_collapsars_1999,siegel_super-kilonovae_2022}. Outflows from the innermost accretion disk are also most likely to be unbound from the system, perhaps with the help of a magnetically dominated jet that may form in the polar regions and help to unbind the outer stellar layers in polar regions.

\subsubsection{The ignition threshold and the breakdown of protonization}
\label{sec:results_break-down}

Figure \ref{fig:Averages_ignition} summarizes results from a sweep over parameter space for the mass-averaged $Y_e$ at $\dot{M}=\dot{M}_{\rm ign}$. As expected from the discussion in Section~\ref{sec:results_proton-rich_regime}, the inner disk is proton-rich with $\langle Y_e\rangle\approx 0.5-0.7$ as a result of neutrino cooling under radiation and $e^\pm$-pressure dominated conditions as well as mild to vanishing electron degeneracy. Slightly neutron-rich conditions can only be achieved for non-rotating or slowly rotating black holes of mass $M_\bullet \lesssim \text{few}\times M_\odot$ at large values of $\alpha \gtrsim 0.1$, as a result of the onset of degeneracy $\langle \eta_e \rangle \gtrsim 0.3$. At sufficiently large viscosities $\alpha\gtrsim0.1$, the opposing scaling $(Y_e-0.5)\propto \alpha^{-1/6} M_\bullet^{1/6}$ with $\alpha$ and $M_\bullet$ as predicted by Equation~\eqref{eq:disk_temperature_scaling} is evident. This trend is reversed, however, toward smaller values of the viscosity and large black hole masses, due to cooling via $e^\pm$-pair annihilation starting to dominate over $e^\pm$ capture in the increasingly radiation and $e^\pm$-pressure dominated accretion flows. Neutrinos created by pair annihilation under non-degenerate conditions do not change the composition and thus $\langle Y_e\rangle$ approaches $\approx\!0.5$ for $M_\bullet\gtrsim 1000 M_\odot$ at $\alpha = 0.001$. Associated with this is the drop in temperature $T\propto \alpha^{1/6} M^{-1/6}$ (Equation~\eqref{eq:disk_temperature_scaling}) with decreasing viscosity and increasing black hole mass below the proton-neutron mass difference of $\approx\!1.3$\,MeV and the dissociation threshold, which gradually suppresses the $e^\pm$-capture reactions. This marks the breakdown of protonization at small $\alpha$-viscosities $\alpha\sim 0.001$ and large black hole masses $M_\bullet \gtrsim \text{few}\times 1000 M_\odot$.

\subsubsection{The opaque threshold and the breakdown of neutronization}

The overall capability of neutronization of the inner accretion disks is well traced by the physical conditions at $\dot{M}\approx \dot{M}_{\nu,\nua}$. Our results shown in Figure~\ref{fig:ye_map} indicate that neutrino-cooled disks must reach accretion rates $\dot{M}_{\rm ign}\lesssim \dot{M} \lesssim \dot{M}_\nu$ in order to efficiently neutronize the nucleon plasma (see also Section~\ref{sec:results_neutron-rich_regime}). This is a necessary but not sufficient condition. Figure~\ref{fig:ye_map} displays a striking rise in the `valley floor' of the rest-mass density averaged $Y_e$ within $r\le 20 r_{\rm g}$ at $\dot{M}\sim \dot{M}_{\nu,\nua}$ from $\langle Y_e \rangle \approx 0.1$ for stellar mass black holes $M_\bullet\lesssim 10 M_\odot$ to $\langle Y_e \rangle \gtrsim 0.45$ at $M_\bullet \gtrsim \text{few} \times 1000 M_\odot$, depending on the spin of the black hole. The disk composition, however, also depends on the $\alpha$-viscosity, and a more comprehensive coverage of the parameter space in terms of $\langle Y_e \rangle$ near $\dot{M}\sim \dot{M}_{\nu,\nua}$ is provided by Figure~\ref{fig:Averages_opaque_trapped}. In this section, we focus mainly on the composition around $\dot{M}\sim \dot{M}_{\nu,\nua}$, as this corresponds to the most neutron-rich part of the parameter space, at least at sufficiently small values of $\alpha$ ($\alpha \lesssim 0.1-0.2)$. In principle, neutronization can break down as it is based on a number of assumptions: i) the dominance of baryon pressure over radiation and $e^\pm$ pressure, ii) the capture of relativistic electrons and positrons on free nucleons dominates neutrino cooling, and iii) local turbulence through the MRI described by an effective $\alpha$-viscosity dominates over instabilities due to self-gravity in setting the accretion rate. We explore the breakdown of neutronization in the `neutron-rich valley' around $\dot{M}\sim \dot{M}_{\nu,\nua}$ in more detail in the following.

As evident from Figure~\ref{fig:Averages_opaque_trapped}, neutronization in the inner accretion disk becomes increasingly suppressed with increasing values of $\alpha$, starting at values $\alpha \gtrsim 0.2$ for non-rotating black holes and $\alpha\gtrsim 0.1$ for $\chi_\bullet = 0.95$. We identify the reason for this behavior as an avoided crossing of the opaque and the ignition thresholds, which occurs precisely around the respective values of $\alpha$ (see Figure~\ref{fig:MdotvsAlpha}). As a result of this, the inner accretion flow becomes necessarily dominated by radiation and $e^\pm$ pressure with increasing $\alpha$ (Section~\ref{sec:results_scaling_relations_Mdot}), which is clearly illustrated by the bottom panel of Figure~\ref{fig:Averages_opaque_trapped}. The ratio of cooling through $e^\pm$ capture relative to pair annihilation scales proportionally (see Equation~\eqref{eq:F_ann_Fcapture}), and thus fewer and fewer emitted neutrinos are associated with neutronization. This results in an increase of $Y_e$ with increasing $\alpha$ to a composition comparable to that at the ignition threshold, as expected for an avoided crossing.

At sufficiently small values of the effective viscosity ($\alpha \lesssim 0.2$ for $\chi_\bullet = 0$ and $\alpha \lesssim 0.1$ for rapidly rotating black holes with $\chi_\bullet = 0.95$), the composition at $\dot{M}=\dot{M}_{\nu}$ approaches a nearly constant value for any given black hole mass. This is mainly regulated by the pressure ratio $p_{\gamma,e^\pm} / p_{\rm b}$, which also approaches an approximately constant value at small $\alpha$ for a given black hole mass (bottom panel of Figure~\ref{fig:Averages_opaque_trapped}). This is expected from Equation~\eqref{eq:pressure_ratio_Mdotnu}, regardless of which pressure component dominates. With increasing black hole mass, the pressure ratio increases as $\propto M_\bullet$ in the baryon-pressure dominated regime for stellar-mass black holes and gradually transitions into the weaker scaling of $\propto M_\bullet^{1/6}$ once disks at $\simeq 1000 M_\odot$ transition into the radiation and $e^\pm$-pressure dominated regime (essentially independent of the black hole spin $\chi_\bullet)$. Note that the temperature in the baryon-pressure dominated regime is approximately independent of $\alpha$ and $M_\bullet$ (see Equation~\eqref{eq:opaque_thresh_step1}) and decreases moderately with black hole mass $\propto M^{-1/6}_\bullet$ (Equation \eqref{eq:disk_temperature_scaling}) once the disks become radiation and $e^\pm$-pressure dominated. As $p_{\gamma,e^\pm} / p_{\rm b}$ increases to an order unity fraction at $M_\odot \gtrsim 100 M_\odot$ (roughly independent of the black hole spin), the ratio of cooling through $e^\pm$ capture relative to pair annihilation (see Equation~\eqref{eq:F_ann_Fcapture}) reaches the percent level and $Y_e$ starts to increase markedly as a result of fewer neutrinos participating in neutronization via the charged current interactions (Equations \eqref{reaction:heatcool1} and \eqref{reaction:heatcool2}).

Much of the parameter space for large black hole masses and accretion rates corresponds to gravitationally unstable disks $Q_{\rm grav} < 1$ (see Figure~\ref{fig:multi_map}), for which our assumption of stationarity and that of a pure Kerr background spacetime does not hold anymore. Gravitationally unstable disks tend to self-regulate to restore $Q_{\rm grav} \approx 1$ by enhancing angular momentum transport through gravitationally driven turbulence \citep{gammie_nonlinear_2001}. This corresponds to an increase in the effective $\alpha$-viscosity parameter, which, at least in a stationary sense, starts to suppress neutronization at values $\alpha \gtrsim 0.1-0.2$ (see above; Figure~\ref{fig:Averages_opaque_trapped}). The average composition of the inner accretion disk reported here thus likely reflects a lower limit in the gravito-turbulent regime.

\subsubsection{The trapping thresholds}

Except for very light black holes with $M_\bullet \lesssim 3 M_\odot$ and $\alpha\approx\text{few}\times 10^{-3}$ ($\chi_\bullet=0.95$) or $\alpha\approx0.1$ ($\chi_\bullet=0$), the inner accretion flow at $\dot{M}=\dot{M}_{\nu-\text{trap}}$ reaches an average composition of $\langle Y_e \rangle >  0.25$ (Figure~\ref{fig:Averages_opaque_trapped}), which closes off the `valley floor' of neutronization toward high accretion rates (Figure~\ref{fig:ye_map}). This increase in $Y_e$ relative to $\dot{M}_\nu$ is the result of an overall suppression of lepton number emission through electron capture under degenerate conditions due to the onset of neutrino trapping and an increase in the ratio of pair annihilation relative to $e^\pm$ captures related to an increase in radiation and $e^\pm$ pressure (Equation~\eqref{eq:F_ann_Fcapture}; Figure~\ref{fig:Averages_opaque_trapped}).

\section{Discussion}
\label{sec:discussion}

\subsection{Comparison with 3D GRMHD simulations}
\label{sec:results_comparison}

We find good agreement between the predictions of our model and the exploration of the ignition threshold across $M_\bullet\sim 80-3000 M_\odot$ with three-dimensional GRMHD simulations of isolated collapsar disks by \citet{agarwal_ignition_2025}. For technical reasons, the ignition threshold in \citet{agarwal_ignition_2025} was defined by the criterion $\langle \eta_e \rangle = 0.5$ when averaged between $(1-3)\times r_{\rm ISCO}$. In both studies, the expected approximate scaling $\dot{M}_{\rm ign}\propto M_\bullet^{4/3}\alpha^{5/3}$ is well reflected by the respective numerical results. 

To further test the absolute normalization of the scaling relation, we analyzed models with the above mentioned criterion $\langle\eta_e\rangle=0.5$ using the parameters $\alpha = 0.009$ for the self-consistent, effective GRMHD viscosity, $\chi_\bullet = 0.8$, and $M_\bullet = (80,\,500,\,1000)M_\odot$ reported by \citet{agarwal_ignition_2025}. We find that the predictions of our model of $\dot{M}_{\rm ign}\approx (0.05,\,0.34,\,1.28)M_\odot\,\text{s}^{-1}$ for these black hole masses either agree within error bars with those of \citet{agarwal_ignition_2025}, or are up to a factor of $\approx 2$ smaller than their uncertainty range in \citet{agarwal_ignition_2025}. This level of agreement is remarkable given that the accretion disks in \citet{agarwal_ignition_2025} do not have infinite spatial extent (they are rather compact) and `sweep across' the ignition threshold (i.e.~they are not truly stationary).

Finally, we also find that the breakdown of neutronization at the ignition threshold and effective $\alpha\approx 0.009$ of \citet{agarwal_ignition_2025} is consistent with the predictions of our model (see Figure~\ref{fig:ye_map} for a rough comparison).

\subsection{PISN events and their super-kilonova signatures}
\label{sec:results_superKNe}

Figure~\ref{fig:ye_map} shows example trajectories of the accretion process in massive collapsars with rapidly rotating progenitor stars just above the pair-instability supernova mass gap ($\gtrsim 130M_\odot$) at the onset of core-collapse, as in \citealt{siegel_super-kilonovae_2022}. We generated these trajectories with the semi-analytical collapsar model of \citealt{siegel_super-kilonovae_2022}, assuming $\alpha=0.06$ for disk viscosity as appropriate for Figure~\ref{fig:ye_map}. We used the progenitor models of \citealt{renzo_predictions_2020} with initial He-core mass of $200 M_\odot$ (red line), $220 M_\odot$ (light blue line), and $250M_\odot$ (pink line), the latter being the exact same model \texttt{250.25} of \citealt{siegel_super-kilonovae_2022} (see their Figure~3). All progenitor models are endowed with a parametrized (broken power-law) angular momentum profile at the time of core-collapse as in \citealt{siegel_super-kilonovae_2022}, assuming the fiducial parameter choices as in their Figure 1.

These rapidly rotating progenitors give rise to collapsar systems, which populate the pair-instability supernova mass gap between $\approx\!50-130\,M_\odot$ with black holes `from above', due to substantial mass loss (typically up to tens of solar masses) through disk outflows during the accretion process. Here, the models yield final black hole masses between $\approx\!80-100M_\odot$ (Figure~\ref{fig:ye_map}).

Based on analogous scaling relations to those derived in Section~\ref{sec:theoretical_scalings}, obtained for Newtonian accretion disks, \citealt{siegel_super-kilonovae_2022} predict the possibility of massive, neutron-rich disk outflows that can escape the collapsing star, carry up to several solar masses of $r$-process generating material, and can thus produce ``super-kilonovae'', scaled-up versions of kilonovae. Although we find slight modifications of the scaling relations to be appropriate in this context (see Section~\ref{sec:results_scaling_relations_Mdot}; dominating baryon pressure at $\dot{M}\sim\dot{M}_{\nu,\nua}$ as compared to radiation and $e^\pm$ pressure dominance), our present numerical results show that such super-collapsar accretion histories can indeed transition through a neutron-rich regime, with the inner accretion disk being dominated by a well-defined, asymptotic, low-$Y_e$ value of $Y_e \lesssim 0.1$ throughout an extensive part of the entire accretion process (Figure~\ref{fig:ye_map}). Whereas direct comparison with the accretion history of the time-dependent models of \citealt{siegel_super-kilonovae_2022} is intrinsically difficult, for the black hole spin evolves during the collapse process, comparison with the inner structure of the accretion disks of our present stationary models confirms the basic assumptions of neutronization underlying the semi-analytic model for super-kilonovae of \citealt{siegel_super-kilonovae_2022}.

\subsection{Implications for r-process in GRB221009A}\label{sec:results_BOAT}

The long-duration gamma-ray burst (GRB) GRB221009A is the brightest GRB ever detected in terms of peak flux and fluence and has the highest isotropic equivalent total energy ($E_{\rm iso}\simeq 1.0\times 10^{55}$\,erg) detected so far as well as an exceptionally large (in the 99th percentile of GRBs) isotropic equivalent peak luminosity of $L_{\rm iso}\simeq 1.0\times 10^{54}\,\text{erg}\,\text{s}^{-1}$ \citep{burns_grb_2023,lesage_fermi-gbm_2023,williams_grb_2023}. If the gamma-ray luminosity of the jet tracks the accretion rate onto the black hole, $L_\gamma = \eta \dot{M}c^2$ with some efficiency $\eta\ll 1$, the large accretion rate implied by GRB221009A suggests that it surpassed the ignition threshold to produce $r$-process elements in disk outflows \citep{siegel_collapsars_2019}. Whereas this together with its proximity (redshift of $z=0.151$, \citealt{malesani_brightest_2025}) makes GRB221009A, in principle, a promising target to search for $r$-process nucleosynthesis, no such $r$-process signatures were identified in this event \citep{blanchard_jwst_2024}.

Should GRB221009A have a wide jet opening angle $\theta_{\rm j}\simeq 27^\circ$ as inferred by \citet{kann_grandma_2023}, we estimate the accretion rate during the GRB phase to
\begin{equation}
    \dot{M} = 6.2\, M_\odot \text{s}^{-1} \left(\frac{L_{\gamma,{\rm iso}}}{10^{54}\text{erg}\,\text{s}^{-1}}\right)\left(\frac{\eta}{0.01}\right)^{-1}\left(\frac{\theta_{\rm j}}{27^\circ}\right)^2,
\end{equation}
and thus, with the results of Section~\ref{sec:results_scaling_relations_Mdot},
\begin{eqnarray}
    \frac{\dot{M}}{\dot{M}_{\nu-\text{trap}}} &\simeq& 3 \left(\frac{L_{\gamma,{\rm iso}}}{10^{54}\text{erg}\,\text{s}^{-1}}\right)\left(\frac{\eta}{0.01}\right)^{-1} \nonumber\\
    &&\times\left(\frac{\theta_{\rm j}}{27^\circ}\right)^2 \left(\frac{M_\bullet}{3M_\odot}\right)^{-1.35}\mskip-5mu\left(\frac{\alpha}{0.01}\right)^{-0.21}.
    \label{eq:Mdot_Mdot-trap_GRB221009A}
\end{eqnarray}
Here, we have assumed a rapidly spinning black hole ($\chi_\bullet = 0.95$) as typical for collapsar black holes. Given the presence of an associated typical Type Ic-BL supernova \citep{blanchard_jwst_2024}, we also normalized to a typical black hole mass during the early GRB phase of an ordinary collapsar \citep{siegel_collapsars_2019}. Finally, we normalize to a typical value of the $\alpha$-viscosity as obtained from three-dimensional GRMHD simulations \citep{balbus_instability_1998,hawley_dynamical_2002,penna_shakura-sunyaev_2013,siegel_collapsars_2019,de_igniting_2021,agarwal_ignition_2025}.

This estimate suggests that collapsar accretion disks in GRBs as luminous as GRB221009A may have surpassed the neutrino trapping threshold, such that the average electron-fraction in the inner disk increases beyond $\langle Y_e\rangle \gtrsim 0.25-0.3$ (see Section~\ref{sec:results_neutron-rich_regime}; Figure \ref{fig:Averages_opaque_trapped}), preventing significant production of lanthanides in the disk outflows. Even if $\langle Y_e\rangle$ is moderately lower, the strong irradiation of the outflows by neutrinos from the inner accretion disk in the regime $\dot{M}\gtrsim\dot{M}_{\nu}$ may suppress lanthanide production. More neutron-rich outflows at larger radii $r>r_{\nu-\text{trap}}$ may still synthesize lanthanides; however, collapsar accretion disks in the early GRB phase are expected to be very compact, as the progenitor layers collapse inside out, and thus it remains unclear whether and how much disk material from $r>r_{\nu-\text{trap}}$ can be unbound from the system at these early stages in the collapse process. A red excess relative to the supernova at late times in the light curves or significantly reddened spectra, as predicted for high-opacity, lanthanide-bearing collapsar outflows by \citealt{siegel_collapsars_2019} and \citealt{siegel_super-kilonovae_2022}, are thus not expected for exceptionally bright GRBs such as GRB221009A. Nevertheless, a search for spectral features of light $r$-process elements such as strontium as in kilonovae from neutron-star mergers \citep{watson_identification_2019,sneppen_emergence_2024} may be promising.

\section{Conclusions}
\label{sec:conclusions}

We present a one-dimensional, general-relativistic, viscous-hydrodynamic model of accretion disks around black holes, targeted at accretion flows in which weak interactions are ignited and neutrino emission cools the plasma. We focus on the compositional properties of the plasma in the inner accretion disk within tens of gravitational radii and explore the self-protonization and self-neutronization of the flow at sufficiently high accretion rates when ordinary nuclei are dissociated into a nucleon plasma, across the vast parameter space of accretion rates $\dot{M}\sim 10^{-6}-10^6 M_\odot \,\text{s}^{-1}$, black hole masses of $M_\bullet\sim 1 - 10^5 M_\odot$ and dimensionless spins of $\chi_\bullet \in [0,1)$, as well as $\alpha$-viscosity values of $\alpha\sim 10^{-3}-1$. 

Such accretion flows can arise in ordinary collapsars, thought to be the progenitors of long GRBs \citep{siegel_collapsars_2019}, in massive collapsars above the PISN mass gap, which may populate the PISN mass gap with black holes and give rise to super-kilonovae \citep{siegel_super-kilonovae_2022}, as well as during the collapse of supermassive stars \citep{agarwal_ignition_2025}. Although we assume equal numbers of neutrons and protons ($Y_e=0.5$) as an outer radial boundary condition of the accreting plasma, targeted to collapsars, the physics of the inner accretion flow becomes largely insensitive to this, and the model (with adapted boundary condition) as well as the concrete properties of the accretion flows discussed here are also broadly applicable to accretion disks that form in the aftermath of neutron-star mergers. 

The compositional properties of the accretion flows explored here can provide some indication for the prospects and expected outcomes of $r$-process nucleosynthesis in disk outflows. The predictions made here need to be checked with multi-dimensional simulations and detailed neutrino transport in future work. Our model allows for a computationally efficient scan of the vast parameter space, currently impossible to cover for multi-dimensional GRMHD or viscous hydrodynamic simulations, which can at most focus on a few selected parameter choices at a time \citep{just_r-process_2022,dean_collapsar_2024,issa_magnetically_2025,shibata_self-consistent_2025,agarwal_ignition_2025}. Furthermore, our model serves as a useful basis for the interpretation and analysis of the physics and dynamics of more detailed, non-stationary, three-dimensional GRMHD simulations by means of a relatively simple physical model. Our main conclusions may be summarized as follows.

\begin{enumerate}
    \item For a given black hole and self-consistent description of turbulence ($\alpha$-viscosity), the accretion state is mainly determined by the accretion rate onto the black hole. We find (Figures~\ref{fig:ye_map} and \ref{fig:multi_map}) that a characterization of the accretion state relative to certain critical thresholds is useful: the minimum accretion rates required to efficiently cool the inner flow ($\dot{M}_{\rm ign}$), to render the inner flow opaque to electron neutrinos and antineutrinos ($\dot{M}_{\nu}$ and $\dot{M}_{\nua}$), and to trap neutrinos and antineutrinos in the inner flow ($\dot{M}_{\nu-\text{trap}}$ and $\dot{M}_{\nua-\text{trap}}$).
    
    \item The accretion flow self-neutronizes due to the onset of electron degeneracy at $\dot{M}_{\rm ign}\lesssim \dot{M} \lesssim\dot{M}_{\nu}$ and enters a self-regulation regime based on electron degeneracy that balances viscous heating and neutrino cooling (Figures \ref{fig:M_dot_nu} (IV), \ref{fig:M_dot_diff} (IV)), entering a `valley floor' with a well defined, small value of $Y_e \lesssim0.1$ at around $\dot{M}\sim\dot{M}_{\nu,\nua}$ that can be analytically estimated (Equation~\eqref{eq:Y_e_trapped}; Figures~\ref{fig:ye_map}, \ref{fig:M_dot_nu} (I), \ref{fig:M_dot_diff} (I)).
    
    \item The inner accretion flow at $\dot{M}\lesssim \dot{M}_{\rm ign}$ as well as the flow in a radial range around the neutronized inner flow at $\dot{M}>\dot{M}_{\rm ign}$ protonizes as $(Y_e-0.5)\propto \alpha^{-1/6}M_\bullet^{1/6}$ (Equation~\eqref{eq:ye_proton_rich_MBH}) due to the proton-neutron mass difference skewing the balance of the charged current interactions toward positron capture onto neutrons in mildly to non-degenerate conditions (Figures~\ref{fig:ye_map} and \ref{fig:Averages_ignition}). The resulting $Y_e\approx 0.5-0.8$ can be well estimated analytically in these regions (Equation~\eqref{eq:ye_proton_rich}). A breakdown of protonization at $\dot{M}_{\rm ign}$ is observed according to the above scaling for $M_\odot\lesssim 3 M_\odot$ and $\alpha\lesssim 1$; a gradual breakdown also occurs at $M_\odot\gtrsim 3000 M_\odot$ for $\alpha \lesssim 0.001$ (Figure~\ref{fig:Averages_ignition}) due to $T\propto \alpha^{1/6}M_\odot^{-1/6}$ dropping below $\simeq\!1$\,MeV required for dissociation and $e^\pm$-capture and increasing dominance of radiation and $e^\pm$ pressure.

    \item With expansion speeds of $\approx 0.03-0.12 c$ set by the recombination of nucleons into $\alpha$-particles and entropies of $\sim 10 k_{\rm B}$ per baryon (e.g., Figure~\ref{fig:multi_map}), disk outflows from the neutron-rich parts of the accretion disks can synthesize $r$-process elements (Section~\ref{sec:results_neutron-rich-nucleosynthesis}). Disks around black holes as massive as $\sim10^3M_\odot$ can give rise to neutron-rich winds; however, the production of lanthanides is disfavored for black holes with $\chi_\bullet = 0-0.95$ heavier than $\sim 200-500 M_\odot$, due to increasing proton-fractions ($Y_e\gtrsim 0.25$) and/or increasing expansion timescales ($\tau\gtrsim 0.5$\,s) that lead to late-time reheating of the ejecta and a restart of the $r$-process at higher $Y_e$. Associated super-kilonovae may thus appear `red', peaking in the infrared, for $M_\bullet\lesssim 200-500 M_\odot$ corresponding to $\chi_\bullet = 0-0.95$, and `blue', peaking in the optical wavelength range, for larger black hole masses.
    
    \item Using a simple disk outflow model (Appendix~\ref{app:outflow_model}), we find that outflows from the proton-rich parts of the accretion disks may give rise to the synthesis of proton-rich isotopes via the $\nu p$-process, specifically for large black hole masses $M_\bullet \gtrsim 100 M_\odot$ and accretion rates $\dot{M}\approx \dot{M}_\nu$ (Figure~\ref{fig:Delta_n}). Given the large uncertainties in our estimates, further investigation is required in this direction to assess in more detail the contribution of such disks to the nucleosynthesis of neutron-deficient p-nuclei.

    \item The characteristic threshold accretion rates approximately follow power laws of the form $\dot{M}_{\rm char} \propto M_\bullet^\beta \alpha^{\gamma}$ (Figures~\ref{fig:MdotvsM} and \ref{fig:MdotvsAlpha}), with exponents (Table~\ref{table:scalings}) in good agreement with approximate analytic scaling relations derived in Section~\ref{sec:theoretical_scalings}. For the opaque thresholds, broken power laws apply due to the presence of `avoided crossings' with the ignition threshold, which enforce a transition from baryon to radiation and $e^\pm$-pressure dominated plasma (Figure~\ref{fig:crossing}).

    \item The onset of a breakdown of neutronization in the low-$Y_e$ valley floor around $\dot{M}\sim\dot{M}_{\nu}$ is observed at $\alpha\gtrsim 0.1-0.2$, depending on the black hole spin (Figure~\ref{fig:Averages_opaque_trapped}). This is a result of an avoided crossing of the opaque and ignition thresholds (Section~\ref{sec:results_break-down}). The nearly constant `valley floor' for a given black hole mass at $\alpha\lesssim 0.1-0.2$ increases from $Y_e\lesssim 0.1$ at $M_\bullet\approx \text{few}\times M_\odot$ to $Y_e\approx 0.5$ at several thousand $M_\odot$ as a result of the disks becoming increasingly radiation and $e^\pm$-pressure dominated.

    \item The predictions of our model for both the absolute normalization and relative scaling of the onset of degeneracy in the inner disk ($\langle\eta_e\rangle=0.5$) are in good agreement with the 3D GRMHD simulations of collapsar accretion disks across $M_\odot =80-3000 M_\odot$ by \citet{agarwal_ignition_2025} (their `ignition threshold'). Our results are also consistent with the 3D GRMHD simulations of post-merger accretion disks of \citet{de_igniting_2021}, who report an ignition threshold at $\approx 10^{-3}M_\odot\,\text{s}^{-1}$ for a remnant black hole of $M_\bullet = 3 M_\odot$ and an effective $\alpha$-viscosity of $\alpha\approx 1 \times 10^{-2}$.
    
    \item Our results confirm the neutronization of accretion disks around black holes of tens of $M_\odot$ at $\dot{M}\gtrsim \dot{M}_{\rm ign}$ as realized in models of massive collapsars. These findings thus support the basic assumptions of neutronization underlying the semi-analytic model for super-kilonovae of \citet{siegel_super-kilonovae_2022}.

    \item We estimate that the accretion flow in the collapsar triggering the brightest GRB of all times, GRB221009A, likely resided in the neutrino-trapped regime while the GRB was powered (Equation~\eqref{eq:Mdot_Mdot-trap_GRB221009A}), suppressing the production of lanthanides (Figure~\ref{fig:Averages_opaque_trapped}). Reddening of the supernova lightcurves at late times due to high-opacity, lanthanide-bearing collapsar outflows as predicted by \citet{siegel_collapsars_2019} and \citet{siegel_super-kilonovae_2022} may thus not be expected for exceptionally bright GRBs such as GRB221009A.
    
\end{enumerate}

\begin{acknowledgements} 

The authors thank Geoffrey Ryan for valuable discussions during the early stages of this project and for providing optimized code to compute Fermi-Dirac integrals (\texttt{fdlo}, \citealt{ryan_geoffryanfdlo_2022}). We thank Aman Agarwal for providing the trajectories of collapsar accretion histories in Figure~\ref{fig:ye_map}. We also thank N.~C.~Stone and Y.~Lerner for discussion and feedback on the disk model, and B.~D.~Metzger for comments on the manuscript. JHM acknowledges the Perimeter Institute for Theoretical Physics for providing the research environment to start this project through their Perimeter Scholars International Program. The authors gratefully acknowledge the computing time made available to them on the high-performance computer ``Lise'' at the NHR Center NHR@ZIB and ``Emmy'' at the NHR Center NHR@Göttingen. These centers are jointly supported by the Federal Ministry of Education and Research and the state governments participating in the NHR (www.nhr-verein.de/unsere-partner).

\end{acknowledgements}

\software{\texttt{Matplotlib} \citep{hunter_matplotlib_2007}, \texttt{NumPy} \citep{harris_array_2020}, \texttt{SciPy} \citep{virtanen_scipy_2020}, \texttt{hdf5} \citep{the_hdf_group_hierarchical_2025} and \texttt{fdlo} \citep{ryan_geoffryanfdlo_2022}.}

\appendix
\section{Disk equations}
\label{app:disk_structure_equations}

In this appendix, we derive all relevant structure equations of an axisymmetric, neutrino-cooled, thin accretion disk in Kerr spacetime. Following the nomenclature of \citet{novikov_astrophysics_1973}, \citet{page_disk-accretion_1974}, and \citet{chen_neutrino-cooled_2007}, we define the following dimensionless auxiliary functions
\begin{equation}
\label{eq:auxiliary_functions}
    \begin{aligned}
    \Acu &= 1 + \frac{\chi_\bullet^2}{x^4} + \frac{2\chi_\bullet^2}{x^6}, \\
    \Dcu & = 1 - \frac{2}{x^2} + \frac{\chi_\bullet^2}{x^4},
    \end{aligned}
    \hspace{5.2mm}
    \begin{aligned}
    \Bcu &= 1 + \frac{\chi_\bullet}{x^3}, \\
    \Fcu & = 1-\frac{2\chi_\bullet}{x^{3}} + \frac{\chi_\bullet^2}{x^4},
    \end{aligned}
    \hspace{5.2mm}
    \begin{aligned}
    \Ccu & = 1 - \frac{3}{x^2} + \frac{2\chi_\bullet}{x^3},\\
    \Gcu & = 1 -  \frac{2}{x^2} + \frac{\chi_\bullet}{x^3},
    \end{aligned}
\end{equation}

\begin{equation}
J  = \frac{x^4 - 4\chi_\bullet x - 3\chi_\bullet^2}{x^4 + 2 \chi_\bullet x - 3x^2},\hspace{5.2mm}S = \frac{\Ccu^{3/2} \Qcu}{\Bcu \Dcu^2},
\end{equation}

as well as
\begin{equation*}
\begin{aligned}
    \Qcu = \frac{\Bcu}{\Ccu^{1/2}} \frac{1}{x}\bigg[&x - x_0 -\frac{3}{2}\chi_\bullet \mathrm{ln}\left(\frac{x}{x_0}\right) - \frac{3(x_1-\chi_\bullet)^2}{x_1(x_1-x_2)(x_1-x_3)}\mathrm{ln}\left(\frac{x-x_1}{x_0-x_1}\right)\\
    &- \frac{3(x_2-\chi_\bullet)^2}{x_2(x_2-x_1)(x_2-x_3)}\mathrm{ln}\left(\frac{x-x_2}{x_0-x_2}\right) - \frac{3(x_3-\chi_\bullet)^2}{x_3(x_3-x_1)(x_3-x_2)}\mathrm{ln}\left(\frac{x-x_3}{x_0-x_3}\right) \bigg],
\end{aligned}
\end{equation*}
where $x = (r/M_\bullet)^{1/2}$ is a normalized radial coordinate, $\chi_\bullet = a_\bullet / M_\bullet \in [0,1)$ the dimensionless black hole spin, and
\begin{eqnarray}
        x_0 &= (r_{\mathrm{ISCO}}/M_\bullet)^{1/2}, \mskip80mu
        x_1 &= 2\cos\left[\frac{1}{3} \arccos(\chi_\bullet) - \frac{\pi}{3}\right], \\
        x_2 &= 2\cos\left[\frac{1}{3} \arccos(\chi_\bullet) + \frac{\pi}{3}\right],\; 
        x_3 &= -2\cos\left[\frac{1}{3} \arccos(\chi_\bullet)\right].
\end{eqnarray}
We transform the Kerr metric $g_{\mu\nu}$ from Boyer-Lindquist coordinates (Equation~\eqref{eq:Kerr_metric_Boyer_Lindquist}) into cylindrical coordinates $x^\mu = (t,r, \phi,z)$. Since we consider thin accretion disks ($H/r \ll 1)$, we only retain leading-order terms in $z/r$. The vertical structure equation requires us to consider metric components up to second order in $z/r$, which we obtain as
\begin{equation}
\label{eq:kerr-metric}
    \begin{aligned}
g_{tt} &= -\left[ 1 - \frac{2}{x^2} + \frac{z_\bullet^{2}}{x^6} {\left(1 + \frac{2 \chi_\bullet^{2}}{x^4} \right)} \right],\\
g_{t \phi} &= - \frac{\chi_\bullet M_\bullet}{x^2} {\left[ 2 - \frac{z_\bullet^{2}}{x^4} {\left( 3 + \frac{2 \, \chi_\bullet^{2}}{x^4} \right)} \right]},\\
g_{\phi \phi} &= M^2 \left[ \Acu x^4 - \frac{\chi_\bullet^2 z_\bullet^2}{x^4} \left( 1+ \frac{5}{x^2} + \frac{2 \chi_\bullet^2}{x^6} \right) \right],
    \end{aligned}
    \hspace{5.2mm}
    \begin{aligned}
g_{rr} &= \frac{1}{\Dcu} \left[ 1 - \frac{z_\bullet^2}{x^6} \left( 2 + \frac{1}{\Dcu} \right) \left(1-\frac{\chi_\bullet^2}{x^2}\right) \right],\\
g_{rz} &= -\frac{z_\bullet}{x^2}\left(1-\frac{1}{\Dcu}\right),\\
g_{zz} &= 1 +\frac{z_\bullet^{2}}{\Dcu x^6} {\left[ 2 - \frac{\chi_\bullet^2}{x^4} \left( 2 - \frac{\chi_\bullet^2}{x^2} \right) \right]},
    \end{aligned}
\end{equation}
with all other components equal to zero. Here, we have introduced $z_\bullet = z/M_\bullet$. The Christoffel symbols relevant for the below derivation, expanded up to first order in $z/r$, read:
\begin{equation}
\label{eq:christoffel-symbols}
    \begin{aligned}
        \Gamma^r_{tt} &= \frac{\Dcu}{x^4 M_\bullet}, \\
        \Gamma^r_{t \phi } &= -\frac{\chi_\bullet\Dcu}{x^4}, \\
        \Gamma^r_{\phi \phi} &= - \Dcu x^2 M_\bullet \left(1 - \frac{\chi_\bullet^2}{x^6}\right),
    \end{aligned}
    \hspace{10mm}
    \begin{aligned}
        \Gamma^z_{tt} &= \frac{z_\bullet}{x^6 M_\bullet} \left( \Dcu + \frac{2\chi_\bullet^2}{x^4}\right),\\
        \Gamma^z_{t \phi }&=  - \frac{\chi_\bullet z_\bullet}{x^6} \left( 2 + \Dcu + \frac{2 \chi_\bullet^2}{x^4}\right),\\
        \Gamma^z_{\phi \phi} &=  \frac{z_\bullet M_\bullet}{x^2} \left[ 2 + \frac{\chi_\bullet^2}{x^4}\left( 4 + \Dcu + \frac{2\chi_\bullet^2}{x^4} \right) \right].
    \end{aligned}
\end{equation}

The energy-momentum tensor of accreting matter in this background spacetime is given by
\begin{equation}\label{eq:Ap_energy_momentum_tensor}
    T^{\mu \nu} = h \rho u^\mu u^\nu + p g^{\mu \nu} + t^{\mu \nu} + u^\mu q^\nu + u^\nu q^\mu,
\end{equation}
where $h = (\rho + e + p)/\rho$ is the relativistic specific enthalpy, with $e$ the internal energy density and $p$ the pressure, and $u^\mu$ is the time-like 4-velocity of the fluid, $u_\mu u^\mu  = -1$. The radiative energy flux vector of neutrinos is denoted by $q^\mu$. Furthermore,
\begin{equation}
    t^{\mu\nu} = - (2\eta \sigma^{\mu\nu} + 3\zeta\mathcal{P}^{\mu\nu} \Theta) = -\eta [\mathcal{P}^{\mu\alpha}\nabla_\alpha u^\nu + \mathcal{P}^{\nu\beta} \nabla_\beta u^\mu] + \mathcal{P}^{\mu\nu}(2\eta-3\zeta)\Theta \label{eq:def_viscous_stress_tensor}
\end{equation}
denotes the viscous stress tensor as measured in the local rest frame of the baryonic fluid. Here, $\eta$ and $\zeta$ are the shear and bulk viscosity, respectively \citep{landau_fluid_2004}, $\sigma^{\mu\nu} = \mathcal{P}^{\mu\alpha}\mathcal{P}^{\nu\beta} [\frac{1}{2} (\nabla_\alpha u_\beta + \nabla_\beta u_\alpha) - g_{\alpha\beta}\Theta]$ is the shear tensor, $\Theta = \frac{1}{3}\nabla_\mu u^\mu$ is the expansion, $\mathcal{P}^{\alpha\beta}=u^\alpha u^\beta + g^{\alpha\beta}$ is the projection tensor onto the dimensions orthogonal to $u^\mu$. Both $t^{\mu \nu}$ and $q^\mu$ are orthogonal to the 4-velocity, i.e. $u_\mu t^{\mu \nu} = 0$ and $u_\mu q^\mu = 0$.

We make repeated use of the following properties or assumptions:
\begin{itemize}
    \item[$(i)$] Due to the axisymmetric and stationary nature of the background spacetime, we neglect all azimuthal and temporal derivatives and set $\partial_\phi(\cdot) =0$,  $\partial_t(\cdot) =0$.
    \item[$(ii)$] Due to the height-integrated nature of our one-dimensional disk model, we do not consider vertical fluid motion and set the four-velocity to $u^\mu = (u^t, u^r, u^\phi, 0)$.
    \item[$(iii)$] We assume quasi-circular motion, $|u^r| \ll |u^\phi|,\,|u^t|$, which implies $|u_z|\simeq |u^r|$. Together with $(i)$, the expansion then vanishes to zero order in small quantities ($\Theta \simeq 0$).
    \item[$(iv)$] We assume the accretion flow to be rest-mass energy dominated, $\rho \gg e, p, |{t^{\mu}}_{\nu}|, |q^\mu|$.
    \item[$(v)$] Due to the thin nature of the disk, we neglect lateral neutrino radiation transport within the disk and only consider neutrino flux escaping from the disk in the $z$-direction. We thus set $q^\mu = (q^t,0,0,q^z)$, with $q^\mu q_\mu = 0$.
    \item[$(vi)$] The normalization of $u^\mu$ and the orthogonality of $S^{\mu\nu}$ and $q^\mu$ with $u^\mu$ directly imply the following relations
    \begin{eqnarray}
        \nabla_\mu (u_\nu u^\nu) &=& 0 \mskip20mu \Longrightarrow \mskip20mu u_\nu \nabla_\mu u^\nu = 0, \label{eq:Ap_velocity_divergence}\\
        \nabla_\mu (u_\nu t^{\mu \nu}) &=& 0 \mskip20mu \Longrightarrow \mskip20mu t^{\mu \nu}\nabla_\mu u_\nu  = - u_\nu \nabla_\mu t^{\mu \nu}, \label{eq:Ap_stress_divergence} \\
        \nabla_\mu (u_\nu q^\nu) &=& 0 \mskip20mu \Longrightarrow \mskip20mu q_\nu \nabla_\mu u^\nu = -  u_\nu \nabla_\mu q^\nu.  \label{eq:Ap_flux_divergence}
    \end{eqnarray}
\end{itemize}
Furthermore, we compute stresses in the co-moving frame of the accreting fluid, which we approximate by the circular, geodesic motion with four-velocity $u^\phi = (u^t,0,u^\phi,0)$, where $u^t$ and $u^\phi$ are given by Equation~\eqref{eq:Ap_ut_uphi}. The tetrad vectors then read \citep{novikov_astrophysics_1973}
\begin{equation}
    e_{\hat t} = u^\mu,\mskip20mu e_{\hat r} = \Dcu^{1/2}\partial_r, \mskip20mu e_{\hat \phi} = \frac{\Fcu}{\Ccu^{1/2} \Dcu^{1/2} x} \partial_t + \left(\frac{\Bcu\Dcu^{1/2}}{\Acu \Ccu^{1/2} x^2 M_\bullet} + \frac{\Fcu}{\Acu\Ccu^{1/2} \Dcu^{1/2}}\frac{2\chi_\bullet}{x^7 M_\bullet} \right)\partial_\phi,  \mskip20mu e_{\hat z} = \partial_z,\label{eq:tetrad}
\end{equation}
and the corresponding one-forms are
\begin{equation}
    e^{\hat t} = \frac{\Gcu}{\Ccu^{1/2}}\, \dt - \frac{\Fcu}{\Ccu^{1/2}} M_\bullet x\,\dd{\phi} ,\mskip20mu e^{\hat r} = \frac{1}{\Dcu^{1/2}}\,\dr, \mskip20mu e^{\hat \phi} = -\frac{\Dcu^{1/2}}{\Ccu^{1/2}x}\,\dt + \frac{\Bcu\Dcu^{1/2}}{\Ccu^{1/2}} M_\bullet x^2\, \dd{\phi},  \mskip20mu e^{\hat z} = \dd{z}. \label{eq:tetrad_oneforms}
\end{equation}
Hats denote components relative to the co-moving frame; they are raised and lowered with the Minkowski metric $\eta_{\hat\mu\hat\nu}=\mathrm{diag}(-1,1,1,1)$. The tetrad frame is orthonormal in the sense that $e^\mu_{\hat\alpha} e^{\hat\beta}_{\mu} = \delta^{\hat\beta}_{\hat\alpha}$.

\subsection{Baryon number conservation}

In the absence of disk winds, baryon number is strictly conserved in the accretion disk,
\begin{equation}
    \nabla_\mu (\nb u^\mu) = 0 \mskip20mu\Longleftrightarrow \mskip20mu\partial_\mu(\sqrt{-g} \,\nb u^\mu)=0. \label{eq:baryon_number_conservation}
\end{equation}
Multiplying the latter expression by the baryon mass $\mb$ and integrating over a spacetime volume $r \in [r,r+\Delta r], \phi \in [0,2\pi],z \in [-H,H]$, and $t\in [t,\Delta t]$, one obtains
\begin{equation}
    \dot{M} \equiv - 4\pi H r \rho u^r = \text{const.} \label{eq:app_Mdot}
\end{equation}
The sign convention adopted here ensures that a positive accretion rate $\dot{M}$ corresponds to matter falling into the black hole.

\subsection{Lepton number conservation}

Due to the loss of neutrinos via neutrino cooling, the lepton number is not strictly conserved in the accretion disk. We represent this loss of neutrinos by a sink term and express the conservation law of lepton number as
\begin{equation}
    \nabla_\mu (\nb \Ylep u^\mu) = R, \label{eq:lepton_number_conservation}
\end{equation}
where
\begin{equation}
    \Ylep= \frac{n_{e^{-}} - n_{e^+} + \nnue - \nnuae}{\nb}= \Ye + \frac{\nnue - \nnuae}{\nb}
\end{equation}
is the lepton number and $R$ is the radiated lepton number per unit volume per unit time in the rest frame of the fluid. Furthermore, $n_{e^+}$ and $n_{e^-}$ are the number densities of positrons and electrons (Equation~\eqref{eq:electron_density}), respectively, and $\nnue$ and $\nnuae$ are the number densities of thermalized electron neutrinos and antineutrinos within the disk (Equation~\eqref{eq:n_nue-n_nuea}).

In terms of disk-height integrated quantities,
\begin{equation}
    R = -\frac{1}{H}(\dot n_\nue - \dot n_\nuae), \label{eq:R_neutrino_emission}
\end{equation}
where $H$ is the half-thickness of the disk as usual and $\dot n_\nue$ and $\dot n_\nuae$ are the (height-integrated) radiated number fluxes of electron neutrinos and antineutrinos per unit area from one face of the disk (Equation~\eqref{eq:n_dot_transition}). The minus sign is chosen such that positive emissivities imply a sink term for the lepton number in Equation~\eqref{eq:lepton_number_conservation}.
Expanding Equation~\eqref{eq:baryon_number_conservation} as
\begin{equation}
    \nabla_\mu (\nb u^\mu) = \nb \nabla_\mu u^\mu + u^r \partial_r \nb = 0, \label{eq:baryon_number_conservation_exp}
\end{equation}
the left-hand side of Equation~\eqref{eq:lepton_number_conservation} can be rewritten as
\begin{eqnarray}
    \nabla_\mu (\nb \Ylep u^\mu) 
    &=& u^\mu \nabla_\mu (\nb \Ye) + \nb \Ye \nabla_\mu u^\mu  + u^\mu \nabla_\mu(\nnue - \nnuae) + (\nnue - \nnuae) \nabla_\mu u^\mu \\
    &=& \nb u^r \partial_r\Ye + u^r \partial_r (\nnue - \nnuae) - \frac{\nnue - \nnuae}{\nb} u^r \partial_r \nb \\
    &=& u^r \left[ \frac{\rho}{\mpr}\ddr{\Ye}  - (\nnue - \nnuae) \frac{1}{\rho} \ddr{\rho} + \ddr{(\nnue - \nnuae)}\right]. 
\end{eqnarray}
Therefore, the total lepton number balance, i.e. the equation of lepton number conservation height-integrated according to Equation~\eqref{eq:def_height_integration}, reads
\begin{equation}
    H u^r \left[ \frac{\rho}{\mpr}\ddr{\Ye}  - (\nnue - \nnuae) \frac{1}{\rho} \ddr{\rho} + \ddr{(\nnue - \nnuae)}\right] = (\dot n_\nuae - \dot n_\nue).
\end{equation}

\subsection{Energy-momentum conservation}

We project the conservation law of energy and momentum, $\nabla_\nu T^{\mu \nu} = 0$, onto $u^\mu$ to obtain an energy balance equation as seen by a comoving observer,
\begin{equation}\label{eq:Ap_energy_balance}
    u_\mu \nabla_\nu T^{\mu \nu} = 0,
\end{equation}
as well as orthogonal to $u^\mu$ using the projection tensor $\mathcal{P}^{\alpha\beta}=u^\alpha u^\beta + g^{\alpha \beta}$ to obtain the momentum balance equation for the dimensions ($i=1,2,3$) orthogonal to $u^\mu$,
\begin{equation}\label{eq:Ap_momentum_balance}
    (u^i u^\nu + g^{i \nu}) \nabla_\mu {T^\mu}_\nu = 0.
\end{equation}

\subsubsection{Energy conservation}

Using baryon number conservation \eqref{eq:baryon_number_conservation} as well as the identities \eqref{eq:Ap_velocity_divergence} and \eqref{eq:Ap_flux_divergence}, we obtain from Equation~\eqref{eq:Ap_energy_balance} the relation
\begin{equation}
        u^\mu\nabla_\mu e + (e+p)\nabla_\mu u^\mu - u_\nu \nabla_\mu t^{\mu \nu} + \nabla_\mu q^\mu + q^\nu u^\mu \nabla_\mu u_\nu = 0.
\end{equation}
Again making use of baryon number conservation, and assuming quasi-geodesic motion (acceleration $a^\nu \equiv u^\mu \nabla_\mu u^\nu \simeq 0$ to leading order), neglecting radial neutrino flux gradients (consistent with vertical transport only, see $(v)$ above), one finds in an axisymmetric and stationary context
\begin{equation}
    u^r\left(\partial_r e - \frac{e+p}{\rho} \partial_r \rho\right) + t^{\mu \nu}\nabla_\mu u_\nu + \frac{1}{\sqrt{|g|}}\partial_z(\sqrt{|g|}q^z) = 0. \label{eq:app_energy_conservation_1}
\end{equation}

We introduce the cooling rate per area of the disk as
\begin{equation}
    F^- \equiv \int_0^H \frac{1}{\sqrt{|g|}}\partial_z(\sqrt{|g|}q^z) \sqrt{g_{zz}}\dz = \int_{0}^H \partial_z q^z \dz = q^z(z=H).
\end{equation}
Furthermore, we write
\begin{equation}
    u_\nu \nabla_\mu t^{\mu \nu} = - \frac{1}{2} t^{\mu \nu} (\nabla_\mu u_\nu + \nabla_\nu u_\mu) = -t^{\mu \nu}\sigma_{\mu\nu} = - t^{\hat{\mu} \hat{\nu}} \sigma_{\hat{\mu}\hat{\nu}} = - 2 t^{\hat{r} \hat{\phi}} \sigma_{\hat{r}\hat{\phi}}, \label{eq:viscous_tensor_identity}
\end{equation}
where hats denotes the components in the co-moving tetrad of the accreting fluid (cf.~Equation~\eqref{eq:tetrad}). The shear tensor for quasi-geodesic motion is given by $\sigma_{\mu\nu}=\frac{1}{2}(\nabla_\mu u_\nu + \nabla_\nu u_\mu)$. In the co-moving tetrad frame, covariant derivatives reduce to partial derivatives, and one obtains with Equations~\eqref{eq:tetrad} and \eqref{eq:tetrad_oneforms} the only non-vanishing components
\begin{eqnarray}
    \sigma_{\hat r \hat \phi} = \sigma_{\hat \phi \hat r} = \frac{1}{2} \partial_{\hat r} u_{\hat\phi} = \frac{1}{2} e^\mu_{\hat r} e^\nu_{\hat\phi} \partial_{\mu} u_{\nu} = \frac{1}{2} e^r_{\hat r} e^{\hat\phi}_{\nu} \partial_{r} u^{\nu} 
    = \gamma^2 r \Acu \partial_r \Omega = -\frac{3}{4}\frac{M^{1/2}}{r^{3/2}}\frac{\Dcu}{\Ccu}, \label{eq:shear_comoving_tetrad}
\end{eqnarray}
where $\gamma=\sqrt{-g_{tt}}u^t=\Dcu^{1/2}\Acu^{-1/2} u^t$ is the Lorentz factor and where we have used Equation~\eqref{eq:Ap_angular_velocity} for the angular velocity. We can thus write viscous heating per unit area of the disk as
\begin{equation}
    F^+ \equiv - \int_0^H t^{\mu\nu}\sigma_{\mu\nu} \sqrt{g_{zz}} \dz = \frac{3}{2}\frac{M_\bullet^{1/2}}{r^{3/2}} \frac{\Dcu}{\Ccu} \int_0^H t^{\hat{r}\hat{\phi}}\dz. \label{eq:app_def_viscou_heating}
\end{equation}

Therefore, integrating Equation~\eqref{eq:app_energy_conservation_1} over height, we find
\begin{equation}
    u^r \left[ \frac{\dd (H e)}{\dr} - \frac{e+p}{\rho } \frac{\dd (\rho H)}{\dr} \right] = F^+ - F^-.
\end{equation}

\subsubsection{Radial momentum balance}

Direct evaluation of radial momentum balance ($i=1$ in Equation~\eqref{eq:Ap_momentum_balance}) leads to leading order in velocity components
\begin{equation}
    h\rho u^\mu\nabla_\mu u^r + g^{r\mu}\partial_\mu p + \nabla_\mu t^{\mu r} + q^\mu \nabla_\mu u^r = 0.
\end{equation}
To leading order in small quantities, and neglecting radial pressure as well as shear gradients, consistent with quasi-circular motion, one thus finds
\begin{equation}
    u^\mu\nabla_\mu u^r = 0 \mskip20mu \Longleftrightarrow \mskip20mu u^\mu(\partial_\mu u^r + \Gamma^r_{\mu \nu} u^\nu) = 0. 
\end{equation}
Again, to leading order in velocity components, this leads to the local balance equation
\begin{equation}
    \Gamma^r_{\mu \nu} u^\mu u^\nu = 0,
\end{equation}
which is identical to its height-integrated version. Together with $u_\mu u^\mu = -1$ to leading order in velocity components, and using the expressions for the Christoffel symbols in Equation~\eqref{eq:christoffel-symbols}, yields explicit expressions for the velocity components,
\begin{equation}
\label{eq:Ap_ut_uphi}
     u^t = \frac{\Bcu}{\Ccu^{1/2}},\mskip20mu u^\phi = \frac{1}{x^3M_\bullet \Ccu^{1/2}}.
\end{equation}
Therefore, the angular velocity is given by 
\begin{equation}
\label{eq:Ap_angular_velocity}
    \Omega = \frac{u^\phi}{u^t} = \frac{1}{x^3M_\bullet\Bcu} = \left(\frac{r^{3/2}}{M_\bullet^{1/2}} + \chi_\bullet M_\bullet \right)^{-1}.
\end{equation}

\subsubsection{Vertical momentum balance}

Vertical momentum balance ($i = 3$ in Equation~\eqref{eq:Ap_momentum_balance}) reads
\begin{equation}
    h\rho u^\mu \nabla_\mu u^z + g^{z\nu} \partial_\nu p + \nabla_\mu t^{\mu z} + q^z \nabla_\mu u^\mu + u^\mu \nabla_\mu q^z + q^\mu \nabla_\mu u^z = 0.
\end{equation}
To lowest order in small quantities, and neglecting radial pressure gradients and shear gradients as above, one obtains
\begin{equation}\label{eq:Ap_hydrostatic_equilibrium_1}
   g^{zz} \partial_z p + \Gamma^z_{\mu \nu} u^\mu u^\nu = 0.
\end{equation}
With the expressions \eqref{eq:christoffel-symbols} and \eqref{eq:Ap_ut_uphi}, this simplifies to linear order in $z/r$ to
\begin{equation}
    \frac{\partial p}{\partial z} = - \frac{\rho z J}{x^6M^2}.
\end{equation}
Upon vertical integration across one half-disk using $p(z=H)=0$, we find the midplane pressure $p_{1/2}$ due to one half-disk as

\begin{equation}
    p_{1/2}(z=0) = \frac{1}{2}\frac{J \rho H^2}{x^3M_\bullet^2}.
\end{equation}
The other half-disk contributes equally to the pressure in the midplane. Thus one obtains the equation of vertical balance as
\begin{equation}
\label{eq:Ap_hydrostatic_equilibrium_2}
    \left( \frac{H}{r} \right)^2 = \frac {p}{\rho} \frac{r }{J M_\bullet}.
\end{equation}

\subsubsection{Azimuthal momentum balance}

Azimuthal momentum balance is expressed by the $i=2$ component of Equation~\eqref{eq:Ap_momentum_balance}, from which we obtain
\begin{equation}\label{eq:Ap_azimuthal_equilibrium_1}
    u_\phi u^r \partial_r p + h\rho u^\mu \nabla_\mu u_\phi + \nabla_\mu {t^\mu}_\phi + u_\phi u_\nu \nabla_\nu t^{\mu \nu} + q^\mu \nabla_\mu u_\phi  + u^\mu \nabla_\mu q_\phi - u_\phi q^\nu u^\mu \nabla_\mu u_\nu= 0.
\end{equation}
Neglecting radial pressure gradients as above, one has to leading order in small quantities
\begin{equation}
\label{eq:Ap_azimuthal_equilibrium_2}
\begin{aligned}
    0 &= \rho u^\mu\nabla_\mu u_\phi + \nabla_\mu {t^\mu}_\phi + u_\phi u_\nu \nabla_\mu t^{\mu \nu} \simeq \rho u^r \partial_r l + \frac{1}{r} \partial_r (r {t^r}_\phi) + l u_\nu \nabla_\mu t^{\mu \nu}.
\end{aligned}
\end{equation}
Identifying $l = u_\phi$ as the specific angular momentum (angular momentum per unit rest-mass) and $r\rho u^r \equiv -\dot{m}/(2\pi)=\text{const.}$ as the accreted mass per height (which follows directly from Equation~\eqref{eq:baryon_number_conservation}), expressing 
\begin{equation}
    {t^r}_\phi = e^r_{\hat r} e^{\hat \phi}_\phi t_{\hat r\hat\phi} = r\Bcu \Ccu^{-1/2}\Dcu t_{\hat{r}\hat{\phi}}
\end{equation}
in terms of the stress components in the tetrad co-moving with the mean fluid flow (cf.~Equations~\eqref{eq:tetrad} and \eqref{eq:tetrad_oneforms}), as well as re-expressing $u_\nu \nabla_\mu t^{\mu \nu}$ as in Equation~\eqref{eq:viscous_tensor_identity}, we obtain the differential equation

\begin{equation}
    \partial_r \left(r^2\frac{\Bcu \Dcu}{\sqrt{\Ccu}}t_{\hat{r}\hat{\phi}}  - \frac{\dot{m}}{2\pi}l \right) + \frac{3}{2} \sqrt{\frac{M}{r}} \frac{\Dcu}{\Ccu}l t_{\hat{r}\hat{\phi}} = 0.
\end{equation}
This equation is mathematically equivalent to the height-integrated equation (5.6.13) of \citet{novikov_astrophysics_1973}. The solution can thus be obtained from the result in closed form of \citet{page_disk-accretion_1974} by translating that to the present case,
\begin{equation}
    t_{\hat{r}\hat{\phi}} = \frac{\dot{m}}{2\pi} \frac{M_\bullet^{1/2}}{r^{3/2}} \frac{\Ccu^{1/2}}{\Bcu \Dcu}\Qcu. \label{eq:trphi_azimuthal momentum balance}
\end{equation}
Furthermore, adopting the usual $\alpha$-viscosity ansatz for accretion disks, $\nu \equiv \eta /\rho = \frac{2}{3}\alpha c_s H$, where $c_s\equiv\sqrt{p/\rho}$ is the isothermal sound speed, we also obtain from Equation~\eqref{eq:def_viscous_stress_tensor} using vertical balance \eqref{eq:Ap_hydrostatic_equilibrium_2}
\begin{equation}
     t_{\hat{r}\hat{\phi}} = -2 \eta  \sigma_{\hat{r}\hat{\phi}} = \frac{3}{2} \eta \frac{M_\bullet^{1/2}}{r^{3/2}} \frac{\Dcu}{\Ccu} = \alpha p \frac{\Dcu}{J^{1/2}\Ccu}. 
\end{equation}
Equating the above two expressions for $t_{\hat{r}\hat{\phi}}$, we obtain
\begin{equation}
    u^r = -\alpha c_s \sqrt{\frac{p}{\rho}} \frac{r^{1/2}}{M_\bullet^{1/2}}\frac{\Bcu \Dcu^2}{J^{1/2}\Ccu^{3/2}\Qcu} = -\alpha c_s \frac{H}{r} S^{-1}.
\end{equation}
Furthermore, integrating Equation~\eqref{eq:trphi_azimuthal momentum balance} over height using Equation~\eqref{eq:app_Mdot}, the viscous heating per unit area of the disk (Equation~\eqref{eq:app_def_viscou_heating}) is given by the following closed expression:
\begin{equation}
    F^+ = \frac{3}{2}\frac{M_\bullet}{r^3} \frac{\Qcu}{\Bcu\Ccu^{1/2}} \int_0^H (-r\rho u^r)\sqrt{g_{zz}} \dz 
    = \frac{3\dot{M}M_\bullet}{8\pi r^3} \frac{\Dcu^2}{\Ccu^2}S. 
\end{equation}

\section{Equation of state and thermodynamic quantities}
\label{app:EOS}

In this appendix, we compute the contributions to the plasma pressure, internal energy density, and the entropy from all species considered here (baryons, photons, electrons and positrons, as well as electron neutrinos and antineutrinos).

\subsection{Baryons}

The Helmholtz free energy of a multi-species ideal fluid can be written as (cf.~\citealt{landau_statistical_1980})
\begin{equation}
    \mathcal{F}_{\rm b} = -k_B T\sum_i N_i \ln \left[\frac{eV}{N_i} \frac{(2\pi m_i k_B T)^{3/2}}{h^3} Z_{{\rm int},i}\right],
\end{equation}
where $Z_{{\rm int},i}=\sum_k g_{ik} \exp(-\epsilon_{ik}/k_B T)$ is the internal partition function of particle species $i$ (the nuclear partition function in this case). Here, $V$ denotes volume, $N_i$ the number of particles of species $i$, and $m_i$ their mass. Assuming a mix of free nucleons and $\alpha$-particles ($i=\text{nuc},\alpha$), each having one energy level (ground state), the spin statistical weights $g_i = 2 I_i + 1$ are $g_{\rm nuc} = 2$ (fermions) and $g_\alpha = 1$ (ground state, $I_\alpha = 0$). Choosing the zero-energy state to be the free nucleon state ($\epsilon_{\rm nuc}=0$, $\epsilon_\alpha = - E_{\rm bind} = -(2m_p + 2 m_n - m_\alpha)c^2 \simeq -28.30 \text{MeV}$), we find the baryon pressure as
\begin{equation}
    p_{\rm b} = -\left(\frac{\partial \mathcal{F}_{\rm b}}{\partial V}\right)_{T,N_i} = \frac{m_e c^2}{m_{\rm b}}\rho\theta \left( X_{\rm f} + \frac{1-X_{\rm f}}{4}\right).
\end{equation}
The entropy per baryon in units of $k_B$ reads
\begin{eqnarray}
    s_{\rm b} = -\left(\frac{\partial \mathcal{F}_{\rm b}}{\partial T}\right)_{V,N_i} 
    \mskip-20mu = X_{\rm f} \left\{\frac{5}{2}-\ln\left[\frac{X_{\rm f}\rho}{2m_{\rm b}} \frac{ h^3}{(2\pi m_{\rm b} m_e c^2 \theta)^{3/2}}\right] \right\} 
    + \frac{1-X_{\rm f}}{4} \left\{\frac{5}{2}-\ln \left[\frac{(1-X_{\rm f})\rho}{4m_{\rm b}}\frac{ h^3/4}{(2\pi m_{\alpha} m_e c^2 \theta)^{3/2}}\right]\right\}. \label{eq:entropy_baryons}
\end{eqnarray}
The energy density is given by
\begin{eqnarray}
    e_{\rm b} = \frac{1}{V} (\mathcal{F}_{\rm b} + T S_{\rm b}) \label{eq:Ub} 
    = \frac{3}{2} \frac{m_e c^2}{m_{\rm b}}\rho\theta \left(X_{\rm f} + \frac{1 - X_{\rm f}}{4}\right) - E_{\rm bind} \frac{\rho}{m_{\rm b}} \frac{1-X_{\rm f}}{4},
\end{eqnarray}
where $S_{\rm b}$ denotes entropy.

\subsection{Radiation}

For photons in thermal equilibrium $\mu = 0$, the Helmholtz free energy equals the grand canonical potential $\pmb{\Omega}$, and one can write (cf.~\citealt{landau_statistical_1980})
\begin{eqnarray}
    \mathcal{F}_{\gamma} = k_B T \frac{V}{\pi^2 c^3} \int_0^\infty \drm\omega\, \omega^2 \ln\left(1-e^{-\hbar \omega / k_B T}\right) 
    = -\frac{1}{3} a_{\rm r} V T^4,
\end{eqnarray}
where $a_{\rm r} = \pi^2 k_B^4 / (15 c^3\hbar^3)$ is the radiation constant.

The entropy per baryon in units of $k_B$ is given by
\begin{equation}
    s_\gamma = -\frac{1}{V}\frac{m_{\rm b}}{\rho} \left(\frac{\partial \mathcal{F}_{\gamma}}{\partial T}\right)_{V} = \frac{4}{3} \frac{a_{\rm r} m_{\rm b} m_e^3c^6}{k_B^4 }\frac{\theta^3}{\rho}, \label{eq:entropy_gamma}
\end{equation}
the pressure reads
\begin{equation}
    p_\gamma = -\left(\frac{\partial \mathcal{F}_\gamma}{\partial V}\right)_T = \frac{1}{3} \frac{a_{\rm r} m_e^4 c^8}{k_B^4} \theta^4, \label{eq:pressure_photons}
\end{equation}
and the radiation energy density is given by
\begin{equation}\label{eq:Ugamma}
    e_\gamma = \frac{1}{V}(\mathcal{F}_\gamma + T S_\gamma)= 3p_\gamma.
\end{equation}

\subsection{Leptons}

Since a wide range of physical regimes are realized in different spatial parts of an accretion disk and across the wide parameter space considered here, we need to take arbitrary levels of relativity and degeneracy of leptons into account, without resorting to limiting cases. The Helmholtz free energy $\mathcal{F} = \pmb{\Omega} + \mu N$ is most easily obtained via the grand canonical potential (cf.~\citealt{landau_statistical_1980})
\begin{equation}
    \pmb{\Omega} = -k_B T \frac{gV}{2\pi^2\hbar^3}\int_0^\infty \drm p\,p^2\ln\left[1+ e^{-(\epsilon-\mu)/k_B T }\right], \label{eq:grand_canonical_potential_quantum_gas}
\end{equation}
where $\epsilon = \sqrt{p^2 c^2 + m c^4}$ is the relativistic particle energy, $g$ is the statistical weight ($g=2$ for electrons and positrons, $g=1$ for neutrinos), and $m$ is the lepton mass. Introducing $\lambda_e \equiv \pi^{2/3} \hbar/m_ec$, and noting that $\pmb{\Omega} = -pV$, one finds for electrons and positrons 
\begin{equation}
    p_{e^\pm} = \frac{1}{3} \frac{m_e c^2}{\lambda_e^3} \int_0^\infty \drm\xi \,\xi^4(\xi^2+1)^{-1/2}f(\sqrt{\xi^2 + 1},\mp \eta_e,\theta). \label{eq:pressure_fermions_massive}
\end{equation}
The number densities are given by
\begin{equation}
    n_{e^\pm} = -\frac{1}{V}\left(\frac{\partial \pmb{\Omega}_{e^\pm}}{\partial \mu_{e^\pm}}\right)_{T,V} \mskip-15mu= \frac{1}{\lambda_{e}^3} \int_0^\infty \mskip-10mu\drm\xi \,\xi^2 f(\sqrt{\xi^2 + 1},\mp\eta_e,\theta), \label{eq:n_e+-_derivation}
\end{equation}
the corresponding energy density by
\begin{eqnarray}
    e_{e^{\pm}} = \frac{m_e c^2}{\lambda_e^3} \int_0^\infty \mskip-15mu\drm\xi \,\xi^2(\xi^2+1)^{1/2}f(\sqrt{\xi^2 + 1},\mp\eta_e,\theta),
\end{eqnarray}
and the entropy per baryon in units of $k_B$ is
\begin{eqnarray}
    s_{e^\pm} = \frac{m_{\rm b}}{\rho} \left(\frac{e_{e^{\pm}} + p_{e^\pm}}{m_e c^2 \theta} \pm n_{e^\pm} \eta_{e}\right). \label{eq:entropy_epm_general}
\end{eqnarray}

For electron neutrinos and antineutrinos, it is necessary to distinguish between the transparent (free-streaming) and opaque (thermalized) regimes. In the thermalized regime, neutrinos and antineutrinos are modeled by a massless ideal Fermi gas, for which one obtains from Equation~\eqref{eq:grand_canonical_potential_quantum_gas}, together with $\pmb{\Omega} = -pV$, the pressure as
\begin{eqnarray}
    p^{\text{opaque}}_{\nue,\nuea} = \frac{1}{6} \frac{m_e c^2}{\lambda_e^3} \theta^4 \int_0^\infty \drm\xi \,\frac{\xi^3}{\exp(\xi \mp \eta_\nu) + 1}.\label{eq:pressure_neutrinos}
\end{eqnarray}
Here, the upper sign refers to electron neutrinos, whereas the lower sign to the antineutrinos. The corresponding energy densities are given by 

\begin{equation}\label{eq:Unu_opaque}
    e^{\text{opaque}}_{\nue,\nuea} = 3 p^{\text{opaque}}_{\nue,\nuea},
\end{equation}
and the number densities are obtained as
\begin{eqnarray}
    n_{\nue,\nuea}^{\text{opaque}} &=& -\frac{1}{V}\left(\frac{\partial \pmb{\Omega}^{\text{opaque}}_{\nue,\nuea}}{\partial \mu_{\nue,\nuea}}\right)_{T,V} 
    = \frac{1}{2} \frac{\theta^3}{\lambda_e^3} \int_0^\infty \drm\xi \,\frac{\xi^2}{\exp(\xi \mp \eta_{\nu})+1}.
    \label{eq:number_density_nue-nuea}
\end{eqnarray}

We include the internal energy density of the thermalized neutrinos and anti-neutrinos in the equation of state as 
\begin{equation}
    e_\nu = x_\nu e^{\text{opaque}}_\nue, \mskip20mu e_{\bar\nu} = x_{\bar\nu} e^{\text{opaque}}_{\bar\nue},
\end{equation}
where $x_\nu$ and $x_{\bar\nu}$ are transition parameters defined in Equation~\eqref{eq:xnu}. The latter smoothly interpolate between zero (transparent regime) and unity (optically thick regime) for the two species. The consistent partial pressures for the equation of state are then given by $p_{\nu,\bar\nu} = e_{\nu,\bar\nu}/3$.

\section{Disk outflow model}
\label{app:outflow_model}

Here, we construct a simplified accretion disk outflow model. It is mainly used to estimate the number of neutrons $\Delta_n$ created per seed during a potential $\nu p$-process (see Equation~\eqref{eq:pruet_nup_parameter}).

First, we estimate as a function of time the distance $R_{\rm ej}$ to the disk, the temperature $T_{\rm ej}$, and the density $\rho_{\rm ej}$ of a fluid element ejected from the disk with the initial distance to the black hole $R_0$, initial (local disk) temperature $T_0$, and (local disk) density at ejection $\rho_0$. We assume that the outflows, once unbound from the black hole-disk potential, approach roughly spherical expansion. The continuity equation $\nabla\cdot (\rho \mathbf{v})=0$ for the outflow with velocity $\mathbf{v}$ then implies $\rho_{\rm ej}\propto R_{\rm ej}^{-2}$ for $v=v_{\rm ej}=\text{const.}$. We assume adiabatic expansion, implying $T_{\rm ej} \propto \rho^{1/3}_{\rm ej}$ if relativistic particles ($e^\pm$, $\gamma$) dominate the specific entropy (Equations~\eqref{eq:entropy_gamma} and \eqref{eq:entropy_epm_general} with \eqref{eq:pressure_neutrinos}) or $T_{\rm ej} \propto \rho^{2/3}_{\rm ej}$ if baryons dominate the specific entropy (Equation~\eqref{eq:entropy_baryons}).

We assume a constant ejection speed $v_{\rm ej} = v_{\infty}$, given by the asymptotic Lorentz factor $\Gamma_\infty = -hu_0$. In launching the disk outflows, we follow the results of numerical simulations \citep{fernandez_delayed_2013,siegel_three-dimensional_2018,agarwal_ignition_2025} and assume that the disk winds are typically marginally unbound at launch ($hu_0 \simeq -1$) and that the outflow speeds are primarily set by $\alpha$-particle recombination, i.e.~we set (compare Appendix~\ref{app:EOS}) $h = 1 + \epsilon_{\rm bind}$, with $\epsilon_{\rm bind} = E_{\rm bind}(1-X_{f,\text{rec}})/(4 m_{\rm b})$ once the outflows are launched. The mass fraction $X_{f,\text{rec}}$ of free nucleons remaining after recombination is determined by the number of neutrons under neutron-rich conditions and by the number of protons under proton-rich conditions, leading to $X_{f,\text{rec}} = |1-2Y_{e0}|$, where we take $Y_{e0}$ as the local disk electron-fraction. For values of $Y_{e0}\sim 0.1-0.7$, one obtains ejecta velocities in the range of $v_{\rm ej}=v_\infty \simeq 0.05-0.12c$. The corresponding expansion timescale is given by $\tau \approx R_\mathrm{ej}/v_\mathrm{ej}$.

To compute $\Delta_n$, one needs to determine the ratio $Y_p/Y_{\rm heavy}$, the reaction rate $\lambda_{\bar{\nu}_e p}$, and the time window in which the ejecta stays in the temperature regime between $1-3$\,GK.

At the end of the recombination process and assuming proton-rich conditions, $X_p = Y_p = X_{f,\text{rec}}=|1-2Y_{e0}|$, since all neutrons are either integrated into $\alpha$-particles or into heavier nuclei. Therefore, we can replace the ratio in Equation \eqref{eq:pruet_nup_parameter} by
\begin{equation}\label{eq:mass_ratio}
    \frac{Y_p}{Y_{\mathrm{heavy}}} = \bar{A}\frac{X_p}{X_{\mathrm{heavy}}},
\end{equation}
where $\bar{A}\approx 60$ is the average mass of the heavy seed particles formed after $\alpha$ recombination (Section~\ref{sec:results_proton-rich-nucleosynthesis}; \citealt{frohlich_neutrino-induced_2006,pruet_nucleosynthesis_2006}). The ratio $X_p/X_{\mathrm{heavy}}$, in turn, strongly depends on entropy ($\propto s^3$; \citealt{hoffman_nucleosynthesis_1997}. Given the typical entropies of $\langle s \rangle \gtrsim 100$ in regions with $Y_e > 0.5$ (Figures \ref{fig:ye_map} and \ref{fig:multi_map}), $X_{\rm heavy} \approx 0.01(1-X_p)$ provides a good approximation to the mass fraction of seeds at the onset of the $\nu$p-process. This estimate is based on results from nucleosynthesis calculations performed with nuclear reaction networks, which show that for entropies $s \sim 100$, only $\lesssim1$\% of the mass is in atoms heavier than $^4\mathrm{He}$ (`$\alpha$-rich freeze-out'; \citealt{hoffman_nucleosynthesis_1997}).

In order to estimate the reaction rate $\lambda_{\bar{\nu}_e p}$, we model the accretion disk as a neutrino `light bulb' that irradiates the outflows using the approximation \citep{qian_nucleosynthesis_1996}
\begin{align}\label{eq:antineutrino_rate}
    \lambda_{\bar{\nu}_e p} &\approx \frac{1+3g_a^2}{4\pi^2}\frac{G_F^2}{(\hbar c)^4} \frac{L_\nuae}{R_{\rm ej}^2}
    \left(1.2\langle E_\nuea\rangle -2Q + \frac{Q^2}{\langle E_\nuea\rangle}\right)\\
    &\approx 0.06 \left(\frac{L_\nuae}{10^{52}\mathrm{erg} \text{ }\mathrm{ s^{-1}}}\right)\left(\frac{R_{\rm ej}}{10^8 \rm cm}\right)^{-2}\left( \frac{\langle E_\nuea\rangle}{10 \rm MeV} - 0.216\right),
\end{align}
where $L_\nuae$ is the antineutrino luminosity of the disk and $\langle E_\nuea\rangle$ is the corresponding average antineutrino energy.

The average antineutrino energy is given by $\langle E_\nuea\rangle = F^-_\nuea / \dot{n}_\nuea$, where the energy and number flux $F_\nuae^-$ and $\dot{n}_\nuea$ are given by the electron antineutrino expressions of Equations \eqref{eq:cooling_transition} and \eqref{eq:n_dot_transition}, respectively. We compute the total disk contribution to the above expression by radial integration,
\begin{equation}
    L_\nuea \left(1.2\langle E_\nuea\rangle -2Q + \frac{Q^2}{\langle E_\nuea\rangle}\right) =\nonumber
    \int_{r_{\rm ISCO}}^{r_{\rm out}}  2\pi r \left[1.2 \frac{(F^-_\nuea(r))^2}{\dot{n}_\nuea(r)} -2QF^-_\nuea(r) + Q^2 \dot{n}_\nuea(r)\right] \dr,
\end{equation}
where we take $r_{\rm out}$ as the outer boundary of the disk. Our results are insensitive to the choice of $r_{\rm out}$ as long as $r_{\rm out}\gg r_{\rm g}$ so that it encompasses the regions that dominate the neutrino flux.

To estimate the prospects for a successful $\nu p$-process, we take ejecta conditions in the most optimistic scenario for a $\nu p$-process: We follow fluid elements ejected at $R_\mathrm{launch}=r(Y_e = \mathrm{max}(Y_e))$ from the disk, i.e.~launched from the region in the disk with the most proton-rich conditions. We use the previously described outflow model to determine the time window in which the ejecta reside between $T=3$\,GK and $T=1.5$\,GK, together with the corresponding distances $R_{\rm ej}$. These time limits are used to solve the integral in Equation~\eqref{eq:pruet_nup_parameter} and the corresponding trajectory $R_{\rm ej}(t)$ is used in Equation~\eqref{eq:antineutrino_rate}.

\bibliography{gr-astro}
\bibliographystyle{aasjournalv7}

\end{document}